\documentclass[longauth]{aa} 

\usepackage{graphicx}
\usepackage{txfonts}
\usepackage{tabularx}
\usepackage{multirow}
\usepackage{booktabs}
\usepackage{float}
\usepackage{xcolor}
\usepackage{comment}
\bibpunct{(}{)}{;}{a}{}{,} 
\usepackage[breaklinks, colorlinks, citecolor=blue, linkcolor=blue]{hyperref}
\makeatletter
\renewcommand*\aa@pageof{, page \thepage{} of \pageref*{LastPage}}
\makeatother

\begin{document}

\title{Characterisation of the TOI-421 planetary system using CHEOPS, TESS, and archival radial velocity data \thanks{Based on data from CHEOPS Guaranteed Time Observations, collected under Programme ID CH\textunderscore PR100024. The raw and detrended photometric time series data are available in electronic form at the CDS via anonymous ftp to cdsarc.cds.unistra.fr (130.79.128.5) or via \url{https://cdsarc.cds.unistra.fr/cgi-bin/qcat?J/A+A/}}}

\subtitle{}

\author{A. F. Krenn\inst{1} $^{\href{https://orcid.org/0000-0003-3615-4725}{\includegraphics[scale=0.5]{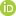}}}$
\and
D. Kubyshkina\inst{1} $^{\href{https://orcid.org/0000-0001-9137-9818}{\includegraphics[scale=0.5]{figures/orcid.jpg}}}$\and 
L. Fossati\inst{1} $^{\href{https://orcid.org/0000-0003-4426-9530}{\includegraphics[scale=0.5]{figures/orcid.jpg}}}$\and 
J. A. Egger\inst{2} $^{\href{https://orcid.org/0000-0003-1628-4231}{\includegraphics[scale=0.5]{figures/orcid.jpg}}}$\and 
A. Bonfanti\inst{1} $^{\href{https://orcid.org/0000-0002-1916-5935}{\includegraphics[scale=0.5]{figures/orcid.jpg}}}$\and 
A. Deline\inst{3}\and 
D. Ehrenreich\inst{3,4} $^{\href{https://orcid.org/0000-0001-9704-5405}{\includegraphics[scale=0.5]{figures/orcid.jpg}}}$\and 
M. Beck\inst{3} $^{\href{https://orcid.org/0000-0003-3926-0275}{\includegraphics[scale=0.5]{figures/orcid.jpg}}}$\and 
W. Benz\inst{2,5} $^{\href{https://orcid.org/0000-0001-7896-6479}{\includegraphics[scale=0.5]{figures/orcid.jpg}}}$\and 
J. Cabrera\inst{6} $^{\href{https://orcid.org/0000-0001-6653-5487}{\includegraphics[scale=0.5]{figures/orcid.jpg}}}$\and 
T. G. Wilson\inst{7} $^{\href{https://orcid.org/0000-0001-8749-1962}{\includegraphics[scale=0.5]{figures/orcid.jpg}}}$\and 
A. Leleu\inst{2,3} $^{\href{https://orcid.org/0000-0003-2051-7974}{\includegraphics[scale=0.5]{figures/orcid.jpg}}}$\and 
S. G. Sousa\inst{8} $^{\href{https://orcid.org/0000-0001-9047-2965}{\includegraphics[scale=0.5]{figures/orcid.jpg}}}$\and 
V. Adibekyan\inst{8} $^{\href{https://orcid.org/0000-0002-0601-6199}{\includegraphics[scale=0.5]{figures/orcid.jpg}}}$\and 
A. C. M. Correia\inst{9} $^{\href{https://orcid.org/0000-0002-8946-8579}{\includegraphics[scale=0.5]{figures/orcid.jpg}}}$\and 
Y. Alibert\inst{5,2} $^{\href{https://orcid.org/0000-0002-4644-8818}{\includegraphics[scale=0.5]{figures/orcid.jpg}}}$\and 
L. Delrez\inst{10,11} $^{\href{https://orcid.org/0000-0001-6108-4808}{\includegraphics[scale=0.5]{figures/orcid.jpg}}}$\and 
M. Lendl\inst{3} $^{\href{https://orcid.org/0000-0001-9699-1459}{\includegraphics[scale=0.5]{figures/orcid.jpg}}}$\and 
J. A. Patel\inst{12}\and 
J. Venturini\inst{3} $^{\href{https://orcid.org/0000-0001-9527-2903}{\includegraphics[scale=0.5]{figures/orcid.jpg}}}$\and 
R. Alonso\inst{13,14} $^{\href{https://orcid.org/0000-0001-8462-8126}{\includegraphics[scale=0.5]{figures/orcid.jpg}}}$\and 
G. Anglada\inst{15,16} $^{\href{https://orcid.org/0000-0002-3645-5977}{\includegraphics[scale=0.5]{figures/orcid.jpg}}}$\and 
J. Asquier\inst{17}\and 
T. Bárczy\inst{18} $^{\href{https://orcid.org/0000-0002-7822-4413}{\includegraphics[scale=0.5]{figures/orcid.jpg}}}$\and 
D. Barrado Navascues\inst{19} $^{\href{https://orcid.org/0000-0002-5971-9242}{\includegraphics[scale=0.5]{figures/orcid.jpg}}}$\and 
S. C. C. Barros\inst{8,20} $^{\href{https://orcid.org/0000-0003-2434-3625}{\includegraphics[scale=0.5]{figures/orcid.jpg}}}$\and 
W. Baumjohann\inst{1} $^{\href{https://orcid.org/0000-0001-6271-0110}{\includegraphics[scale=0.5]{figures/orcid.jpg}}}$\and 
T. Beck\inst{2}\and 
N. Billot\inst{3} $^{\href{https://orcid.org/0000-0003-3429-3836}{\includegraphics[scale=0.5]{figures/orcid.jpg}}}$\and 
X. Bonfils\inst{21} $^{\href{https://orcid.org/0000-0001-9003-8894}{\includegraphics[scale=0.5]{figures/orcid.jpg}}}$\and 
L. Borsato\inst{22} $^{\href{https://orcid.org/0000-0003-0066-9268}{\includegraphics[scale=0.5]{figures/orcid.jpg}}}$\and 
A. Brandeker\inst{12} $^{\href{https://orcid.org/0000-0002-7201-7536}{\includegraphics[scale=0.5]{figures/orcid.jpg}}}$\and 
C. Broeg\inst{2,5} $^{\href{https://orcid.org/0000-0001-5132-2614}{\includegraphics[scale=0.5]{figures/orcid.jpg}}}$\and 
S. Charnoz\inst{23} $^{\href{https://orcid.org/0000-0002-7442-491X}{\includegraphics[scale=0.5]{figures/orcid.jpg}}}$\and 
A. Collier Cameron\inst{7} $^{\href{https://orcid.org/0000-0002-8863-7828}{\includegraphics[scale=0.5]{figures/orcid.jpg}}}$\and 
Sz. Csizmadia\inst{6} $^{\href{https://orcid.org/0000-0001-6803-9698}{\includegraphics[scale=0.5]{figures/orcid.jpg}}}$\and 
P. E. Cubillos\inst{1,24}\and 
M. B. Davies\inst{25} $^{\href{https://orcid.org/0000-0001-6080-1190}{\includegraphics[scale=0.5]{figures/orcid.jpg}}}$\and 
M. Deleuil\inst{26} $^{\href{https://orcid.org/0000-0001-6036-0225}{\includegraphics[scale=0.5]{figures/orcid.jpg}}}$\and 
O. D. S. Demangeon\inst{8,20} $^{\href{https://orcid.org/0000-0001-7918-0355}{\includegraphics[scale=0.5]{figures/orcid.jpg}}}$\and 
B.-O. Demory\inst{2,5} $^{\href{https://orcid.org/0000-0002-9355-5165}{\includegraphics[scale=0.5]{figures/orcid.jpg}}}$\and 
A. Erikson\inst{6}\and 
A. Fortier\inst{2,5} $^{\href{https://orcid.org/0000-0001-8450-3374}{\includegraphics[scale=0.5]{figures/orcid.jpg}}}$\and 
M. Fridlund\inst{27,28} $^{\href{https://orcid.org/0000-0002-0855-8426}{\includegraphics[scale=0.5]{figures/orcid.jpg}}}$\and 
D. Gandolfi\inst{29} $^{\href{https://orcid.org/0000-0001-8627-9628}{\includegraphics[scale=0.5]{figures/orcid.jpg}}}$\and 
M. Gillon\inst{10} $^{\href{https://orcid.org/0000-0003-1462-7739}{\includegraphics[scale=0.5]{figures/orcid.jpg}}}$\and 
M. Güdel\inst{30}\and 
M. N. Günther\inst{17} $^{\href{https://orcid.org/0000-0002-3164-9086}{\includegraphics[scale=0.5]{figures/orcid.jpg}}}$\and 
J. Hasiba\inst{1}\and 
A. Heitzmann\inst{3}\and 
C. Helling\inst{1,31}\and 
S. Hoyer\inst{26} $^{\href{https://orcid.org/0000-0003-3477-2466}{\includegraphics[scale=0.5]{figures/orcid.jpg}}}$\and 
K. G. Isaak\inst{17} $^{\href{https://orcid.org/0000-0001-8585-1717}{\includegraphics[scale=0.5]{figures/orcid.jpg}}}$\and 
L. L. Kiss\inst{32,33}\and 
K. W. F. Lam\inst{6} $^{\href{https://orcid.org/0000-0002-9910-6088}{\includegraphics[scale=0.5]{figures/orcid.jpg}}}$\and 
J. Laskar\inst{34} $^{\href{https://orcid.org/0000-0003-2634-789X}{\includegraphics[scale=0.5]{figures/orcid.jpg}}}$\and 
A. Lecavelier des Etangs\inst{35} $^{\href{https://orcid.org/0000-0002-5637-5253}{\includegraphics[scale=0.5]{figures/orcid.jpg}}}$\and 
C. Lovis\inst{3} $^{\href{https://orcid.org/0000-0001-7120-5837}{\includegraphics[scale=0.5]{figures/orcid.jpg}}}$\and 
D. Magrin\inst{22} $^{\href{https://orcid.org/0000-0003-0312-313X}{\includegraphics[scale=0.5]{figures/orcid.jpg}}}$\and 
P. F. L. Maxted\inst{36} $^{\href{https://orcid.org/0000-0003-3794-1317}{\includegraphics[scale=0.5]{figures/orcid.jpg}}}$\and 
C. Mordasini\inst{2,5}\and 
V. Nascimbeni\inst{22} $^{\href{https://orcid.org/0000-0001-9770-1214}{\includegraphics[scale=0.5]{figures/orcid.jpg}}}$\and 
G. Olofsson\inst{12} $^{\href{https://orcid.org/0000-0003-3747-7120}{\includegraphics[scale=0.5]{figures/orcid.jpg}}}$\and 
R. Ottensamer\inst{30}\and 
I. Pagano\inst{37} $^{\href{https://orcid.org/0000-0001-9573-4928}{\includegraphics[scale=0.5]{figures/orcid.jpg}}}$\and 
E. Pallé\inst{13,14} $^{\href{https://orcid.org/0000-0003-0987-1593}{\includegraphics[scale=0.5]{figures/orcid.jpg}}}$\and 
G. Peter\inst{38} $^{\href{https://orcid.org/0000-0001-6101-2513}{\includegraphics[scale=0.5]{figures/orcid.jpg}}}$\and 
G. Piotto\inst{22,39} $^{\href{https://orcid.org/0000-0002-9937-6387}{\includegraphics[scale=0.5]{figures/orcid.jpg}}}$\and 
D. Pollacco\inst{40}\and 
D. Queloz\inst{41,42} $^{\href{https://orcid.org/0000-0002-3012-0316}{\includegraphics[scale=0.5]{figures/orcid.jpg}}}$\and 
R. Ragazzoni\inst{22,39} $^{\href{https://orcid.org/0000-0002-7697-5555}{\includegraphics[scale=0.5]{figures/orcid.jpg}}}$\and 
N. Rando\inst{17}\and 
H. Rauer\inst{6,43,44} $^{\href{https://orcid.org/0000-0002-6510-1828}{\includegraphics[scale=0.5]{figures/orcid.jpg}}}$\and 
I. Ribas\inst{15,16} $^{\href{https://orcid.org/0000-0002-6689-0312}{\includegraphics[scale=0.5]{figures/orcid.jpg}}}$\and 
M. Rieder\inst{5,45}\and 
N. C. Santos\inst{8,20} $^{\href{https://orcid.org/0000-0003-4422-2919}{\includegraphics[scale=0.5]{figures/orcid.jpg}}}$\and 
G. Scandariato\inst{37} $^{\href{https://orcid.org/0000-0003-2029-0626}{\includegraphics[scale=0.5]{figures/orcid.jpg}}}$\and 
D. Ségransan\inst{3} $^{\href{https://orcid.org/0000-0003-2355-8034}{\includegraphics[scale=0.5]{figures/orcid.jpg}}}$\and 
A. E. Simon\inst{2} $^{\href{https://orcid.org/0000-0001-9773-2600}{\includegraphics[scale=0.5]{figures/orcid.jpg}}}$\and 
A. M. S. Smith\inst{6} $^{\href{https://orcid.org/0000-0002-2386-4341}{\includegraphics[scale=0.5]{figures/orcid.jpg}}}$\and 
M. Stalport\inst{46}\and 
M. Steller\inst{1} $^{\href{https://orcid.org/0000-0003-2459-6155}{\includegraphics[scale=0.5]{figures/orcid.jpg}}}$\and 
Gy. M. Szabó\inst{47,48}\and 
N. Thomas\inst{2}\and 
S. Udry\inst{3} $^{\href{https://orcid.org/0000-0001-7576-6236}{\includegraphics[scale=0.5]{figures/orcid.jpg}}}$\and 
B. Ulmer\inst{38}\and 
V. Van Grootel\inst{11} $^{\href{https://orcid.org/0000-0003-2144-4316}{\includegraphics[scale=0.5]{figures/orcid.jpg}}}$\and 
E. Villaver\inst{13,14}\and 
V. Viotto\inst{22} $^{\href{https://orcid.org/0000-0001-5700-9565}{\includegraphics[scale=0.5]{figures/orcid.jpg}}}$\and 
N. A. Walton\inst{49} $^{\href{https://orcid.org/0000-0003-3983-8778}{\includegraphics[scale=0.5]{figures/orcid.jpg}}}$\and 
T. Zingales\inst{22,39} $^{\href{https://orcid.org/0000-0001-6880-5356}{\includegraphics[scale=0.5]{figures/orcid.jpg}}}$
}

\authorrunning{A. F. Krenn et al.}

\institute{
\label{inst:1} Space Research Institute, Austrian Academy of Sciences, Schmiedlstrasse 6, A-8042 Graz, Austria \\
\email{andreas.krenn@oeaw.ac.at} \and
\label{inst:2} Weltraumforschung und Planetologie, Physikalisches Institut, University of Bern, Gesellschaftsstrasse 6, 3012 Bern, Switzerland \and
\label{inst:3} Observatoire Astronomique de l'Université de Genève, Chemin Pegasi 51, 1290 Versoix, Switzerland \and
\label{inst:4} Centre Vie dans l’Univers, Faculté des sciences, Université de Genève, Quai Ernest-Ansermet 30, 1211 Genève 4, Switzerland \and
\label{inst:5} Centre for Space and Habitability, University of Bern, Gesellschaftsstrasse 6, 3012 Bern, Switzerland \and
\label{inst:6} Institute of Planetary Research, German Aerospace Centre (DLR), Rutherfordstrasse 2, 12489 Berlin, Germany \and
\label{inst:7} Centre for Exoplanet Science, SUPA School of Physics and Astronomy, University of St Andrews, North Haugh, St Andrews KY16 9SS, UK \and
\label{inst:8} Instituto de Astrofisica e Ciencias do Espaco, Universidade do Porto, CAUP, Rua das Estrelas, 4150-762 Porto, Portugal \and
\label{inst:9} CFisUC, Departamento de F\'isica, Universidade de Coimbra, 3004-516 Coimbra, Portugal \and
\label{inst:10} Astrobiology Research Unit, Université de Liège, Allée du 6 Août 19C, B-4000 Liège, Belgium \and
\label{inst:11} Space sciences, Technologies and Astrophysics Research (STAR) Institute, Université de Liège, Allée du 6 Août 19C, 4000 Liège, Belgium \and
\label{inst:12} Department of Astronomy, Stockholm University, AlbaNova University Centre, 10691 Stockholm, Sweden \and
\label{inst:13} Instituto de Astrofisica de Canarias, Via Lactea s/n, 38200 La Laguna, Tenerife, Spain \and
\label{inst:14} Departamento de Astrofisica, Universidad de La Laguna, Astrofísico Francisco Sanchez s/n, 38206 La Laguna, Tenerife, Spain \and
\label{inst:15} Institut de Ciencies de l'Espai (ICE, CSIC), Campus UAB, Can Magrans s/n, 08193 Bellaterra, Spain \and
\label{inst:16} Institut d’Estudis Espacials de Catalunya (IEEC), Gran Capità 2-4, 08034 Barcelona, Spain \and
\label{inst:17} European Space Agency (ESA), European Space Research and Technology Centre (ESTEC), Keplerlaan 1, 2201 AZ Noordwijk, The Netherlands \and
\label{inst:18} Admatis, 5. Kandó Kálmán Street, 3534 Miskolc, Hungary \and
\label{inst:19} Depto. de Astrofisica, Centro de Astrobiologia (CSIC-INTA), ESAC campus, 28692 Villanueva de la Cañada (Madrid), Spain \and
\label{inst:20} Departamento de Fisica e Astronomia, Faculdade de Ciencias, Universidade do Porto, Rua do Campo Alegre, 4169-007 Porto, Portugal \and
\label{inst:21} Université Grenoble Alpes, CNRS, IPAG, 38000 Grenoble, France \and
\label{inst:22} INAF, Osservatorio Astronomico di Padova, Vicolo dell'Osservatorio 5, 35122 Padova, Italy \and
\label{inst:23} Université de Paris Cité, Institut de physique du globe de Paris, CNRS, 1 Rue Jussieu, F-75005 Paris, France \and
\label{inst:24} INAF, Osservatorio Astrofisico di Torino, Via Osservatorio, 20, I-10025 Pino Torinese To, Italy \and
\label{inst:25} Centre for Mathematical Sciences, Lund University, Box 118, 221 00 Lund, Sweden \and
\label{inst:26} Aix Marseille Univ, CNRS, CNES, LAM, 38 rue Frédéric Joliot-Curie, 13388 Marseille, France \and
\label{inst:27} Leiden Observatory, University of Leiden, PO Box 9513, 2300 RA Leiden, The Netherlands \and
\label{inst:28} Department of Space, Earth and Environment, Chalmers University of Technology, Onsala Space Observatory, 439 92 Onsala, Sweden \and
\label{inst:29} Dipartimento di Fisica, Universita degli Studi di Torino, via Pietro Giuria 1, I-10125, Torino, Italy \and
\label{inst:30} Department of Astrophysics, University of Vienna, Türkenschanzstrasse 17, 1180 Vienna, Austria \and
\label{inst:31} Institute for Theoretical Physics and Computational Physics, Graz University of Technology, Petersgasse 16, 8010 Graz, Austria \and
\label{inst:32} Konkoly Observatory, Research Centre for Astronomy and Earth Sciences, 1121 Budapest, Konkoly Thege Miklós út 15-17, Hungary \and
\label{inst:33} ELTE E\"otv\"os Lor\'and University, Institute of Physics, P\'azm\'any P\'eter s\'et\'any 1/A, 1117 Budapest, Hungary \and
\label{inst:34} IMCCE, UMR8028 CNRS, Observatoire de Paris, PSL Univ., Sorbonne Univ., 77 av. Denfert-Rochereau, 75014 Paris, France \and
\label{inst:35} Institut d'astrophysique de Paris, UMR7095 CNRS, Université Pierre \& Marie Curie, 98bis blvd. Arago, 75014 Paris, France \and
\label{inst:36} Astrophysics Group, Lennard Jones Building, Keele University, Staffordshire, ST5 5BG, United Kingdom \and
\label{inst:37} INAF, Osservatorio Astrofisico di Catania, Via S. Sofia 78, 95123 Catania, Italy \and
\label{inst:38} Institute of Optical Sensor Systems, German Aerospace Centre (DLR), Rutherfordstrasse 2, 12489 Berlin, Germany \and
\label{inst:39} Dipartimento di Fisica e Astronomia "Galileo Galilei", Universita degli Studi di Padova, Vicolo dell'Osservatorio 3, 35122 Padova, Italy \and
\label{inst:40} Department of Physics, University of Warwick, Gibbet Hill Road, Coventry CV4 7AL, United Kingdom \and
\label{inst:41} ETH Zurich, Department of Physics, Wolfgang-Pauli-Strasse 2, CH-8093 Zurich, Switzerland \and
\label{inst:42} Cavendish Laboratory, JJ Thomson Avenue, Cambridge CB3 0HE, UK \and
\label{inst:43} Zentrum für Astronomie und Astrophysik, Technische Universität Berlin, Hardenbergstr. 36, D-10623 Berlin, Germany \and
\label{inst:44} Institut fuer Geologische Wissenschaften, Freie Universitaet Berlin, Maltheserstrasse 74-100,12249 Berlin, Germany \and
\label{inst:45} Weltraumforschung und Planetologie, Physikalisches Institut, University of Bern, Sidlerstrasse 5, 3012 Bern, Switzerland \and
\label{inst:46} Université de Liège, Allée du 6 Août 19C, 4000 Liège, Belgium \and
\label{inst:47} ELTE E\"otv\"os Lor\'and University, Gothard Astrophysical Observatory, 9700 Szombathely, Szent Imre h. u. 112, Hungary \and
\label{inst:48} MTA-ELTE Exoplanet Research Group, 9700 Szombathely, Szent Imre h. u. 112, Hungary \and
\label{inst:49} Institute of Astronomy, University of Cambridge, Madingley Road, Cambridge, CB3 0HA, United Kingdom
}

\date{Received 13.11.2023; accepted 27.03.2024}

 
  \abstract
   {The TOI-421 planetary system contains two sub-Neptune-type planets ($P_{\rm b} \sim5.2$ days, $T_{\rm eq,b} \sim900$ K, and $P_{\rm c} \sim16.1$ days, $T_{\rm eq,c} \sim650$ K) and is a prime target to study the formation and evolution of planets and their atmospheres. The inner planet is especially interesting as the existence of a hydrogen-dominated atmosphere at its orbital separation cannot be explained by current formation models without previous orbital migration.}
   {We aim to improve the system parameters to further use them to model the interior structure and simulate the atmospheric evolution of both planets, to finally gain insights into their formation and evolution. We also investigate the possibility of detecting transit timing variations (TTVs).}
   {We jointly analysed photometric data of three TESS sectors and six CHEOPS visits as well as 156 radial velocity data points to retrieve improved planetary parameters. We also searched for TTVs and modelled the interior structure of the planets. Finally, we simulated the evolution of the primordial H-He atmospheres of the planets using two different modelling frameworks.}
   {We determine the planetary radii and masses of TOI-421\,b and c to be $R_{\rm b} = 2.64 \pm 0.08 \, R_{\oplus}$, $M_{\rm b} = 6.7 \pm 0.6 \, M_{\oplus}$, $R_{\rm c} = 5.09 \pm 0.07 \, R_{\oplus}$, and $M_{\rm c} = 14.1 \pm 1.4 \, M_{\oplus}$. Using these results we retrieved average planetary densities of $\rho_{\rm b} = 0.37 \pm 0.05 \, \rho_{\oplus}$ and $\rho_{\rm c} = 0.107 \pm 0.012 \, \rho_{\oplus}$. We do not detect any statistically significant TTV signals. Assuming the presence of a hydrogen-dominated atmosphere, the interior structure modelling results in both planets having extensive envelopes. While the modelling of the atmospheric evolution predicts for TOI-421\,b to have lost any primordial atmosphere that it could have accreted at its current orbital position, TOI-421\,c could have started out with an initial atmospheric mass fraction somewhere between $10$ and $35\%$.}
   {We conclude that the low observed mean density of TOI-421\,b can only be explained by either a bias in the measured planetary parameters (e.g. driven by high-altitude clouds) and/or in the context of orbital migration. We also find that the results of atmospheric evolution models are strongly dependent on the employed planetary structure model.}


\maketitle
%

\section{Introduction}
\label{sec_introduction}

The exact mechanisms of planet formation and evolution are still highly debated. For example, the classical core-accretion model \citep[e.g.][]{Bodenheimer1986,Pollack1996} has been challenged by more recent theories predicting much faster planet formation, such as the pebble accretion model \citep{Johansen2010,Ormel2010,Lambrechts2012,Johansen2015}. A large sample of diverse planets with a mass and radius measured at at least $3\sigma$ significance is needed to test and compare different formation and evolution models. Efforts to compile such a sample have recently been led by the Transiting Exoplanet Survey Satellite \citep[TESS;][]{Ricker2014} and the CHaracterising ExOPlanet Satellite \citep[CHEOPS;][]{Benz2021}, in close collaboration with large ground-based observing campaigns conducted with high-resolution spectrographs such as the High Accuracy Radial velocity Planet Searcher \citep[HARPS;][]{Mayor2003}, its twin HARPS-N \citep{cosentino2012}, the Calar Alto high-Resolution search for M dwarfs with Exoearths with Near-infrared and optical Echelle Spectrographs \citep[CARMENS;][]{quirrenbach2014}, and the Echelle SPectrograph for Rocky Exoplanet and Stable Spectroscopic Observations \citep[ESPRESSO;][]{Pepe2021}. More than 5500 exoplanets have been confirmed so far and over 950 of them have both radii and masses measured at more than 3$\sigma$ confidence. 

To test different formation models, we must also understand how planets evolve. Processes such as atmospheric mass loss and orbital migration can have a significant impact on planet parameters such as mass and radius over the course of the evolution. Multi-planet systems are especially important in the context of constraining planet formation and evolution as they allow us to compare planets with different sizes and compositions in the same system, thus they are illuminated by the same star. The prime targets for investigating the evolution of primordial planetary atmospheres are planets that have retained some part of their initial hydrogen-dominated envelope accreted during their formation, while also having been subject to strong mass loss driven by internal thermal energy and/or absorption of high-energy -- X-ray and extreme ultraviolet (EUV; together XUV) -- stellar radiation. The almost 11 Gyrs old system hosting two sub-Neptune-like planets TOI-421 \citep{Carleo2020} checks all of this criteria and therefore has become a gold-standard target to study planetary atmospheric evolution and its implications for planet formation scenarios. 

\citet{Carleo2020} presented a first simulation of the atmospheric evolution using the tool presented by \citet{Kubyshkina2019a,Kubyshkina2019b} and assuming a hydrogen-dominated atmosphere and a formation of the planets at the orbital separation at which they are currently observed. The tool simultaneously constrains the initial atmospheric mass fraction and the evolution of the stellar rotation rate, where the latter acts as a proxy for the XUV stellar emission. They find that their models provide no additional constraint on the stellar rotation history of TOI-421 upon that retrieved from observations of young open cluster stars. For planet c, they find an initial atmospheric mass fraction of roughly 30\% and a present-day atmospheric mass fraction of roughly 10\%, meaning the planet would have lost about two-thirds of its primordial atmosphere over the course of its evolution due to the absorption of XUV radiation. Instead, for TOI-421\,b they find that the existence of a planet as observed by \citet{Carleo2020} is not possible within the framework of the employed model. Given the measured planetary parameters, the planet should have completely lost any hydrogen-dominated atmosphere within the first gigayear, independent of the rotational evolution of the host star. According to their simulations, the planet would either have to orbit about twice as far out than measured or could not host a hydrogen-dominated atmosphere.

TOI-421\,b is not the first planet for which a discrepancy is observed between the low measured bulk density, compatible with the presence of a hydrogen-dominated atmosphere, and the predicted mass loss rates, which are too large to enable the presence of a low-density envelope. \citet{Lammer2016} find the same problem for CoRoT-24b and \citet{Cubillos2017} find that about 15\% of mini-Neptunes out of a large sample detected mostly by the Kepler satellite also share this peculiar property. \citet{Carleo2020} already mention a few possible solutions to this problem. Most importantly, the considered atmospheric evolution framework assumes that planets form within the protoplanetary nebula at the observed orbital separation. A formation further out and a subsequent orbital migration to the present-day orbit would allow for both more hydrogen gas being accreted during the time before the dispersal of the nebula and possibly lower XUV fluxes during the earlier stages of planet evolution (particularly in the case of post-disk migration). Another possible explanation is the presence of high-altitude aerosols, which could lead to an overestimate of the planetary radius \citep{Lammer2016,Cubillos2017,Gao2020}. This possibility finds some support in both observations \citep[e.g.][]{libby2020,chachan2020} and modelling \citep[e.g.][]{wang2019,ohno2021}. There might also be a replenishment of light gases released from the crust into the atmosphere, which would in turn counteract the effect of escape \citep[e.g.][]{Kite2019}. Finally, they also mention the possibility that there are biases in the observed planetary parameters; for example, undetected planets in the system might affect the mass measurements. 

\citet{kubyshkina2022_tois_I,kubyshkina2022_tois_II} use the planet parameters derived by \citet{Carleo2020} to model the atmospheric evolution of the planets using more sophisticated models. \citet{kubyshkina2022_tois_I} first show that the effects of stellar wind on the overall mass loss of the planets are negligible, justifying the use of evolutionary models that do not account for stellar winds. \citet{kubyshkina2022_tois_II} then expand the atmospheric evolution framework already employed in \citet{Carleo2020} by additionally accounting for planetary internal thermal evolution. They find similar results as \citet{Carleo2020}: while within their model planet c has started out with an initial atmospheric mass fraction of $0.30 \pm 0.07$, planet b would have had to start out with a very significant initial atmospheric mass fraction lying somewhere between $0.48$ and $0.71$. However, as mentioned before in \citet{Carleo2020}, \citet{kubyshkina2022_tois_II} also conclude that such a high initial atmospheric mass fraction is unlikely for such a planet. Using a model by \citet{mordasini2020} to estimate the amount of hydrogen a planetary core could accrete before the dispersal of the protoplanetary nebula as a function of mass and orbital separation, they find that planet b could not have formed with such an extensive initial envelope within the ice line, implying a formation scenario beyond the ice line. \citet{Lopez2017} show that in the case of a formation beyond the ice line, the presence of water vapour in the atmosphere could significantly reduce the heating efficiency and thus the mass loss rates. This would also be consistent with most formation models, which predict sub-Neptunes to be water-rich \citep{Alibert2013,Raymond2018,Venturini2020,Izidoro2021b,Emenshuber2021}. 

Finally, it must also be noted that none of the current atmospheric evolution models have taken the effects of the possible presence of a planetary magnetic field into account. The consequence of the presence of a planetary magnetic field for atmospheric mass loss is not trivial. It is believed that the escape is reduced by the confinement of ionised atmospheric species within the closed magnetic field lines, while it is enhanced by the escape of atmospheric ions through the polar regions of the open magnetic lines and re-connection on the night-side \citep{Khodachenko2015,Sakai2018,Carolan2021}.

In this work, we present new space-based photometric data of both planets obtained by CHEOPS and TESS and include them in a global analysis of both photometric and radial velocity datasets. By the addition of data obtained by a second, independent photometric instrument, the high precision photometric measurements obtained by CHEOPS, and the additional transits observed by TESS, we aim to decrease the uncertainties on the system parameters. We investigate the possibility of detecting TTVs. We model the interior structures of the planets within an Bayesian framework and use the updated planetary parameters to constrain different kinds of atmospheric evolution models that assume that both planets host a hydrogen-dominated atmosphere. We compare the performance and results of the different methods to both highlight their advantages and disadvantages as well as better understand the possible past evolution of TOI-421\,b, a sub-Neptune in an orbit that would imply catastrophic hydrodynamic escape of any kind of primordial hydrogen-dominated atmosphere. 


\section{Stellar characterisation}
\label{sec_stellar}

The spectroscopic stellar parameters ($T_{\mathrm{eff}}$, $\log g$, microturbulence velocity, [Fe/H]) were estimated using the ARES+MOOG methodology, which is described in detail in \citet[][]{Sousa-21, Sousa-14, Santos-13}. This was done by using the latest version of ARES \footnote{The last version, ARES v2, can be downloaded at \url{https://github.com/sousasag/ARES}} \citep{Sousa-07, Sousa-15} to consistently measure the equivalent widths (EW) of selected iron lines on the combined HARPS spectrum of TOI-421. We used the list of iron lines presented in \citet[][]{Sousa-08}. A minimisation process is then used in this analysis to find the ionisation and excitation equilibrium to finally obtain the atmospheric parameters $T_{\mathrm{eff}}$, $\log g$, and microturbulence velocity. This process makes use of a grid of Kurucz model atmospheres \citep{Kurucz1993} and the radiative transfer code MOOG \citep{Sneden-73}. We also derived a more accurate trigonometric surface gravity using recent GAIA data following the same procedure as described in \citet[][]{Sousa-21}, which provided a consistent value when compared with the spectroscopic surface gravity.

Using the aforementioned stellar atmospheric parameters, we determined the abundances of Na, Mg, Si, Ti, and Ni  following the classical curve-of-growth analysis method described in \citet{Adibekyan-12, Adibekyan-15}. Similar to the stellar parameter determination, we used ARES to measure the EWs of the spectral lines of these elements, and again used a grid of Kurucz model atmospheres along with the radiative transfer code MOOG to convert the EWs into abundances, assuming local thermodynamic equilibrium. Abundances of the elements are presented in Table~\ref{table_stellar}.

Using the results of our spectral analysis as priors, we built spectral energy distributions (SEDs) using stellar atmospheric models from two catalogues \citep{Kurucz1993,Castelli2003} to compute the stellar radius of TOI-421 via a MCMC modified infrared flux method \citep[IRFM;][]{Blackwell1977,Schanche2020}. From these SEDs, we compared synthetic and observed broadband photometry in the {\it Gaia} $G$, $G_\mathrm{BP}$, and $G_\mathrm{RP}$, 2MASS $J$, $H$, and $K$, and \textit{WISE} $W1$ and $W2$ \citep{Skrutskie2006,Wright2010,GaiaCollaboration2022} bandpasses to derive the stellar bolometric flux that is converted into effective temperature and angular diameter via the Stefan-Boltzmann law. Thus, we obtain the stellar radius by combining the angular diameter with the offset-corrected \textit{Gaia} parallax \citep{Lindegren2021}. As our SEDs were constructed using two catalogues, we account for model uncertainties using a Bayesian modelling averaging of the radius posterior distributions and report a weighted average in Table~\ref{table_stellar}.

The stellar effective temperature $T_{\mathrm{eff}}$, metallicity [Fe/H], and radius $R_{\star}$ along with their uncertainties constitute the basic input set for deriving both the stellar mass $M_{\star}$ and age $t_{\star}$ from stellar evolutionary models. To get a first pair of mass and age estimates, we applied the isochrone placement algorithm \citep{bonfanti2015,bonfanti2016}, which interpolates the input parameters within pre-computed grids of PARSEC\footnote{\textsl{PA}dova and T\textsl{R}ieste \textsl{S}tellar \textsl{E}volutioary \textsl{C}ode: \url{http://stev.oapd.inaf.it/cgi-bin/cmd}} v1.2S \citep{marigo2017} isochrones and tracks. In addition to the basic input set, we further use $v\sin{i}=1.8\pm1.0$ km\,s$^{-1}$ \citep{Carleo2020} as input to let the isochrone fitting work in synergy with gyrochronology as detailed in \citet{bonfanti2016} to better constrain mass and age. Additionally, a second pair of mass and age values was obtained with the Code Liègeois d'Évolution Stellaire \citep[CLES;][]{scuflaire2008}, which builds the evolutionary track best fitting the stellar input parameters following a Levenberg-Marquadt minimisation scheme \citep{salmon2021}. After checking the mutual consistency of the two respective pairs of mass and age values via the $\chi^2$-based criterion outlined in \citet{Bonfanti2021}, we finally merged the results and obtained $M_{\star}=0.833_{-0.054}^{+0.048}\,M_{\odot}$ and $t_{\star}=10.9_{-5.2}^{+2.9}$ Gyr. See \citet{Bonfanti2021} for further details about the specific statistical treatment.

\begin{table}
\caption{Adopted stellar parameters of TOI-421}             
\centering                          
\begin{tabular}{cccc}        
\toprule
\toprule                
Parameter & Value & Unit & Source \\    
\midrule                   
Gmag & $9.78$ & - & SWEET-Cat \\
Spectral Type & G9V & - & C2020 \\
$T_{\mathrm{eff}}$ & $5291\pm64$ & K & This work \\
$\log g$  & $4.48\pm0.03$ & - & This work \\
Radius & $0.866\pm0.006$ & $R_{\odot}$ & This work \\
Mass & $0.833^{+0.048}_{-0.054}$ & $M_{\odot}$ & This work \\
Density &$1.29\pm 0.15$ &  $\rho_\odot$ & This work \\
Age &  $10.9_{-5.2}^{+2.9}$ & Gyr & This work \\
Rotation Period &  $39.6 \pm 1.6$ &  days & This work \\
$\left[\mathrm{Fe/H}\right]$ & $-0.044\pm0.04$ & - & This work \\
$\left[\mathrm{Ni/H}\right]$ & $-0.04\pm0.03$ & - & This work \\
$\left[\mathrm{Mg/H}\right]$ & $0.02\pm0.04$ & - & This work \\
$\left[\mathrm{Si/H}\right]$ & $-0.05\pm0.05$ & - & This work \\
$\left[\mathrm{Ti/H}\right]$ & $0.04\pm0.06$ & - & This work \\
$\left[\mathrm{Na/H}\right]$ & $-0.05\pm0.04$ & - & This work \\
$v\sin{i}$ & $1.8\pm1.0$ & km\,s$^{-1}$ & C2020 \\
\bottomrule                               
\end{tabular}
\label{table_stellar}      
\tablefoot{The Gaia EDR3 ID of TOI-421 is 2984582227215748864. C2020 refers to \citet{Carleo2020}}
\end{table}


\begin{table*}
    \caption{Log of CHEOPS observations.}
    \centering
    \begin{tabular}{cccccc}
    \toprule
    \toprule
    Visit & Start Date & End Date & Duration & Efficiency & File Key \\
    No.    & [UTC]      & [UTC]  & [hours] & [\%]     & - \\
    \midrule
    &&&Planet b &\\
    1 & 2021-11-19 01:13 & 2021-11-19 13:18 & 12.09& 66.0& CH\textunderscore PR100024\textunderscore TG013601\textunderscore V0300 \\
    2 & 2021-12-10 03:32 & 2021-12-10 10:00 & 6.47& 90.2 & CH\textunderscore PR100024\textunderscore TG014701\textunderscore V0300 \\
    3 & 2022-01-10 08:00 & 2022-01-10 15:05& 7.09& 67.6 & CH\textunderscore PR100024\textunderscore TG014702\textunderscore V0300 \\
    4 & 2022-11-28 14:27 & 2022-11-29 00:10& 9.72& 80.0 & CH\textunderscore PR100024\textunderscore TG014703\textunderscore V0300 \\
    \midrule
    &&&Planet c &\\
    5 & 2022-12-04 15:17 & 2022-12-05 02:41& 11.41& 83.4 & CH\textunderscore PR100024\textunderscore TG015301\textunderscore V0300 \\
    6 & 2022-12-20 17:02 & 2022-12-21 0433& 11.52& 77.5 & CH\textunderscore PR100024\textunderscore TG015302\textunderscore V0300 \\
    \bottomrule
    \end{tabular}
    \tablefoot{The file keys can be used to retrieve data from the CHEOPS archive.}
    \label{cheopsobslog}
\end{table*}

\begin{table}
\centering
\caption{Log of TESS photometric observations of TOI-421.}
\label{TESSobslog}
\begin{tabular}{cccc}
\toprule
\toprule
Sector & Start date & End date     & Number of PDCSAP\\
No.    & [UTC]      & [UTC]        & data points\\
\midrule
05 & 2018-07-25 & 2018-08-22 & $17~323$\\
06 & 2018-08-22 & 2018-09-20 & $14~610$\\
32 & 2020-11-19 & 2020-12-17 & $17~423$\\
\bottomrule
\end{tabular}
\end{table}

\begin{table}
\centering
\caption{Log of radial velocity observations of TOI-421.}
\label{RVobslog}
\begin{tabular}{cccc}
\toprule
\toprule
Instrument & Start date & End date     & Number of RVs\\
     & [UTC]      & [UTC]        & data points\\
\midrule
FIES & 2019-02-02 & 2019-03-13 & $9$\\
HARPS & 2019-02-14 & 2020-01-23 & $105$\\
PFS & 2019-02-18 & 2019-10-11 & $9$\\
HIRES & 2019-09-17 & 2020-03-20 & $33$\\
\bottomrule
\end{tabular}
\end{table}

\section{Description of the acquired data}
\label{sec_data}

\subsection{CHEOPS data}
\label{subsec_cheops_data}

CHEOPS performed a total of six observations (visits) of TOI-421 between November 2021 and December 2022 (see Table \ref{cheopsobslog}). Each visit contains a single transit event and also includes data being acquired both before and after the transit. Each individual visit comprises several CHEOPS orbits and a single orbit covers roughly 100 minutes. However, the target cannot be observed throughout an entire CHEOPS orbit, due to Earth occultations and South Atlantic Anomaly (SAA) crossings. This leads to gaps in the observations with a width that varies from visit to visit depending on target coordinates and observation time. The duration and the observing efficiency (i.e. the relative amount of time in which the target is visible to CHEOPS throughout a visit) of each visit are also listed in Table \ref{cheopsobslog}.

The CHEOPS observations are available as sub-array data products \citep{Benz2021} at a cadence equal to the exposure time of 60\,s. The sub-arrays contain a circular region around the target with a radius of 100 pixels. We processed the data with PSF photometry utilising PIPE\footnote{\url{http://github.com/alphapsa/PIPE}}\citep[see also][]{2021A&A...651L..12M, 2021A&A...654A.159S, Brandeker2022}. In general, CHEOPS observations are affected by instrumental noise such as stray light from the Earth and Moon (Moon glint), smearing effects, or spacecraft jitter. The flux measurements usually show a particularly strong correlation with the spacecraft roll angle \citep[see also][]{Lendl2020,Bonfanti2021}. The spacecraft is designed to rotate around itself exactly once every orbit around the Earth. Therefore, the roll angle parameter is directly linked to the orbital position of the spacecraft. Instrumental noise must be accounted for during data analysis in order to identify and measure the transit signals of the planets (see Section \ref{sec_analysis}). Prior to performing data analysis, we removed all points that were flagged by PIPE, which for example include those contaminated by cosmic rays. We performed a sigma clipping and removed all points with the median absolute deviation (MAD) higher than five to discard outliers. We also removed points with peculiar high backgrounds by removing any point with a background larger than four times the median background value, as well as points with a peculiar high pointing offset by removing all points with a centroid-offset of more than one pixel. 

\subsection{TESS data}
\label{subsec_tess_data}

TOI-421 was observed by TESS in Sectors 5, 6 and 32 at 2 min cadence. The TESS photometric baseline, therefore, spans from July 2018 to December 2020 (see Table \ref{TESSobslog} for further details). In our analysis, we used the \texttt{Pre-search Data Conditioning Simple Aperture Photometry} (PDCSAP) flux, provided by the Science Processing Operations Center (SPOC) pipeline \citep{Smith2012,Stumpe2014,2016SPIE.9913E..3EJ}. This light curve type is subject to more treatment than the \texttt{Simple Aperture Photometry} (SAP) flux and is specifically intended for detecting exoplanets. The pipeline attempts to remove systematic artefacts, while keeping planetary transits intact. It also already accounts for dilution caused by nearby contaminating stars. The data were downloaded from the Mikulski Archive for Space Telescopes\footnote{See \url{https://mast.stsci.edu/portal/Mashup/Clients/Mast/Portal.html}.} (MAST). We did not apply outlier removal in the TESS dataset. To save computation time and simplify the treatment of long-term trends caused by stellar activity in the dataset we only used data points, which we expected to be within a 0.25 day window centred around the expected transit mid-point. We used the following transit timing parameters to define this window after having performed a preliminary analysis of the data to retrieve these values: $T_{\rm 0,b} = 2459189.7336$ BJD, $P_{\rm b} = 5.1975736$ days, and $T_{\rm 0,c} = 2459195.30753$ BJD, $P_{\rm c} = 16.0675425$ days.

\subsection{Radial velocity}
\label{subsec_rv_data}

\citet{Carleo2020} published a total of 156 radial velocity (RV) measurements of TOI-421 obtained between February 2019 and March 2020 by four different instruments: The FIbre-fed Échellé Spectrograph at La Palma, Spain \citep[FIES;][]{Fransen1999,Telting2014}; the High Accurarcy Radial velocity Planet Searcher at La Silla, Chile \citep[HARPS;][]{Mayor2003}; the HIgh Resolution Échelle Spectrometer on Hawaii, USA \citep[HIRES;][]{Vogt1994}; and the Planet Finder Spectograph at Las Campanas, Chile \citep[PFS;][]{Crane2006,Crane2008,Crane2010}. An overview of the observation dates and the number of acquired RV points is listed in Table \ref{RVobslog}. A detailed description of the data can be found in \citet{Carleo2020}.

In the case of the FIES, HIRES, and PFS data we used the reduced RV data published by \citet{Carleo2020} and made available at the Centre Données astronomiques de Strasbourg (CDS) astronomical data centre\footnote{\url{https://cdsarc.cds.unistra.fr/viz-bin/cat/J/AJ/160/114}}. The corresponding data products provide only the time stamp of the observation and the measured RV value including its uncertainty. The HARPS data reduced using the HARPS data reduction software (DRS) are also available at the CDS and accompanied with line profile asymmetry indicators, namely the FWHM, the bisector inverse slope of the cross-correlation function (CCF) and the CaII H and K lines activity indicator $\log{R'_{HK}}$.

We did not use the published HARPS radial velocities reduced using the DRS, but employed a novel reduction approach, using the CCFs of the published HARPS spectra, along with their $\log{R'_{HK}}$ activity indicators. Namely, we performed a Skew Normal fit onto the CCFs \citep{simola2019}, and for each epoch of observation we derived the radial velocity, the Skew-Normal full width half maximum (FHWM$_{\mathrm{SN}}$), the contrast ($A$), and the asymmetry parameter ($\gamma$) along with their uncertainties. Details of this reduction process are described in \citet{Bonfanti2023} and Fridlund et al. (2023, accepted).


\section{Data analysis}
\label{sec_analysis}

\subsection{CHEOPS pre-fit detrending}
\label{subsec_detrending_photometry}
We used the \texttt{PyCHEOPS} python package \citep{Maxted2021} to correct the CHEOPS data for instrumental effects and long-term temporal trends. To this end we excluded all data points that were observed during a transit event and then fitted coefficients of detrending models on the out-of-transit data only. The detrending models were chosen independently for each visit by individually adding detrending basis vectors one by one and retaining an additional vector only when supported by a higher Bayes factor computed from the Bayesian information criterion \citep[BIC;][]{schwarz1978}. In this way we used up to second-order harmonics of the roll angle, up to second-order polynomials in PSF centroid position and time, and a first-order linear model on background. Additionally, a 18-segment spline was fit on top of these models. The spline fits small-scale flux changes as a function of roll angle to remove glint effects. After fitting the coefficients of the detrending models on the out-of-transit data, the models were also applied to the in-transit-data and the uncertainties of the individual data-points were inflated to account for the uncertainties of the detrending coefficients. In our joint analysis we then used the detrended datasets only and no additional detrending models. 

\begin{table*}
    \caption{Overview of the results of different baseline models for the HARPS datasets}
    \centering
    \begin{tabular}{cccccccc}
    \toprule
    \toprule
     Dataset & Baseline Model & $R_{\rm b}$ [$R_{\oplus}$] & $M_{\rm b}$ [$M_{\oplus}$] &  $e_{\rm b}$ & $R_{\rm c}$ [$R_{\oplus}$] & $M_{\rm c}$ [$M_{\oplus}$] &  $e_{\rm c}$\\
    \midrule
    \multirow{3}{5mm}{DRS}& \citet{Carleo2020} & $2.68 \pm 0.19$ & $7.2 \pm 0.7$ & $0.16\pm 0.08$ & $5.09 \pm 0.16$ & $16.4 \pm 1.1$ & $0.15 \pm 0.04$\\
    \noalign{\smallskip}
    & Polynomials & $2.69 \pm 0.11$ & $6.6 \pm 0.8$ & $0.12 \pm 0.07$ & $5.03 \pm 0.07$ & $15.0 \pm 1.3$ & $0.23 \pm 0.04$\\ 
    \noalign{\smallskip}
    & SHO-GP & $2.61 \pm 0.08$ & $7.2 \pm 0.7$ & $0.18 \pm 0.04$ & $5.06 \pm 0.07$ & $12.7 \pm 1.4$ & $0.22 \pm 0.04$\\ 
    \noalign{\smallskip}
    \midrule 
    \multirow{2}{5mm}{SNF}& Polynomials & $2.72 \pm 0.11$ & $7.0 \pm 0.9$ & $0.10 \pm 0.06$ & $5.00 \pm 0.08$ & $14.7 \pm 1.4$ & $0.25 \pm 0.05$\\ 
    \noalign{\smallskip}
    & SHO-GP & $2.61 \pm 0.07$ & $7.3 \pm 0.6$ & $0.17 \pm 0.05$ & $5.01 \pm 0.08$ & $12.4 \pm 1.3$ & $0.25 \pm 0.05$\\ 
    \bottomrule
    \end{tabular}
    \tablefoot{\citet{Carleo2020} used a sinusoidal function as baseline and the \texttt{pyaneti} code \citep{barragan2019}. Results employing polynomials as baseline were retrieved using the \texttt{MCMCI} code, while results employing a SHO-GP as baseline were retrieved using the \texttt{allesfitter} code.}
    \label{table_rvanalysis}
\end{table*}

\subsection{Mitigating stellar activity in radial velocity data}
\label{subsec_rvanalysis}

\citet{Carleo2020} have shown that the radial velocity data are affected by stellar activity. Specifically, they performed an in-depth frequency analysis of the HARPS data and found a significant peak in the RV residuals following the subtraction of the planetary signals corresponding to a period of $\sim42$ days and an RV semi-amplitude of $\sim2.4 $ m/s. They also recovered this peak in the periodogram of the $\log{R'_{HK}}$ and FWHM activity indicators in the HARPS DRS data. They attribute this peak to a plage-dominated activity signal after carefully analysing a contour map of the HARPS CCF residuals and also searching for rotational modulation in the available WASP-South and TESS photometry. In their subsequent global analysis, \citet{Carleo2020} account for this activity induced signal by fitting a third Keplerian to the RV data, resulting in an additional sinusoidal signal at a period of $43.24^{+0.57}_{-0.55}$ days and with a semi-amplitude of $2.4 \pm 0.3$ m/s. The epoch of this signal is however fairly unconstrained, with an uncertainty of $\sim5$ days, as reported by \citet{Carleo2020}. Furthermore, stellar activity in general is not strictly sinusoidal \citep[e.g. ][]{Lanza2001} and adding sinusoidal baselines has been shown to sometimes introduce spurious harmonics that bias the final result \citep{Pont2011,Tuomi2014,Rajpaul2015}. Because the epoch is unconstrained and the known problems of adding a simple sinusoidal model, we aim to provide alternative baseline models. To compare different approaches, we analysed the HARPS data, which is the only RV dataset that also provides activity indicators, obtained through two different extractions methods and using two different approaches, as explained below. To help constraining both the ephemeris and the eccentricity of the orbits, we also included all photometric data in this particular analysis. We used the stellar parameters derived in Section \ref{sec_stellar}. To show that the corresponding results are not an artefact of the data reduction process, we performed the analysis once using the DRS-reduced data published in \citet{Carleo2020} and once using the data reduced with the novel approach of performing a Skew Normal fit to the CCFs (see Section \ref{subsec_rv_data}). The later reduction approach will be denoted as SNF.

The first baseline model we used is a model fitting polynomial functions to the activity indicators acompanying both the HARPS DRS and HARPS SNF data. The analysis was performed using the \texttt{MCMCI} code \citep{Bonfanti2020}. As previously done with the CHEOPS data, we choose the detrending models independently for both HARPS datasets by individually adding detrending basis vectors one by one and retaining an additional vector only when supported by a higher Bayes factor computed from the BIC. In this way, we ended up using a second-order polynomial in FWHM and a first-order polynomial in $\log{R'_{HK}}$ for the DRS data and first-order polynomials in FWHM$_{\mathrm{SN}}$, contrast, and $\log{R'_{HK}}$ for the SNF data. We did also perform a breakpoint analysis \citep{simola2022,Bonfanti2023} to determine possible parts of the RV time series that would require different detrending basis vectors, but did not find any breakpoints indicating that the whole time series can be adequately detrended with the same detrending model. The 14 individual transit events present in the TESS data were treated as 14 individual light curves. For each of them a second-order polynomial in time was used as a baseline model.

The second baseline model we used is a Gaussian process \citep[GP;][]{rasmussen05} using a critically damped Simple-Harmonic-Oscillator (SHO) kernel \citep{Foreman-Mackey2017}. The kernel ($\kappa$) is defined by an amplitude of variations $S_0$ and an angular frequency of the variations $\omega_0$, which can be translated to a period of variations by $P = 2\pi / \omega_0$:

\begin{equation}
    \kappa(\tau) = S_0 \, \omega_0 \, e^{- \frac{1}{\sqrt{2}} \, \omega_0 \,\tau} \rm cos \left( \frac{\omega_0 \, \tau}{\sqrt{2}} - \frac{\pi}{4} \right)
\end{equation}

We further constrained the period of the variations by fitting a Gaussian function to the strongest peak in the periodogram of the FWHM$_{\mathrm{SN}}$ activity indicator in the HARPS SNF data, which resulted in a prior on the period of variations at $41.2 \pm 2.1$ days. We also observed a similar peak in the periodogram of the $\log{R'_{HK}}$ activity indicator. The analysis was performed using the \texttt{allesfitter} Python package \citep{guenther2021}. The hybrid spline baseline option available in \texttt{allesfitter} was applied to the TESS photometry.

Table \ref{table_rvanalysis} shows the retrieved planetary parameters for all combinations of RV baseline models and datasets, as well as the values reported in \citet{Carleo2020}. We find that as expected the retrieved planetary radii are not dependent on the chosen RV baseline model. We do not find significant differences in the retrieved mass and eccentricity of planet b, with all results agreeing with each other well within their $1\sigma$ confidence intervals. For the mass of planet c, we find that our new analyses result in a smaller planetary mass compared to the value reported in \citet{Carleo2020}. The masses retrieved when using the SHO-GP baseline are smaller than those obtained using the polynomials baseline, but all four new model approaches are consistent with each other within their $1\sigma$ confidence intervals. In the case of the eccentricity of planet c, we find that all of the new analyses find a higher eccentricity than reported in \citet{Carleo2020}. Again, we also find that all four new approaches agree with each other well within their $1\sigma$ confidence intervals.

Regardless of the adopted technique to extract the RV time series and the indicators of stellar activity (either DRS or SNF), we obtain consistent results in terms of planetary masses and eccentricities after applying the polynomial detrending baseline based on the activity indicators. This proves that the SNF-based technique enables us to infer reliable RV-related parameters, and thus we adopt the SNF-based HARPS RVs also in the following joint analysis (Sec.~\ref{subsec_transit_analysis}). 

\begin{figure}
    \centering
    \includegraphics[width=\hsize]{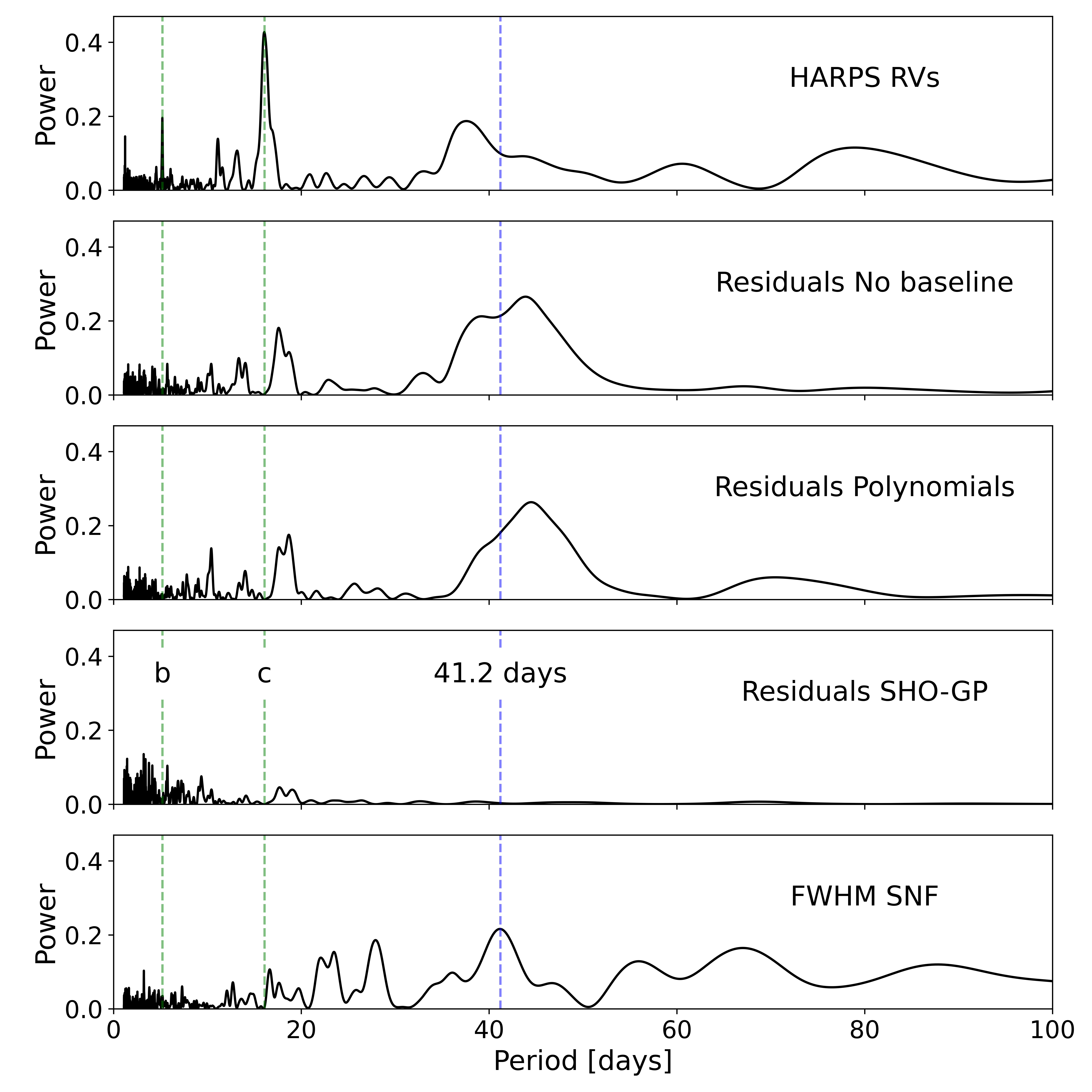}
    \caption{Periodograms of the HARPS SNF data and residuals (i.e. RV - planet signals) when using different baseline models. 1st panel: Raw HARPS SNF data. 2nd panel: Residuals when fitting only for the planetary signals, without any baseline. 3rd panel: Residuals when fitting for the planetary signals and for linear detrending vectors of the activity indicators. 4th panel: Residuals when fitting for the planetary signals and employing a SHO-GP as baseline. 5th panel: FWHM$_{\mathrm{SN}}$. The dashed green lines represent the orbital periods of the two known planets. The dashed blue line represents the most significant peak in the periodogram of the FWHM$_{\mathrm{SN}}$ activity indicator.}
    \label{fig_periodogram}
\end{figure}

Because two completely independent approaches (i.e. the polynomial-based and GP-based detrending) retrieve consistent results independently on how the data were reduced, we believe that both approaches are justified. To choose a model for our final global analysis, we also looked at the periodogram of the residuals when applying the different baseline models (see Figure \ref{fig_periodogram}). We find that the approach using an SHO-GP baseline performs best at removing the activity-induced signal. We therefore chose this approach for our global analysis using all of the available radial velocity datasets.

\begin{figure}
    \centering
    \includegraphics[width=\hsize]{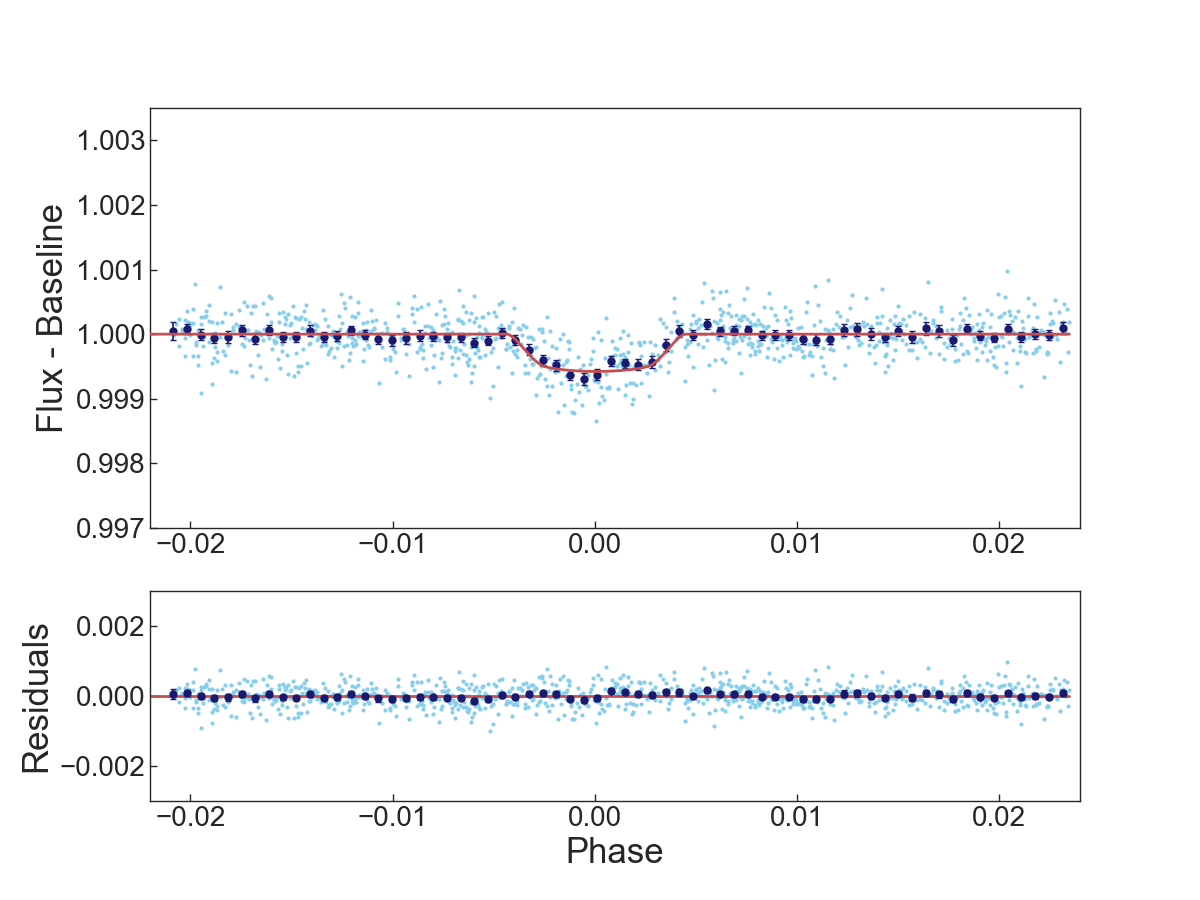}
    \includegraphics[width=\hsize]{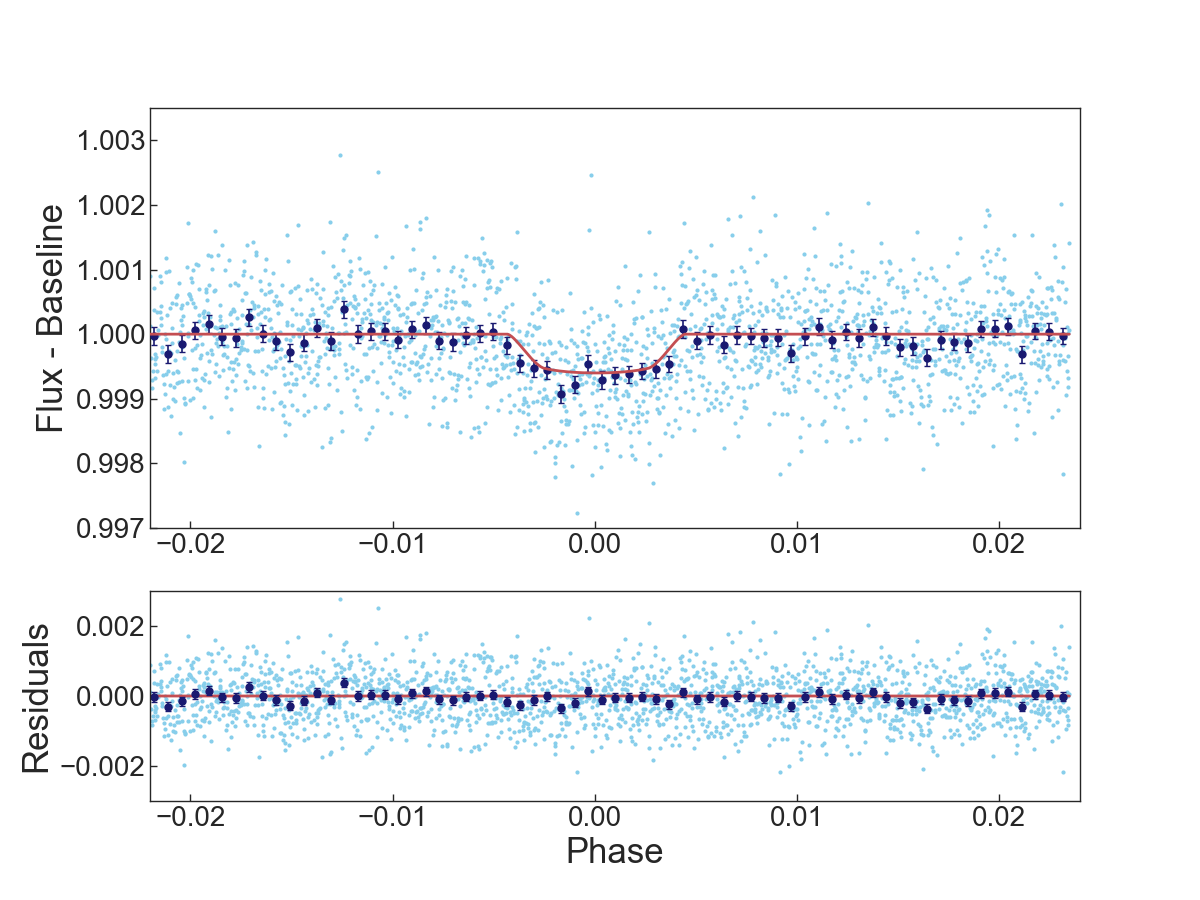}
    \includegraphics[width=\hsize]{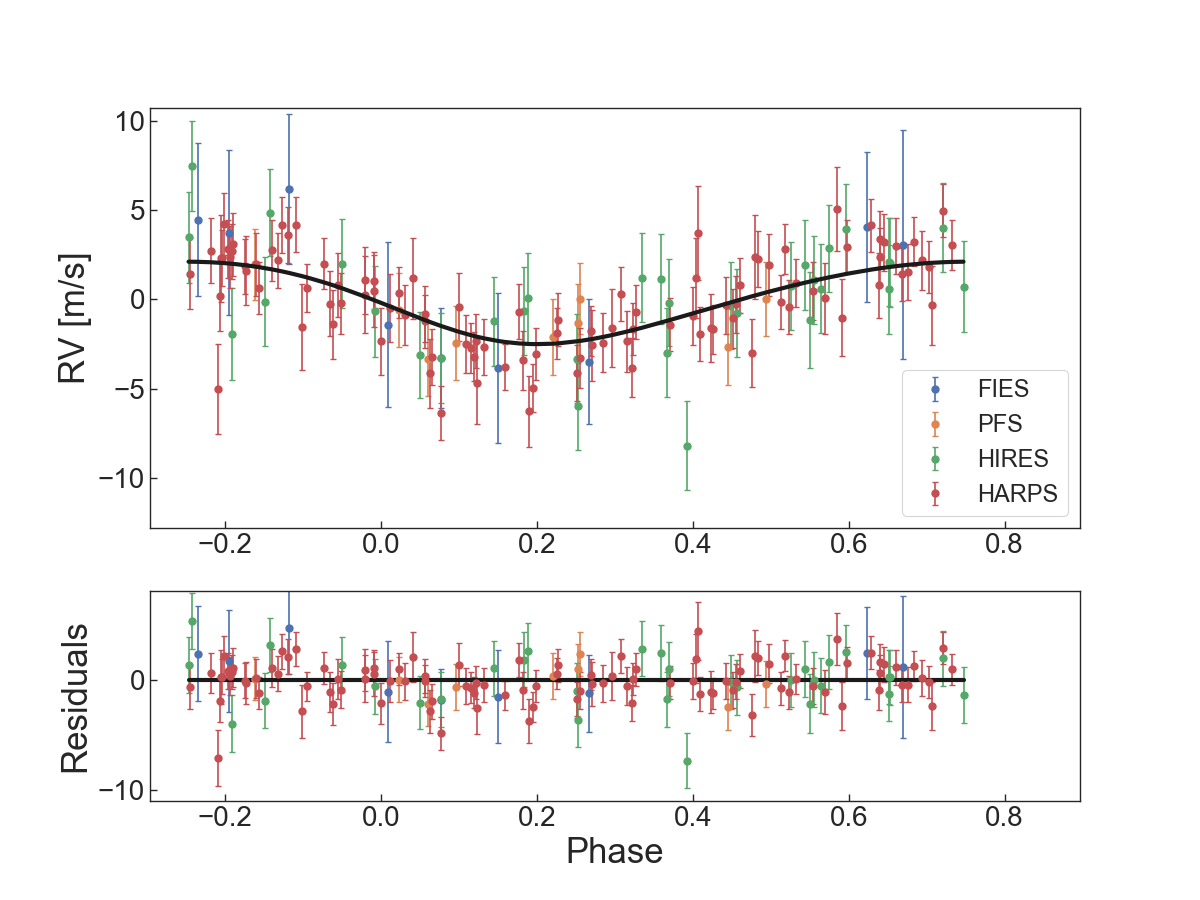}
    \caption{Phase-folded CHEOPS, TESS, and RV data of TOI-421\,b. \textit{Top panel}: Phase-folded and detrended transit light curve of TOI-421\,b using four CHEOPS visits. The bright and dark blue points are the original data and the 5-min binned data points, respectively. The red line is the median fitted model and the residuals of the fit are shown in the panel below the light curve. \textit{Middle panel}: Same as the \textit{top panel}, but using TESS data observed during Sectors 5, 6, and 32. \textit{Bottom panel}: Phase-folded and detrended RV data for TOI-421\,b using FIES (blue), PFS (yellow), HIRES (green) and HARPS (red). The black line is the median fitted model and the residuals of the fit are shown in the panel below the RV curve. The median fitted model of TOI-421\,c has been subtracted from all RV data points.}
    \label{fig:b}
\end{figure}

\begin{figure}
    \centering
    \includegraphics[width=\hsize]{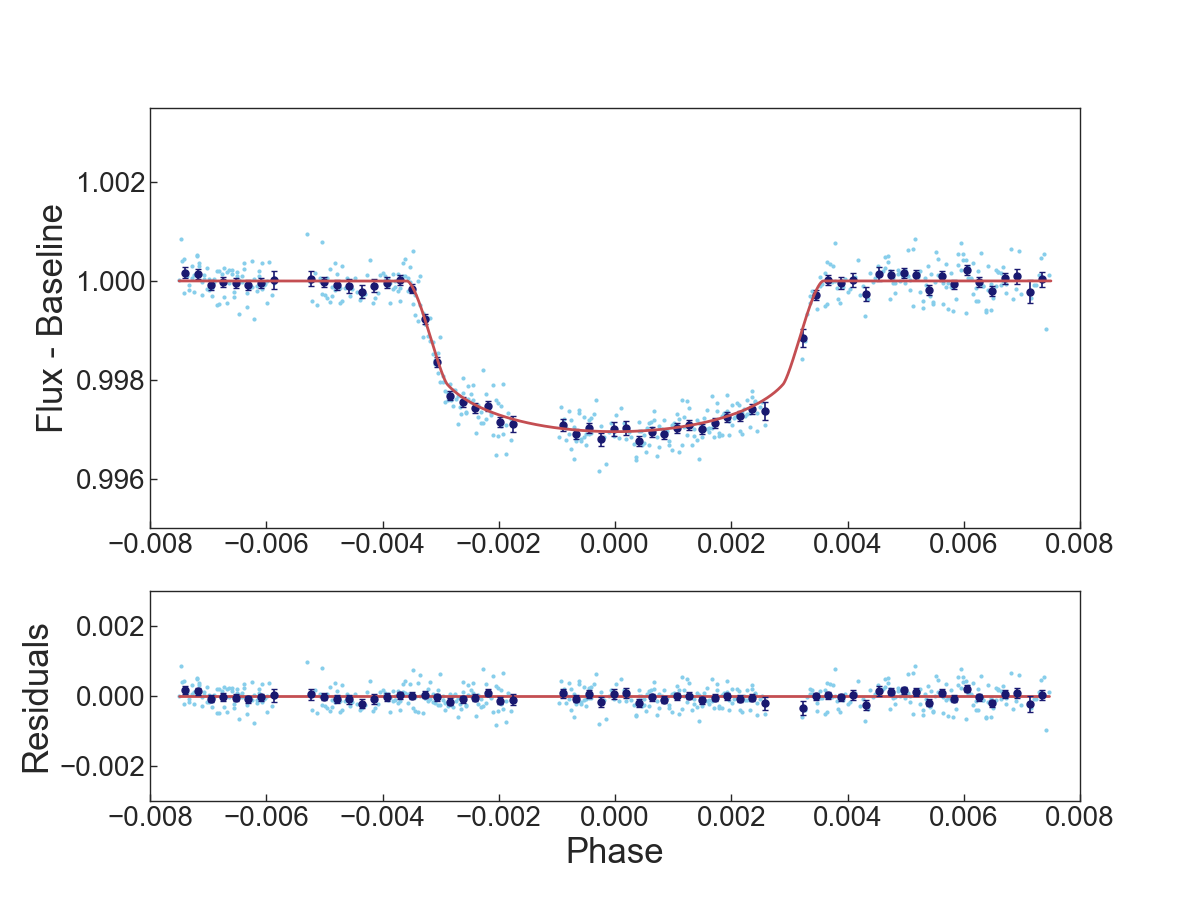}
    \includegraphics[width=\hsize]{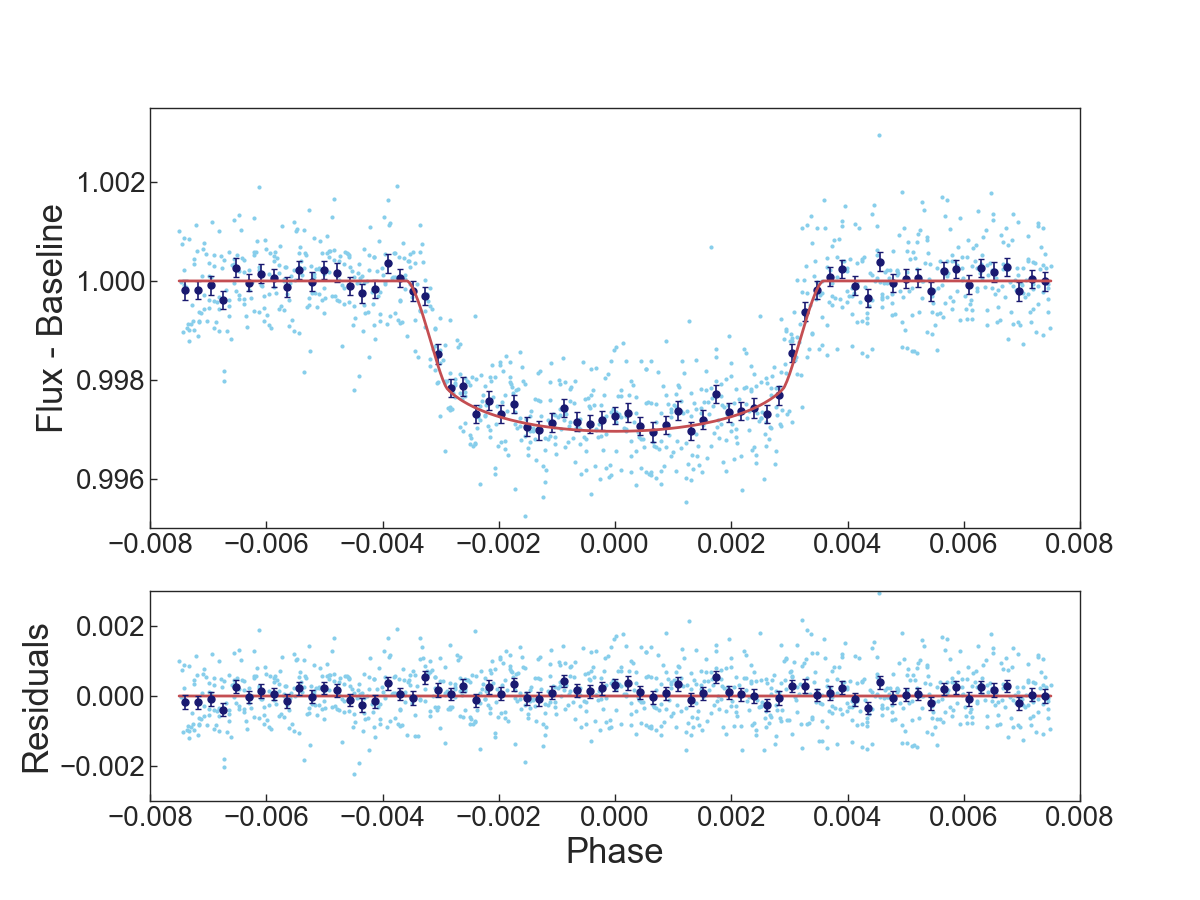}
    \includegraphics[width=\hsize]{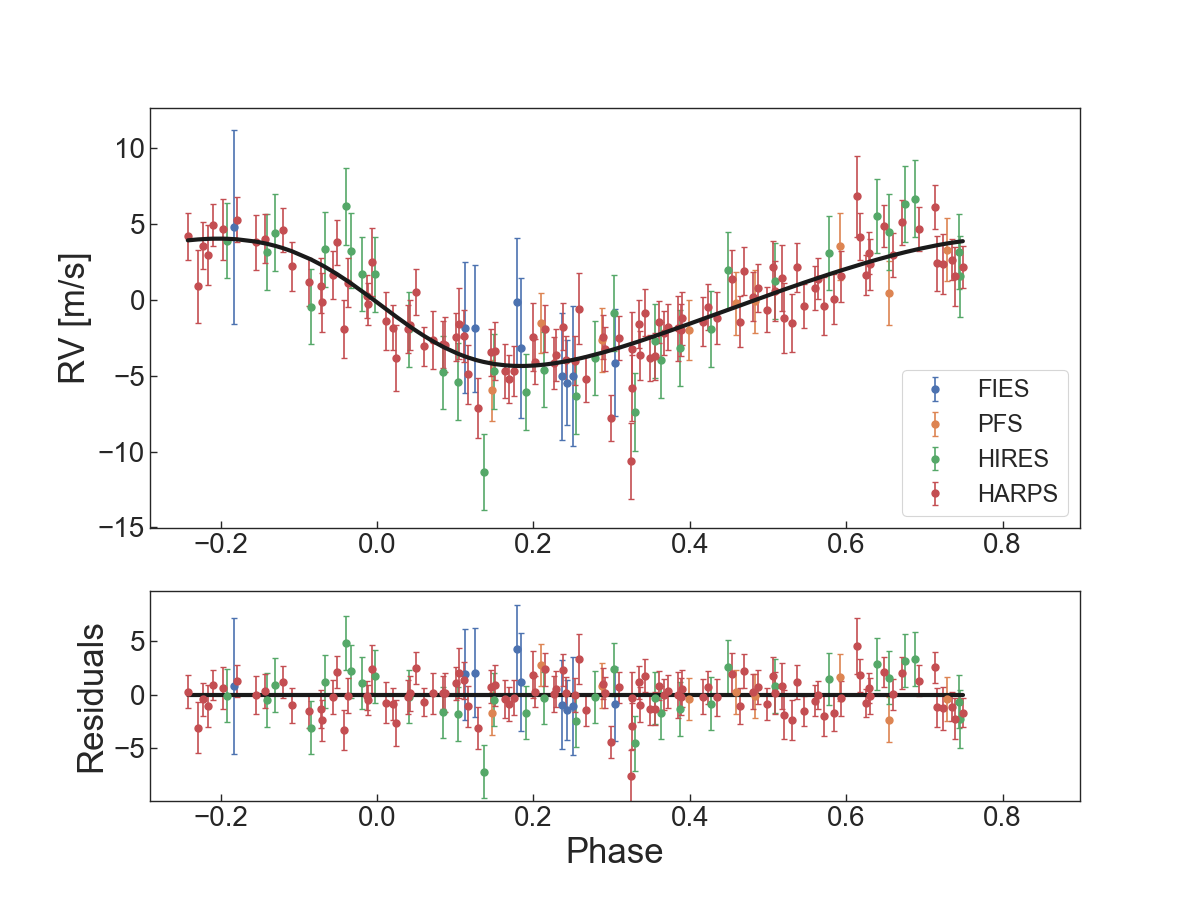}
    \caption{Same as Figure \ref{fig:b}, but for TOI-421\,c and using 2 CHEOPS visits.}
    \label{fig:c}
\end{figure}

\begin{table*}
    \caption{Retrieved planetary and stellar parameters}
    \centering
    \begin{tabular}{lcccr}
    \toprule
    \toprule
    Parameters & Symbols & Priors & Values &  Units \\
    \midrule
    \textbf{Planet b}&&&& \\
    \quad Orbital period & $P_{\rm b}$ & $\mathcal{U}(5.1975, 5.1977)$ & $5.197576 \pm 0.000005$ &  days \\
    \quad Transit time & $T_{\rm 0,b}$ & $\mathcal{U}(2459189.7,2459189.76)$ & { $2459189.7341 \pm 0.0005$} &  BJD\\
    \quad Planet-to-star radius ratio & $R_{\rm b}/R_\star$ & $\mathcal{U}(0.02,0.04)$ &  $0.0279 \pm 0.0008$ &  -\\
    \quad Impact parameter & $b_{\rm b}$ & $\mathcal{U}(0.0,1.1)$ & $0.942 \pm 0.011$ &  - \\
    \quad Eccentricity & $e_{\rm b}$ & $\mathcal{U}(0.0,1.0)$\tablefootmark{(b)} &  $0.13 \pm 0.05$ &  - \\
    \quad Argument of periapsis & $\omega_{\rm b}$ & $\mathcal{U}(0,360)$\tablefootmark{(b)} &  $140 \pm 30$ &  deg \\
    \quad RV semi-amplitude & $K_{\rm b}$ & $\mathcal{U}(1,4)$ &  $2.83 \pm 0.18$ &  m/s \\
    \quad Transit duration & $\rm dur_b$ & - &  $0.0454 \pm 0.0014$ & days \\
    \quad Semi-major axis & $a_{\rm b}$ & - &  $0.0554 \pm 0.0010$ & AU \\
    \quad Radius & $R_{\rm b}$ & - &  $2.64 \pm 0.08$ &  $R_{\oplus}$ \\
    \quad Mass & $M_{\rm b}$ & - &  $6.7 \pm 0.6$ &  $M_{\oplus}$ \\
    \quad Mean density & $\rho_{\rm b}$ & - &  $0.37 \pm 0.05$ &  $\rho_{\oplus}$ \\
    \quad Equilibrium Temperature\tablefootmark{(a)}  & $T_{\rm eq,b}$ & - &  $922 \pm 14$ &  K \\
    \midrule
    \textbf{Planet c}&&&& \\
    \quad Orbital period & $P_{\rm c}$ & $\mathcal{U}(16.0672,16.0678)$ & $16.067541 \pm 0.000004$ &  days \\
    \quad Transit time & $T_{\rm 0,c}$ & $\mathcal{U}(2459195.305,2459195.310)$ & $2459195.30741 \pm 0.00018$ &  BJD\\
    \quad Planet-to-star radius ratio & $R_{\rm c}/R_\star$ & $\mathcal{U}(0.04,0.07)$ &  $0.0540 \pm 0.0006$ &  -\\
    \quad Impact parameter & $b_{\rm c}$ & $\mathcal{U}(0.0,1.1)$ &  $0.70 \pm 0.03$ &  - \\
    \quad Eccentricity & $e_{\rm c}$ & $\mathcal{U}(0.0,1.0)$\tablefootmark{(b)} &  $0.19 \pm 0.04$ &  - \\
    \quad Argument of periapsis & $\omega_{\rm c}$ & $\mathcal{U}(0,360)$\tablefootmark{(b)} &  $102 \pm 14$ &  deg \\
    \quad RV semi-amplitude & $K_{\rm c}$ & $\mathcal{U}(3,6)$ &  $4.1 \pm 0.3$ &  m/s \\
    \quad Transit duration & $\rm dur_c$ & - & $0.1148 \pm 0.0009$ & days \\
    \quad Semi-major axis & $a_{\rm c}$ & - &  $0.1170 \pm 0.0018$ & AU \\
    \quad Radius & $R_{\rm c}$ & - &  $5.09 \pm 0.07$ &  $R_{\oplus}$ \\
    \quad Mass & $M_{\rm c}$ & - &  $14.1 \pm 1.4$ &  $M_{\oplus}$ \\
    \quad Mean density & $\rho_{\rm c}$ & - &  $0.107 \pm 0.012$ &  $\rho_{\oplus}$ \\
    \quad Equilibrium Temperature \tablefootmark{(a)}  & $T_{\rm eq,c}$ & - &  $635 \pm 9$ &  K \\
    \midrule
    \textbf{Stellar Parameters} &&&& \\
    \quad Stellar Density & $\rho_{\star}$ &$\mathcal{N}(1.28, 0.09)$ & $1.28 \pm 0.06$ &  $\rho_\odot$ \\
    \quad Limb darkening coefficients &&&& \\
    \multirow{2}{*}{\qquad CHEOPS passband} & $q_{1_{\mathrm{CHEOPS}}}$ &$\mathcal{N}(0.487, 0.048)$ & $0.44 \pm 0.04$ &  - \\
    & $q_{2_{\mathrm{CHEOPS}}}$ & $\mathcal{N}(0.359, 0.015)$ &$0.363 \pm 0.010$ &  - \\
    \multirow{2}{*}{\qquad TESS passband} & $q_{1_{\mathrm{TESS}}}$ & $\mathcal{N}(0.377, 0.040)$ & $0.378 \pm 0.03$ &  - \\
    & $q_{2_{\mathrm{TESS}}}$ & $\mathcal{N}(0.311, 0.015)$ &$0.316 \pm 0.011$ &  - \\
    \midrule
    \textbf{Instrumental Parameters} &&&& \\
    \quad CHEOPS White-noise & $\sigma_{\rm CHEOPS}$ &$\mathcal{U}(6, 2500)$ & $305 \pm 5$ &  \rm ppm \\
    \quad TESS White-noise & $\sigma_{\rm TESS}$ &$\mathcal{U}(17, 6740)$ & $722 \pm 7$ &  \rm ppm \\
    \quad HARPS Jitter & $\sigma_{\rm HARPS}$ &$\mathcal{U}(0.05, 135)$ & $1.0 \pm 0.4$ &  \rm m/s \\
    \quad HIRES Jitter & $\sigma_{\rm HIRES}$ &$\mathcal{U}(0.05, 135)$ & $2.2 \pm 0.4$ &  \rm m/s \\
    \quad PFS Jitter & $\sigma_{\rm PFS}$ &$\mathcal{U}(0.05, 135)$ & $2.0 \pm 0.7$ &  m/s \\
    \quad FIES Jitter & $\sigma_{\rm FIES}$ &$\mathcal{U}(0.001, 135)$ & $0.02 \pm 0.16$ &  m/s \\
    \quad HARPS offset & $\gamma_{\rm HARPS}$ &$\mathcal{U}(-50, 50)$ & $1.8 \pm 0.9$ &  m/s \\
    \quad HIRES offset & $\gamma_{\rm HIRES}$ &$\mathcal{U}(-50, 50)$ & $-0.8 \pm 0.4$ &  m/s \\
    \quad PFS offset & $\gamma_{\rm PFS}$ &$\mathcal{U}(-50, 50)$ & $2.3 \pm 1.3$ &  m/s \\
    \quad FIES offset & $\gamma_{\rm FIES}$ &$\mathcal{U}(-50, 50)$ & $-7.1 \pm 1.3$ &  m/s \\
    \quad HARPS Amplitude of variations & $\log(S_{0,\rm HARPS})$ &$\mathcal{U}(-15, -5)$ & $-8.9 \pm 0.3$ &  - \\
    \quad HIRES Amplitude of variations & $\log(S_{0,\rm HIRES})$ &$\mathcal{U}(-15, -5)$ & $-11.3 \pm 1.1$ &  - \\
    \quad PFS Amplitude of variations & $\log(S_{0,\rm PFS})$ &$\mathcal{U}(-15, -5)$ & $-9.5 \pm 0.8$ &  - \\
    \quad FIES Amplitude of variations & $\log(S_{0,\rm FIES})$ &$\mathcal{U}(-15, -5)$ & $-12.3 \pm 1.6$ &  - \\
    \quad RV angular frequency of variations & $\omega_{0,\rm RV}$ &$\mathcal{N}(0.153, 0.008)$ & $0.159 \pm 0.007$ &  1/days \\
    \quad RV Period of variations & $P_{\rm rot,\star}$ &$\mathcal{N}(41.2, 2.1)$ & $39.6 \pm 1.6$ &  days \\
    \bottomrule
    \end{tabular}
    \tablefoot{
    The Gaussian priors with mean $\mu$ and standard deviation $\sigma$ are displayed as $\mathcal{N}(\mu, \sigma)$. $\mathcal{U}$ represents an uniform bounded prior.\\
    \tablefoottext{a}{Assuming albedo equal to zero.}\\
    \tablefoottext{b}{The uniform priors on $e_{\rm c}$ and $\omega_{\rm c}$ were imposed on $\sqrt{e_{\rm c}}\, \rm cos(\omega_{\rm c})$ and $\sqrt{e_{\rm c}} \, \rm sin(\omega_{\rm c})$}
    }
    \label{tab:fitted_transit_param}
\end{table*}

\begin{figure*}
    \centering
\includegraphics[width=\textwidth]{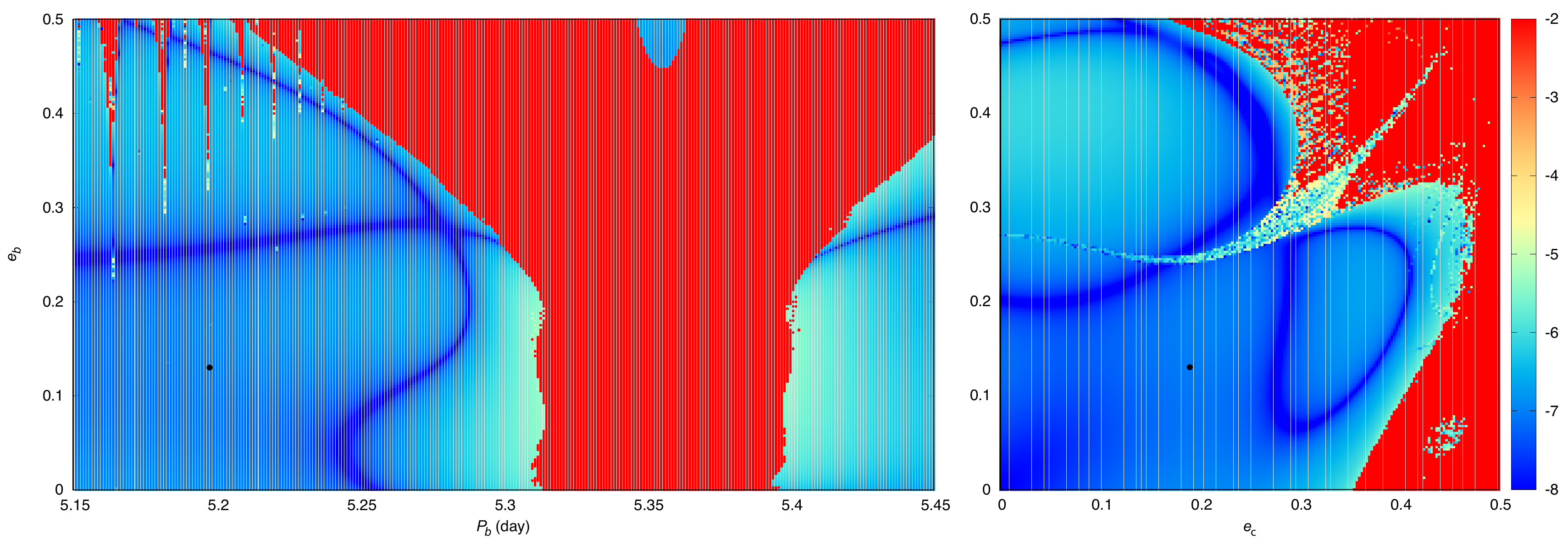}    \caption{Stability analysis of the TOI-421 planetary system. For fixed initial conditions (Table~\ref{tab:fitted_transit_param}), the parameter space of the system was explored by varying the orbital period $P_\mathrm{b}$ and the eccentricity $e_\mathrm{b}$ of planet-b (left panel) and the eccentricities of both planets (right panel). The step size was $10^{-3}$~day in orbital period and $0.0025$ in eccentricity. For each initial condition, the system was integrated over 5000~yr and a stability indicator was calculated, which involved a frequency analysis of the mean longitude of the inner planet. The chaotic diffusion was measured by the variation in the frequency (see text). Red points correspond to highly unstable orbits, while blue points correspond to orbits which are likely to be stable on a billion-years timescale. The black dots show the values of the best fit solution (Table~\ref{tab:fitted_transit_param}).}
    \label{fig:SF1}
\end{figure*}

\subsection{Joint analysis}
\label{subsec_transit_analysis}
We performed a joint analysis of all photometric and RV datasets using \texttt{allesfitter}. For each of the planets we fitted for the radius-ratio $R_{\rm p}/R_{\star}$, the cosine of the inclination $i$, the term $(R_{\rm p} + R_{\star})/a$, the transit mid-time, the orbital period, the RV semi-amplitude, and the parameters $\sqrt{e} \, \rm cos(\omega)$ and $\sqrt{e} \, \rm sin(\omega)$, which determine the eccentricity $e$ and the argument of periapsis $\omega$.

While using uniform priors for the planetary parameters, we did apply Gaussian priors on the stellar radius and mass according to the values presented in Section \ref{sec_stellar}. This resulted in a prior on the mean stellar density, which in turn constrains the transit model. We also applied a quadratic limb-darkening (LD) law using the parametrisation proposed by \citet{Kipping2013} and used Gaussian priors on the LD coefficients according to theoretically computed values including their uncertainties using the \texttt{LDCU} Python package (see Table \ref{tab:fitted_transit_param}). \texttt{LDCU}\footnote{\url{https://github.com/delinea/LDCU}} is a modified version of the python routine implemented by \citet{espinoza2015} that computes the LD coefficients and their corresponding uncertainties using a set of stellar intensity profiles accounting for the uncertainties on the stellar parameters. The stellar intensity profiles are generated based on two libraries of synthetic stellar spectra: ATLAS \citep{Kurucz1979} and PHOENIX \citep{husser2013}. To check the validity of this additional prior, we performed the same analysis presented here also without constraining the LD coefficients except for their physical boundaries. Although in this case the LD coefficients are only poorly constrained, all of the retrieved planetary parameters are consistent with those presented in this work within their $1\sigma$ confidence intervals.

As discussed in \citet{Carleo2020}, the TESS light curves are contaminated by a M dwarf companion (GAIA-ID 2984582227215748224) at an angular distance of $29.4 ''$. The contamination was computed by \citet{Carleo2020} to be $1.8 \pm 0.4 \%$. The TESS PDSCAP light curves already account for the dilution of the light curve caused by possible contaminants via the contamination estimate in the \texttt{TESS Input Catalog} (TIC). We note that the estimated contamination ratio of 0.0246 is consistent with the value computed by \citet{Carleo2020}. Because we used PSF and not aperture photometry for the extraction of the CHEOPS photometry, the CHEOPS data are not affected by contamination. 

For the photometric observations, we did fit for an additional white-noise term per instrument. For the RV data, we also added a further jitter-term and a constant offset for each instrument. In terms of baseline models, we used the already detrended CHEOPS data (see Section \ref{subsec_detrending_photometry}) and applied the hybrid spline baseline option in \texttt{allesfitter} to the TESS photometry. As discussed in Section \ref{subsec_rvanalysis}, for the baselines of the RV data we used a GP with an SHO kernel. While fitting for individual amplitudes of variations $S_0$ for each instrument, we jointly fit for a single angular frequency (and therefore period) of variations. We again imposed a prior on this parameter defined by the strongest peak in the periodogram of the FWHM$_{\mathrm{SN}}$ activity indicator in the HARPS SNF data.

The global analysis was performed using the dynamic nested sampling algorithm \citep[e.g.][]{feroz2008,feroz2019} implemented in \texttt{allesfitter} via the \texttt{dynesty} Python package \citep{speagle}. The results of our fit and the successively computed planetary parameters are listed in Table \ref{tab:fitted_transit_param}. We note that both planetary radii and masses are consistent with the results presented in \citet{Carleo2020} at the $1\sigma$ level. The uncertainties on the radius of both planets decreased by a factor of $\sim2.5$. Despite the more conservative uncertainty on the stellar mass, the uncertainty on the mass of planet b also decreased. Noteworthy is also the difference in the median values of the transit timing. There is a difference of 74 seconds when comparing the period of planet b as derived by us and as derived by \citet{Carleo2020}. This adds up to a difference of 87 minutes ($\sim1.5$ hours) per year. In the case of planet c, the two periods differ by 56 seconds, which amounts to 21 minutes per year. Considering the prospects of future follow-up spectroscopic observations (see Section \ref{sec:discussion}), these differences of several hours since the publication of the original ephemeris in \citet{Carleo2020} are especially relevant. Finally, we note that we have also been able to significantly detect the eccentricity of both planets, which is especially important in the context of planet formation models and possible observations of the planetary eclipses for atmospheric studies.


\subsection{Stability analysis}
\label{subsec_stability_analysis}

The TOI-421 system is composed by two close-in low mass planets ($M_b \approx 6.7\,M_\oplus$, $M_c \approx 14.1\,M_\oplus$) in moderate eccentric orbits ($e_b \approx 0.13$, $e_c \approx 0.19$) and near a 3:1 mean motion resonance ($P_c / P_b \approx 3.091$, Table~\ref{tab:fitted_transit_param}).
Because the masses of the planets are in the super-Earth regime, the stability should be assured \citep[eg.][]{Elser_etal_2013}.
Nevertheless, the significant eccentricity of the planets can introduce many higher order mean motion resonances that can disturb the system.

In order to get a reliable and comprehensive view of the stability of the system, we performed a global frequency analysis \citep{Laskar_1990, Laskar_1993PD} in the same way as achieved for other planetary systems \citep[e.g.][]{Correia_etal_2005, Correia_etal_2010}.
The system was integrated on a regular 2D mesh of initial conditions in the vicinity of the best fit (Table~\ref{tab:fitted_transit_param}).
We used the symplectic integrator SABAC4 \citep{Laskar_Robutel_2001}, with a step size of $5 \times 10^{-4} $~yr and general relativity corrections.
Each initial condition was integrated for 5000~yr, and a stability indicator, $\Delta = |1-n_b'/n_b|$, was computed. 
Here, $n_b$ and $n_b'$ are the main frequency of the mean longitude of the planet over 2500~yr and 5000~yr, respectively, calculated via the frequency analysis \citep{Laskar_1993PD}. In Figure~\ref{fig:SF1},
the results are reported in colour, where orange and red represent strongly chaotic trajectories with $\Delta > 10^{-2}$, while extremely stable systems for which $\Delta < 10^{-8}$ are shown in cyan and blue. 
Yellow indicates the transition between the two, with $\Delta \sim10^{-4}$.

In a first experiment, we explored the stability of the system by varying the orbital period and the eccentricity of the inner planet (Fig.~\ref{fig:SF1}, left).
We confirm that the best fit solution is stable, despite the existence of some higher order mean motion resonances for eccentricities $e_b > 0.3$.
We also note the presence of a large unstable region at low eccentricity on the right hand side of the figure, corresponding to the 3:1 mean motion resonance.
The inner planet is close enough to the star to undergo strong tidal interactions that drive the period ratio to a value above the exact resonance, as it was already reported for many other near resonant systems comprising low-mass planets \citep[e.g.][]{Lissauer_etal_2011K, Delisle_Laskar_2014}.

As the eccentricity plays an important role in the stability of the system, in a second experiment, we varied the eccentricities of both planets (Fig.~\ref{fig:SF1}, right).
Though the eccentricities are not very well constrained in the best fit solution (Table~\ref{tab:fitted_transit_param}), we observe that the system is stable as long as $e_b \lesssim 0.2$ and $e_c \lesssim 0.3$, that is, the system remains stable even if one takes the $3 \sigma$ maximum for both eccentricities.
We conclude that the TOI-421 planetary system as presented in Table~\ref{tab:fitted_transit_param} is reliable and resilient to the uncertainties in the determination of the orbital parameters.


\section{TTV analysis}
\label{sec_ttv_analysis}

\citet{Carleo2020} investigated the possibility of detecting transit timing variations (TTVs). They found an expected TTV signal with a period of $\sim180$ days and an amplitude of $\sim4$ minutes. However, they identify two main issues preventing a TTV detection in their analysis: Their photometric baseline covers less than half of the expected TTV period and they observe large uncertainties in the individual transit centre times of the TESS observations ($\sim1$ minute for the outer and $\sim4$ minutes for the inner planet). With the addition of a new TESS sector and additional high precision photometric observations by CHEOPS, both issues can be addressed. The photometric baseline now spawns $\sim1500$ days, although it is sparsely sampled.

To attempt a detection of TTVs we employed a TTV analysis using \texttt{allesfitter}. We fixed the orbital periods and mid-transit times for both planets according to our previously retrieved results. We fitted for the times of all individual transit events by fitting for the differences to the expected transit time. For all other parameters we imposed Gaussian priors according to the values listed in Table \ref{tab:fitted_transit_param}. 

\begin{figure}
    \centering
    \includegraphics[width=\hsize]{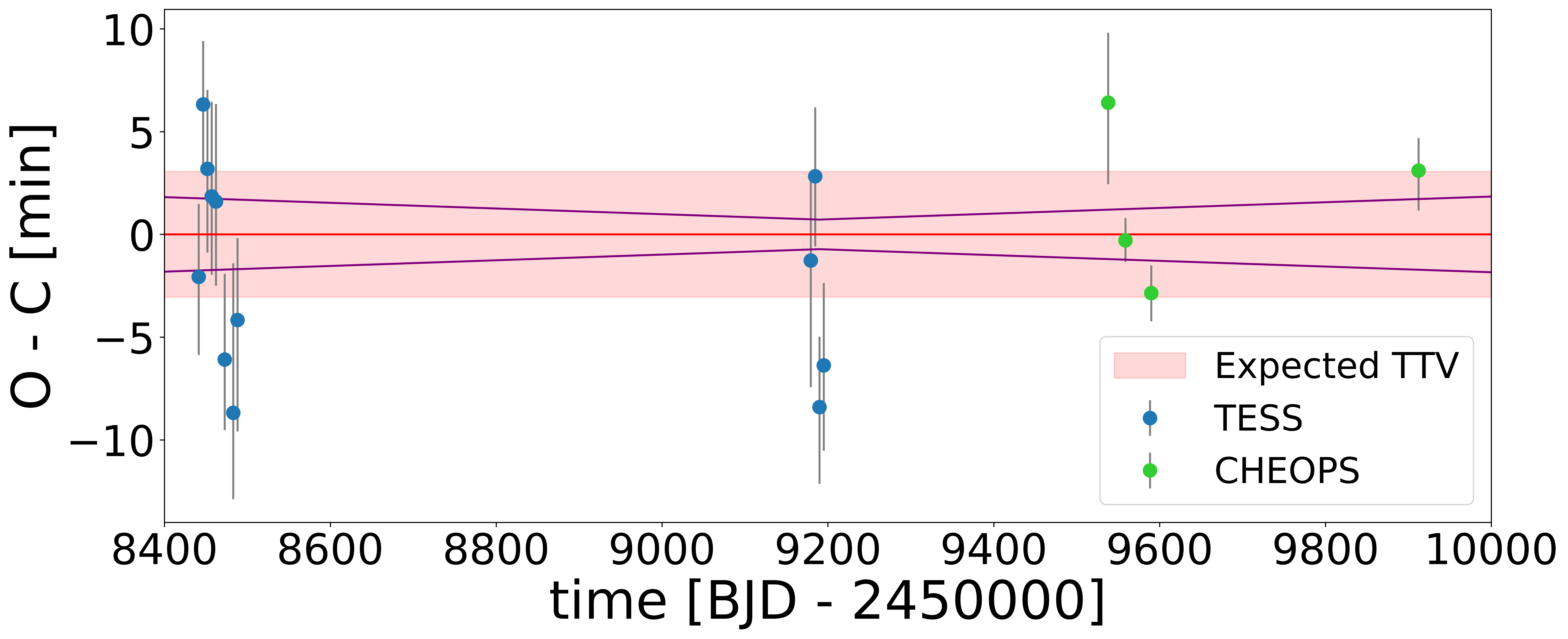}
    \includegraphics[width=\hsize]{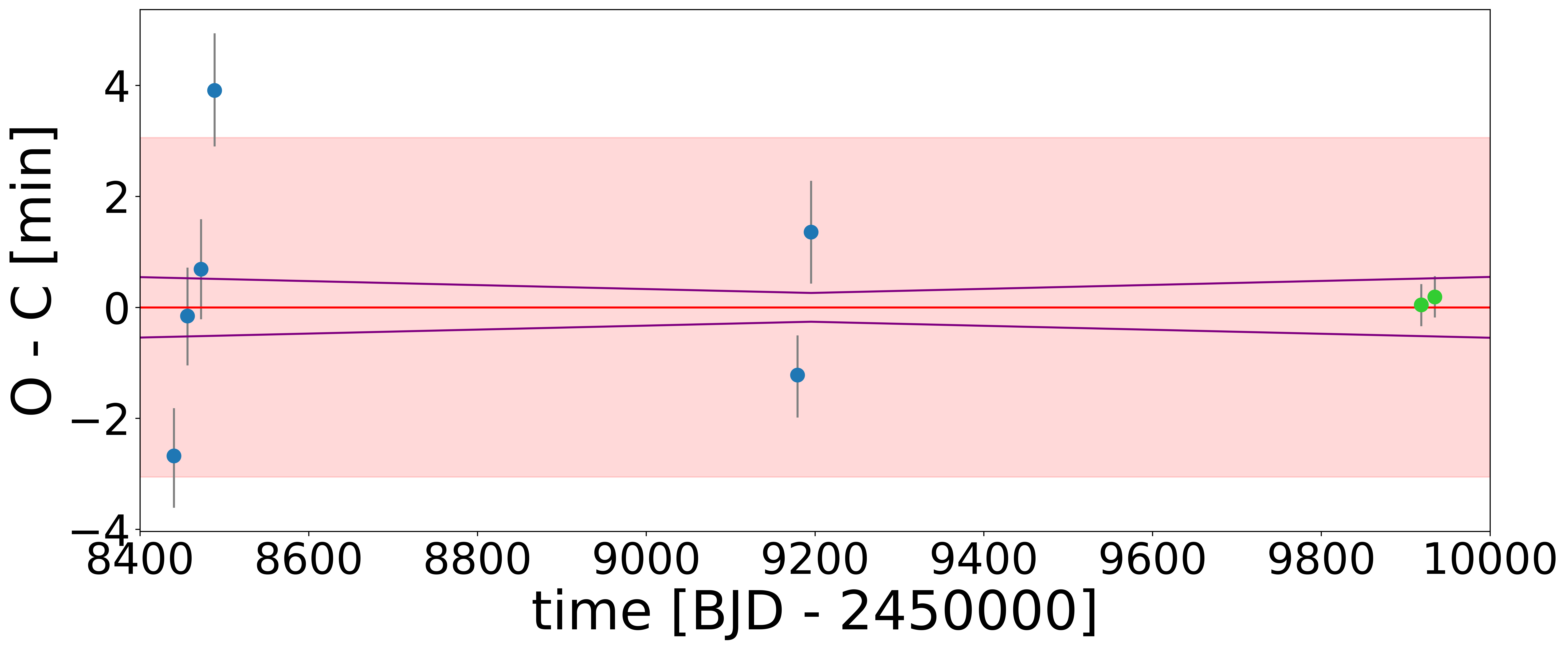}
    \caption{Difference between observed (O) and calculated (C) mid-transit times for TOI-421\,b (top) and TOI-421\,c (bottom) using observations by TESS (blue) and CHEOPS (green). The red line indicates a difference of zero, which corresponds to no TTVs. The purple lines indicate the $1\sigma$ uncertainties of the linear ephemeris. The red area represents the amplitude of the expected variations considering the dynamical interactions of the two planets as computed from the \texttt{TTVfast} code. See text for further details.}
    \label{fig:ttv}
\end{figure}

The differences between the observed and calculated mid-transit times are shown in Figure \ref{fig:ttv}. We also re-computed the amplitude of the TTVs, which would be expected for each planet due to the presence of the other planet. For doing this, we used 300 randomly selected posterior samples from the previous joint analysis (see Section \ref{subsec_transit_analysis}) using the \texttt{TTVFast} code \citep{Deck2014}. The expected TTV amplitude is also shown in Figure \ref{fig:ttv}. We do find hints of TTVs of the order of a few minutes for both planets, however only at low statistical significance. We confirm the observed uncertainties of $\sim4$ and $\sim1$ minutes for the TESS observations reported in \citet{Carleo2020}. The uncertainty on the transit timing of the first CHEOPS observation of planet b is also $\sim4$ minutes. The remaining three transit events of planet b observed by CHEOPS still show uncertainties of $\sim1.5$ minutes, despite the increased photometric precision compared to TESS. Considering also the uncertainty on the linear ephemeris (shown in purple in Figure \ref{fig:ttv}) a statistically significant detection of the expected TTVs would therefore only be possible for the two transit events of planet c observed with CHEOPS, which both show uncertainties of $\sim22$ seconds. However, both of these observations are consistent with no TTVs within their $1\sigma$ confidence intervals. Therefore, despite the longer photometric baseline and the high photometric precision of the CHEOPS observations, we also fail at statistically significantly detecting the expected TTV signal. Additionally, from the TTV analysis we do not find any evidence supporting the existence of additional planets.

\section{Interior structure modelling}
\label{sec_interior}

Using the results from the joint analysis described in Section \ref{subsec_transit_analysis}, we inferred the internal structure of both TOI-421\,b and\,c. We followed the procedure introduced in \cite{Leleu2021}, which is based on the work of \cite{Dorn2017}. In the following, we outline the most important aspects of the used method and present the results.

Our approach uses a full-grid Bayesian inference scheme and models the entire planetary system simultaneously. To make this computationally feasible, we use a deep neural network with six hidden layers trained on our forward model, which models the radius of a planetary structure with a given mass and composition, assuming the planet is spherically symmetric and consists of four fully distinct layers. We use equations of state from \cite{Hakim2018}, \cite{Sotin2007}, and \cite{Haldemann2020} to model the inner iron core (Fe with up to 19\% S), the silicate-rich mantle (Si, Mg, Fe), and a condensed water layer. On top of this structure, we separately model an H-He envelope according to the model of \cite{LopezFortney2014}. For the Si/Mg/Fe ratios of the planets, we assume that they match the ones of the host star exactly \citep{Thiabaud2015}.

In our inference scheme, we use priors that are uniform on a simplex for the mass fractions of the iron core, silicate mantle, and water layer, which ensures that they always add up to one. All these mass fractions are calculated with respect to the condensed part of the planet without the H-He envelope. Additionally, we assume a maximal water mass fraction of 0.5 \citep{Thiabaud2014, Marboeuf2014}. For the H-He layer, our used prior is uniform in log. We note that the results of our analysis do depend to a certain extent on the chosen priors, as this is a highly degenerate problem.

The results are visualised in the corner plots in Figures \ref{internalstructure_b} and \ref{internalstructure_c}. For both planets, the presence of a water layer remains unconstrained. Meanwhile, the posteriors for the mass and radius of the H-He layer as well as the present-day atmospheric mass fraction show median values of $0.03^{+0.04}_{-0.02}$ $M_{\oplus}$, $0.73^{+0.24}_{-0.22}$ $R_{\oplus}$, and $0.0038 \pm 0.0013$ for TOI-421\,b and $2.17^{+0.72}_{-0.67}$ $M_{\oplus}$, $2.96^{+0.25}_{-0.25}$ $R_{\oplus}$, and $0.155 \pm 0.012$ for TOI-421\,c.


\begin{figure}
  \includegraphics[width=\linewidth]{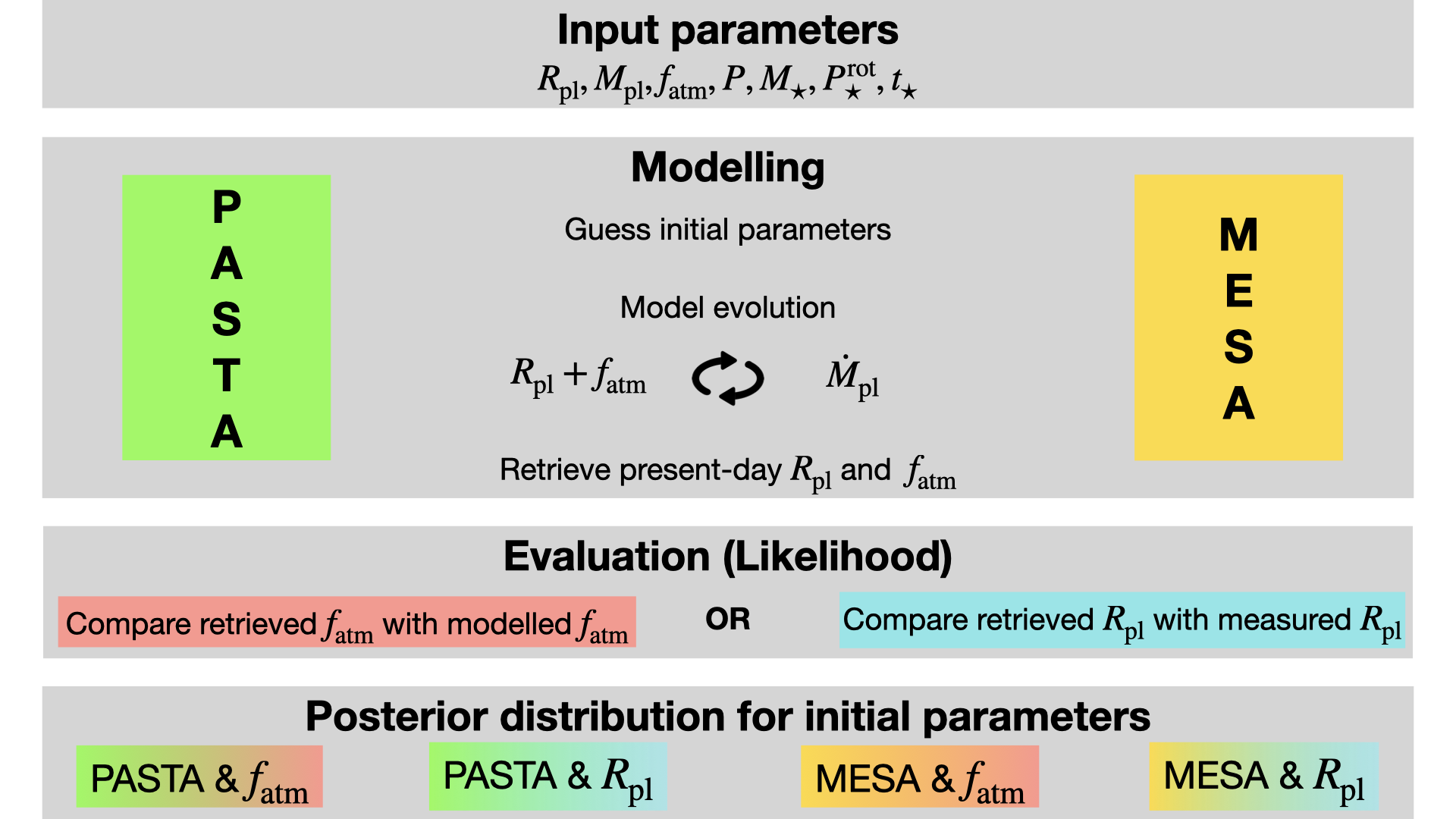}
  \caption{Schematic overview of the different atmospheric evolution modelling approaches compared in this work.}
  \label{fig:atmosmodel}
\end{figure}

\section{Atmospheric evolution}
\label{sec_atmosphere}

In the following sections (\ref{sec:pasta}--\ref{sec_atmosphere_mesa}), we present modelling results of the atmospheric evolution of the planets in the TOI-421 system aiming at constraining their primordial parameters. To have an insight into which physical mechanisms and model assumptions are of particular relevance for such analysis, we employed two different theoretical models and, furthermore, probed two different approaches to fitting the present-day parameters of the planets. First, we employed the \textit{P}lanetary \textit{A}tmospheres and \textit{S}tellar Ro\textit{T}ation r\textit{A}tes~\citep[\texttt{PASTA};][]{bonfanti2021_pasta} algorithm, which has the advantage of a probabilistic approach enabling one to accurately estimate uncertainties, but omits some of the potentially relevant physics (Section \ref{sec:pasta}). For the second model, we employed the framework developed by \citet{kubyshkina2021mesa,kubyshkina2022_tois_II} making use of MESA \citep[Modules for Experiments in Stellar Astrophysics;][]{paxton2011,paxton2013,paxton2018,paxton2019} to track the thermal evolution and atmospheric structure of the planets (later on referred to as 'MESA-based' model; Section~\ref{sec_atmosphere_mesa}). This model includes more advanced physics compared to \texttt{PASTA}, but does not allow one to employ Bayesian statistics due to its long computation time.

In both cases, we run the evolution models attempting to reproduce the present-day parameters of TOI-421\,b and c to constrain the primordial parameters of the system (i.e. the free parameters of our models). However, both of our evolution models employ specific atmospheric structure models (relating radius, mass, and atmospheric mass fraction) different from the model presented in Section\,\ref{sec_interior}. For \texttt{PASTA}, it is the model based on accretion simulations for pure hydrogen atmospheres by \citet{Johnstone2015}, while MESA relies on physics similar to that of \citet[][accounting for thermal evolution]{LopezFortney2014} and initial conditions extrapolated from stellar birth lines. Therefore, when fitting for the present-day radii given in Table~\ref{tab:fitted_transit_param}, both evolution models implicitly fit for atmospheric mass fractions different from that given in Section\,\ref{sec_interior}, which has been obtained using a more sophisticated structure model. Recent literature \citep[e.g.][]{Delrez2021,Cabrera2023} has suggested fitting for the modelled present-day atmospheric mass fractions instead of the radii in order to include the information gained from more sophisticated structure models. In this case, the evolution models implicitly fit for radii different from those obtained from the light curve analysis. In fact, the fitted radii depend on the specific model within the evolution tool that convert $f_{\rm atm}$ into $R_{\rm pl}$. To assess the validity of this alternative approach and identify potential caveats in both approaches, we simulated the evolution for both models once fitting for the observed radii and once fitting for the present-day atmospheric mass fractions from Section\,\ref{sec_interior}. Figure \ref{fig:atmosmodel} provides a schematic overview of the employed workflow and the different model approaches compared in this work. We summarise assumptions and results of each modelling run and fitting approach in Table\,\ref{tab:atm_evo_fit}.

\begin{figure}
  \includegraphics[width=\linewidth]{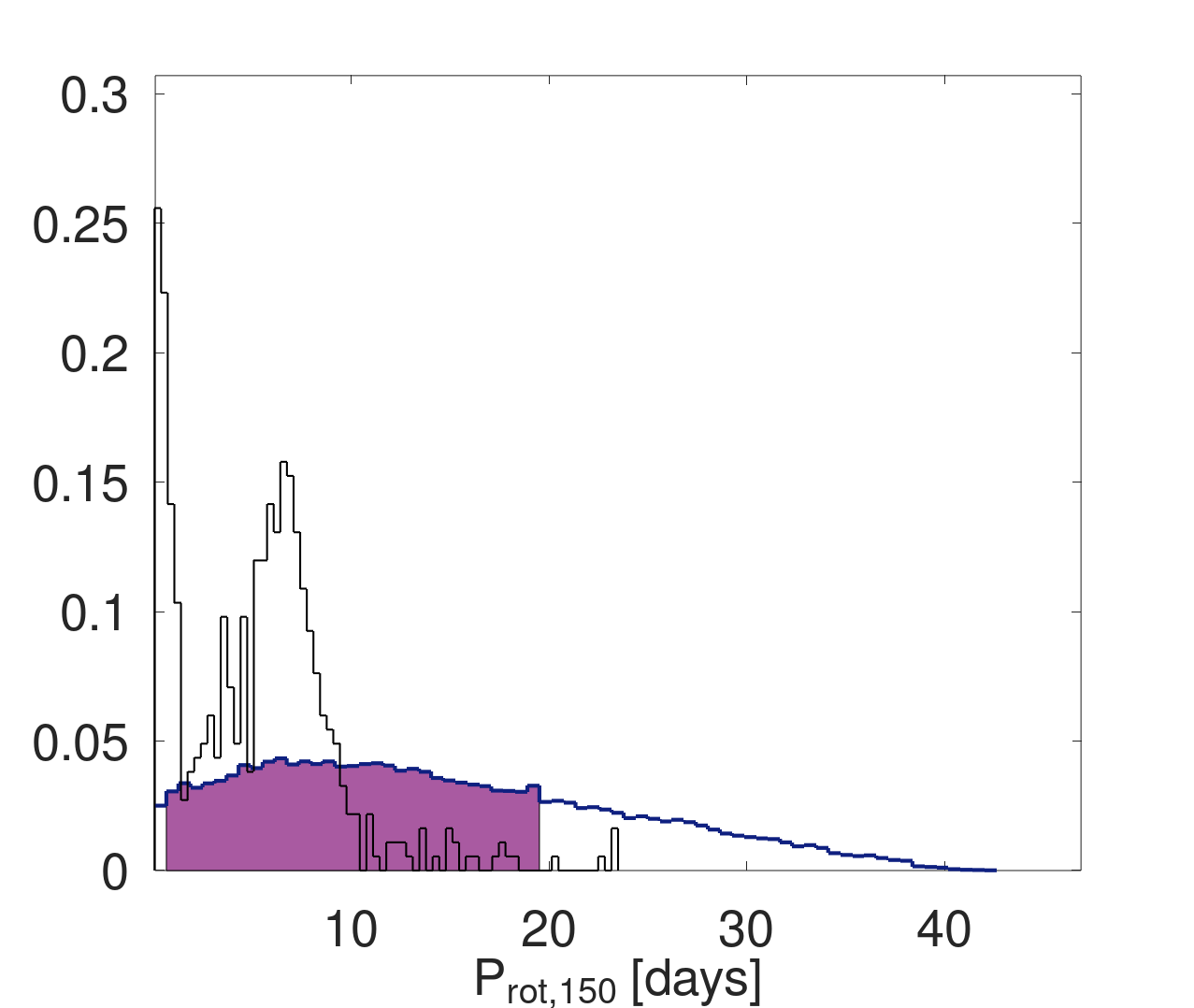}
  \caption{Posterior distribution (dark blue) of the stellar rotation rate of TOI-421 after 150 Myr derived by \texttt{PASTA}. The purple area represents the HPD interval of the distribution. The black line represents the distribution of the stellar rotation rate of young open cluster stars with masses comparable to that of TOI-421 based on the collection of data provided by \citet{johnstone2015_stII}.}
  \label{fig:atmosphericEvolutionStellarRotation}
\end{figure}

\begin{figure*}
  \includegraphics[width=\linewidth]{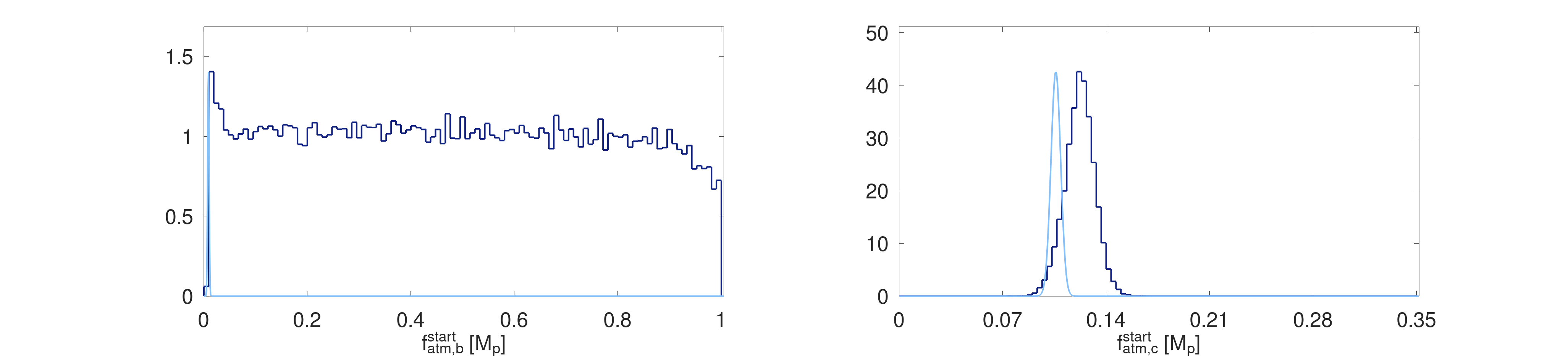}
  \caption{Posterior distribution (dark blue) for the mass of the planetary atmosphere of TOI-421\,b (left) and TOI-421\,c (right) at the time of the dispersal of the protoplanetary disk derived by \texttt{PASTA} when fitting for the planetary radius (i.e. using the interior structure model by \citet{Johnstone2015} to convert the observed radius to the present-day atmospheric mass fraction). The light blue line represents the distribution of the estimated present-day atmospheric mass fraction.}
  \label{fig:atmosphericEvolutionMassFraction_2}
\end{figure*}

\begin{figure*}
  \includegraphics[width=\linewidth]{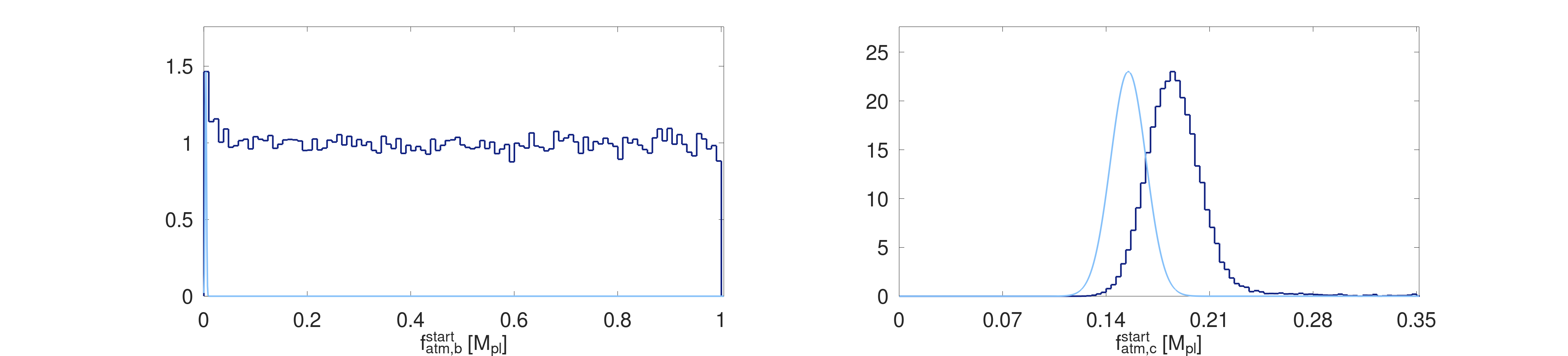}
  \caption{Same as Figure \ref{fig:atmosphericEvolutionMassFraction_2}, but when fitting for the present-day atmospheric mass fraction (i.e. using the interior structure model described in Section \ref{sec_interior} to convert the observed radius to the present-day atmospheric mass fraction).}
  \label{fig:atmosphericEvolutionMassFraction}
\end{figure*}

\subsection{Atmospheric evolution with PASTA}\label{sec:pasta}
PASTA is a planetary atmospheric evolution code based on the original code presented by \citet{Kubyshkina2019a,Kubyshkina2019b}. To account for the wide spread in possible stellar rotation rates (hence, L$_{\rm XUV}$ values) at early ages \citep[e.g.][]{tu2015}, \texttt{PASTA} simultaneously constrains the evolution of planetary atmospheres and of the stellar rotation rate by combining a model predicting planetary atmospheric escape rates based on hydrodynamic simulations -- this has the advantage over other commonly used analytical estimates in that it accounts for both XUV-driven and core-powered mass loss \citep{kubyshkina2018grid} --, a model of the evolution of the stellar XUV flux \citep[using a broken power law to model the evolution of the stellar rotation period;][]{bonfanti2021_pasta}, a model relating planetary parameters and atmospheric mass~\citep{Johnstone2015}, and stellar evolutionary tracks~\citep{Choi2016}.

\texttt{PASTA} works under two main assumptions: (1) planet migration did not occur following the dispersal of the protoplanetary disk; and (2) the planets hosted at some point in the past or still host a hydrogen-dominated atmosphere. The free parameters (i.e. subject to uniform priors) are the planetary initial atmospheric mass fractions at the time of the dispersal of the protoplanetary disk ($f_{\rm atm}^{\rm start}$), which we assume occurs at an age of 5\,Myr \citep[see for example][]{Alexander2014,Kimura2016,Gorti2016}, and the stellar rotation period at 150 Myr as well as the index of the power law controlling the stellar rotation period of stars older than 2 Gyr, which both are used as a proxy for the stellar XUV emission. The code returns constraints on the free parameters and on their uncertainties by implementing the atmospheric evolution algorithm \citep[for more details on the algorithm see][]{bonfanti2021_pasta} in a Bayesian framework (namely the MC3 code~\citealt{Cubillos2017}), using the system parameters with their uncertainties as input priors. \texttt{PASTA} can be set up to either (1) fit the observed planetary radius \citep[e.g.][]{bonfanti2021_pasta} or alternatively (2) the modelled present-day atmospheric mass fraction \citep[e.g.][]{Delrez2021,Cabrera2023}.

As a proxy for the evolution of the stellar rotation period, Fig.~\ref{fig:atmosphericEvolutionStellarRotation} displays the posterior distribution (including the high posterior density, HPD) of the stellar rotation period at an age of 150\,Myr ($P_{\mathrm{rot,150}}$) when fitting for the present-day atmospheric mass fraction. This distribution is then compared to that of stars member of young open clusters that are of comparable mass extracted from~\citet{johnstone2015_stII}. We find that \texttt{PASTA} is unable to constrain the rotation history of the host star, as indicated by the rather flat posterior distribution. We find an identical result (not shown here) when fitting for the planetary radius instead of the present-day atmospheric mass fraction, which is to be expected as the rotation history of the star is unconstrained and therefore independent of the adopted present-day planetary parameters. 

Figures \ref{fig:atmosphericEvolutionMassFraction_2} and \ref{fig:atmosphericEvolutionMassFraction} show the posterior distributions of the initial atmospheric mass fraction for both planets in comparison to the present-day atmospheric mass fraction obtained when fitting for the observed planetary radius and the modelled present-day atmospheric mass fraction, respectively. In both cases \texttt{PASTA} is unable to constrain the initial atmospheric mass fraction of TOI-421\,b. For planet c, both approaches return a well-constrained Gaussian posterior distribution for the initial atmospheric mass fraction, indicating that TOI-421\,c has lost significant amounts of its primary atmosphere to hydrodynamic escape over the course of its evolution.


\begin{figure}
	\centering
	\includegraphics[width=\hsize]{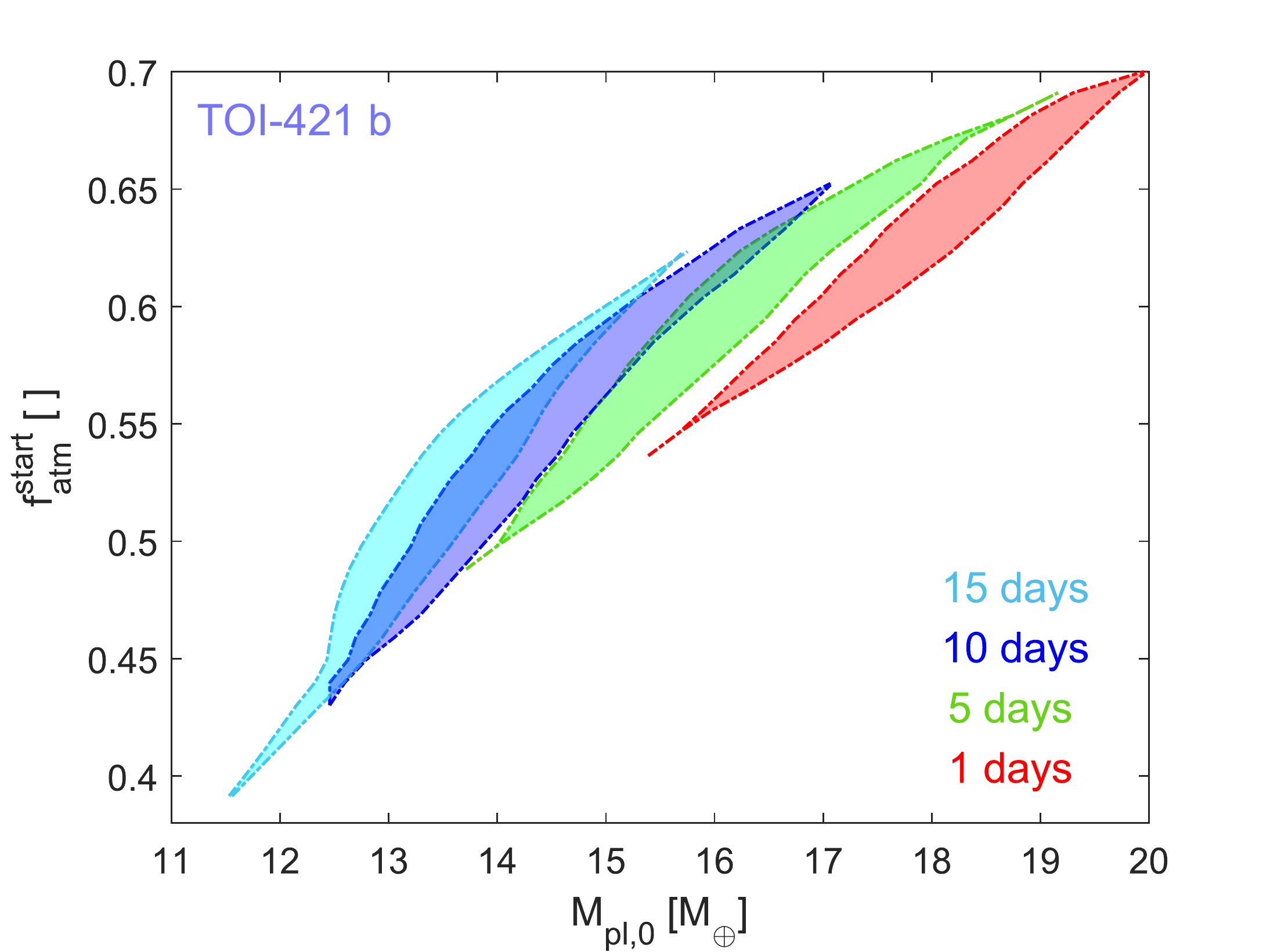}
        \includegraphics[width=\hsize]{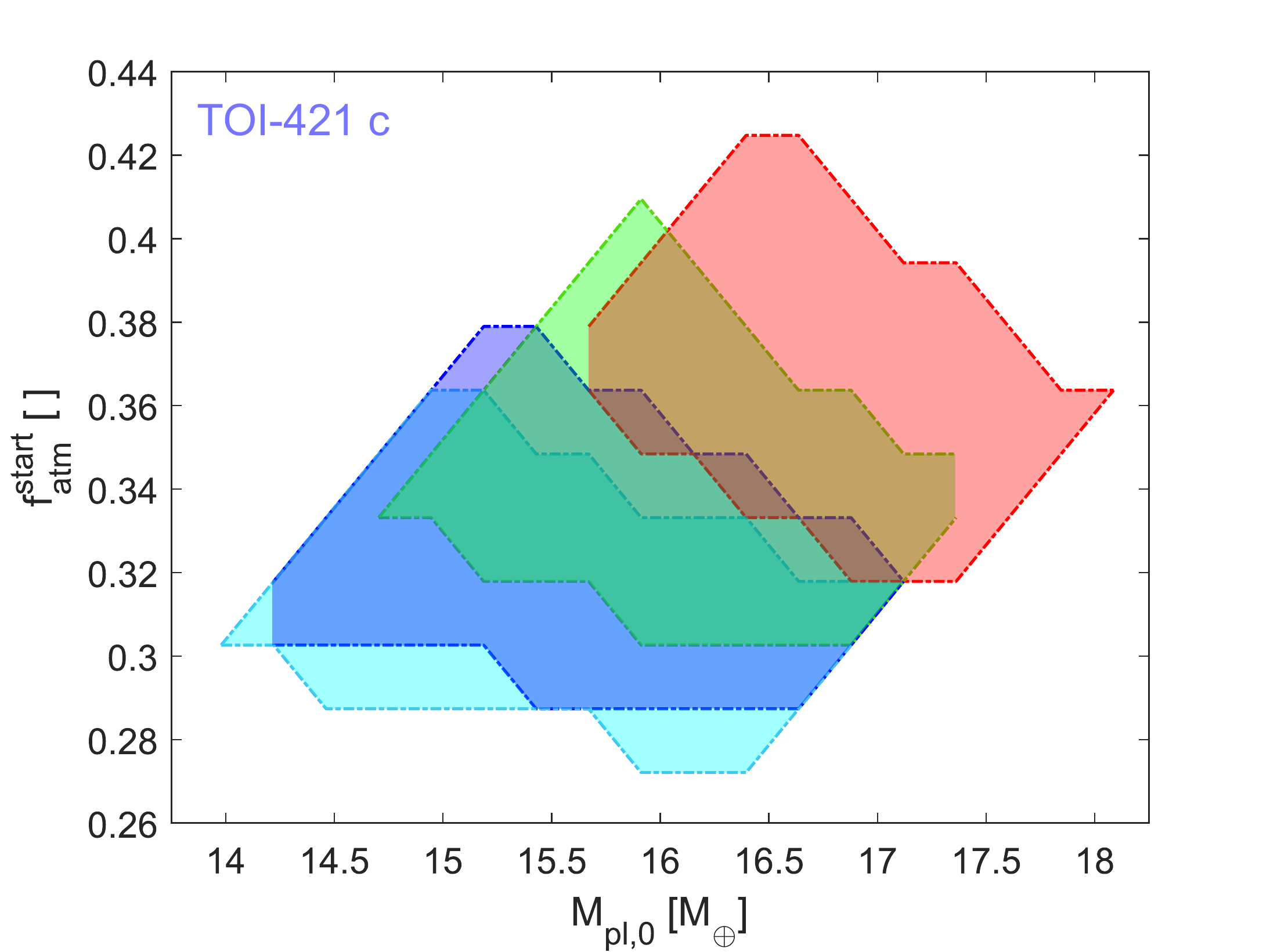}
	\caption{Results of the atmospheric evolution modelling with MESA. \textit{Top panel}: Initial atmospheric mass fraction against the initial mass of the planet allowing us to reproduce the present-day parameters of TOI-421\,b according to our atmospheric evolution models. The admissible combinations of $f_{\rm atm}^{\rm start}$ and $M_{\rm pl,0}$ are shown by the shaded areas, and different colours correspond to different assumptions on the stellar rotation history (i.e. on the rotation period at 150\,Myr), as indicated in the legend. \textit{Bottom panel}: Same as \textit{top}, but for TOI-421\,c.}
	\label{fig:fat0_mpl0}
\end{figure}

\begin{figure}
	\centering
	\includegraphics[width=\hsize]{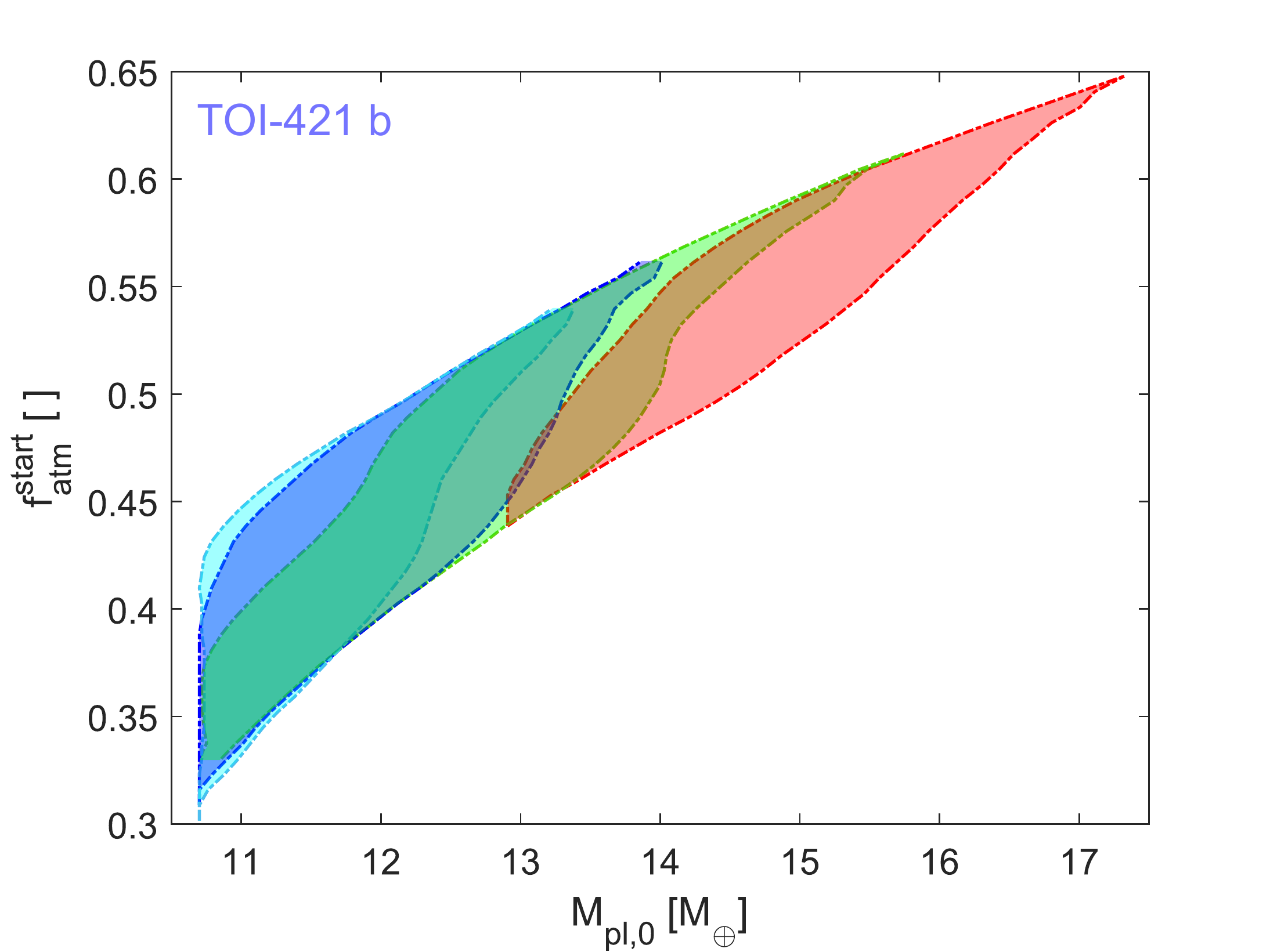}
        \includegraphics[width=\hsize]{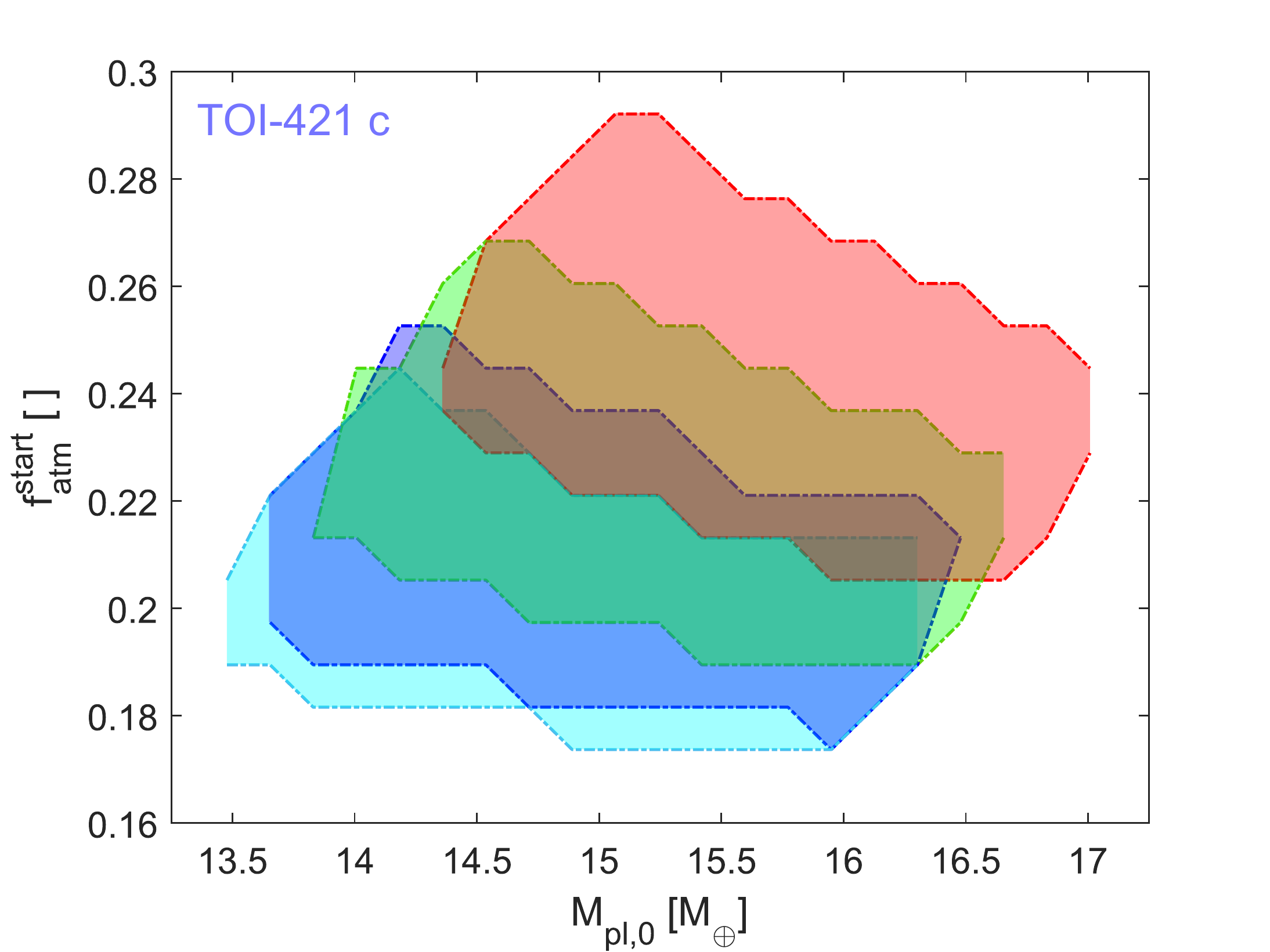}
	\caption{Initial atmospheric mass fraction against the initial mass of the planet allowing us to reproduce the present-day parameters of TOI-421\,b (top) and TOI-421\,c (bottom) obtained employing the atmospheric mass fraction estimated in Section\,\ref{sec_interior}, instead of the observed radius. Notations are the same as in Figure\,\ref{fig:fat0_mpl0}.}
	\label{fig:fat0_mpl0_fatver}
\end{figure}

\subsection{Atmospheric evolution based on MESA}\label{sec_atmosphere_mesa}
As an alternative to \texttt{PASTA}, we also employed the planetary atmosphere evolution model developed by \citet{kubyshkina2021mesa,kubyshkina2022_tois_II} relying on the same basic assumptions (no migration and accretion of a primary atmosphere), but employing more sophisticated physics. In detail, the model combines the thermal evolution of a planet hosting a hydrogen-dominated envelope performed with MESA \citep{paxton2011,paxton2013,paxton2018,paxton2019} and atmospheric escape based on hydrodynamic modelling \citep{kubyshkina2018grid,kubyshkina2021grid} to track the evolution of planetary atmospheric parameters with time. In terms of thermal evolution, the model accounts for both the relaxation of the post-formation luminosity of a planet and changes in planetary equilibrium temperature due to the evolution of the host star. The atmospheric mass loss rates are obtained by direct interpolation on a grid of hydrodynamic upper atmosphere models \citep{kubyshkina2021grid}.

To account for both the evolution of $L_{\rm XUV}$ and of internal stellar parameters, the model employs the \texttt{Mors} stellar evolution code \citep{johnstone2021mors,spada2013}. The stellar parameters predicted by \texttt{Mors} for a 0.833\,$M_{\odot}$ star at the age of $10.9^{+2.9}_{-5.2}$\,Gyr are consistent with the present-day parameters of TOI-421 within $1\sigma$. In terms of X-ray luminosity, the model prediction lies about 15\% below the constraint made by \citep{Carleo2020} based on the $\log{R'_{HK}}$ index as outlined in \citet{Linsky2013ApJ...766...69L,Linsky2014ApJ...780...61L,Fossati2017A&A...601A.104F}. For the specific internal parameters of the model, such as planetary core parameters and the lifetime of the protoplanetary disk, we followed the approach used in \citet{kubyshkina2022_tois_II}, but these parameters have a minor impact on the results.

As the MESA-based evolutionary model is too computationally expensive to run within a Bayesian framework, for the parameters of the system, we consider either a single value (if the parameter is well constrained, further considering that variations within 1$\sigma$ have a minor impact on the results) or a discrete grid of values (for the key parameters). Thus, we considered single values of stellar mass and orbital separations, but a range of values for the initial stellar rotation rate (controlling $L_{\rm XUV}$) and the initial planetary parameters. Namely, we adopted rotation periods of 1, 5, 10, and 15 days at 150 Myr to span the expected range of stellar rotation periods in young open clusters (see for example Fig.\,\ref{fig:atmosphericEvolutionStellarRotation}).
Furthermore, for planet b, we considered initial masses spanning from 10.5 to 24\,$M_{\oplus}$ and, for each planetary mass, initial atmospheric mass fractions varying between {15\%} and 70\%. {The grid steps for both parameters vary depending on the gradient of the predicted values. Thus, for the planetary mass, the step varies from $0.2$\,$M_{\oplus}$ (low-mass end) to $1.0$\,$M_{\oplus}$ (at high masses); for the atmospheric mass fraction the step varies between 1\% and 5\%. For the presentation of the results, we interpolate the outputs onto a regular grid.} Differently from \citet{kubyshkina2022_tois_II}, we discarded $f_{\rm atm}^{\rm start}$ values above 70\%, as they are unlikely from the point of view of formation models. Instead, we extended our grid to lower mass and atmospheric mass fraction values to account for the changes in the observational constraints. For planet c, we considered $M_{\rm pl,0}$ and $f_{\rm atm}^{\rm start}$ lying in the 12.7--22.2\,$M_{\oplus}$ and 15--44\% range, respectively. The grid steps are 0.2--0.7\,$M_{\oplus}$ for the initial mass and 1--2\% for the initial atmospheric mass fraction. After performing forward modelling, we checked a-posteriori, which combinations of initial parameters reproduce the estimated present-day planetary properties accounting for their 1$\sigma$ uncertainties.

\begin{table*}
\renewcommand{\arraystretch}{1.2}
\centering
\caption{Fitting parameters of the evolution models and their constraints on the initial parameters of TOI-421 system.}
\label{tab:atm_evo_fit}
\begin{tabular}{c|c|c|cc|cc|c}
\hline
\hline
Planet & Model\tablefootmark{(a)} & Fitting parameter &  $R_{\rm p}$ [$R_{\oplus}$]\tablefootmark{(b)}  & $f_{\rm atm}$ \tablefootmark{(b)} & $f_{\rm atm}^{\rm start}$ \tablefootmark{(c)} & $M^{\rm start}_{\rm p}$ [$M_{\oplus}$] \tablefootmark{(c)} & $f_{\rm atm}^{\rm loss}$ [\%] \tablefootmark{(d)}\\
\hline
b & \texttt{PASTA} & $R_{\rm p}$   & $\mathbf{2.64 \pm {0.08}}$ & $0.0087\pm0.0014$ & -- & -- & -- \\
b & \texttt{PASTA} & $f_{\rm atm}$ & $2.32 \pm 0.13$     & $\mathbf{0.0038\pm{0.0013}}$ & -- & -- & -- \\
c & \texttt{PASTA} & $R_{\rm p}$   & $\mathbf{{5.09\pm 0.07}}$ & $0.106\pm0.003$ & $0.122\pm0.010$ & $14.4\pm 1.4$ & $ 15 \pm 8$ \\
c & \texttt{PASTA} & $f_{\rm atm}$ & $6.47 \pm 0.7$     & $\mathbf{0.155 \pm{0.012}}$ & $0.19\pm0.02$ & $14.7 \pm 1.5$ & $ 22 \pm 12$\\
b & MESA-based     & $R_{\rm p}$   & $\mathbf{2.64 \pm {0.08}}$ & $0.026\pm0.008$ & $0.55 \pm 0.15$ & $14.5\pm5.0$ & $98 \pm 2$ \\
b & MESA-based     & $f_{\rm atm}$ & $2.13\pm0.13$ & $\mathbf{0.0038\pm{0.0013}}$ & $0.47\pm0.17$ & $12.6\pm4.2$ & $\sim99 $ \\
c & MESA-based     & $R_{\rm p}$   & $\mathbf{5.09\pm 0.07}$ & $0.242\pm0.014$ & $0.35\pm0.08$ & $16.4\pm2.6$ & $41 \pm 21$ \\
c & MESA-based     & $f_{\rm atm}$ & $4.3\pm0.2$ & $\mathbf{0.155\pm{0.012}}$ & $0.23\pm0.06$ & $15.5\pm2.0$ & $39 \pm 22$\\
\hline
\end{tabular}
\tablefoot{Unconstrained values are given as --. \\
\tablefoottext{a}{\texttt{PASTA} models both planets and the stellar rotation history simultaneously, while MESA-based models consider each planet individually, further assuming a fixed stellar rotation history.} \\
\tablefoottext{b}{$R_{\rm p}$ (planetary radius) and $f_{\rm atm}$ (present-day atmospheric mass fraction) are constraining parameters, with the value in bold being the one used to constrain the model, while the other one is derived from the respective model, as listed in column three.} \\
\tablefoottext{c}{$f_{\rm atm}^{\rm start}$ (initial atmospheric mass fraction) and $M^{\rm start}_{\rm p}$ (planetary mass at the beginning of the evolution) are parameters retrieved from the model.} \\
\tablefoottext{d}{Percentage of atmosphere lost during evolution given as $M_{\rm atm}^{\rm loss} = \frac{M_{\rm atm}^{\rm start} - M_{\rm atm}}{M_{\rm atm}^{\rm start}}$}
}
\end{table*}

Figures \ref{fig:fat0_mpl0} and \ref{fig:fat0_mpl0_fatver} summarise our results by showing the $M_{\rm pl,0}$ and $f_{\rm atm}^{\rm start}$ pairs that enable us to reproduce the present-day mass and radius and atmospheric mass fraction assuming different stellar rotation evolution scenarios. At its short orbit, TOI-421\,b experiences extreme atmospheric escape with predicted mass loss rates at the present time as high as $\sim10^{11}-10^{12}$\,g\,s$^{-1}$ \citep{Carleo2020,kubyshkina2022_tois_I,kubyshkina2022_tois_II,Berezutsky2022}. Therefore, assuming that the planet did not migrate since the time of protoplanetary disk dispersal, we conclude that to achieve its present-day parameters, TOI-421\,b had to start its evolution 
with a mass at least a factor of 1.6 larger than measured and with an initial atmospheric mass fraction larger than 30\%, even if the host star evolved as a very slow rotator. There is also a strong correlation between $M_{\rm pl,0}$ and $f_{\rm atm}^{\rm start}$ indicating that the planet has undergone a strong atmospheric boil-off phase at the beginning of its evolution.

Similar to when employing \texttt{PASTA}, we find that TOI-421\,c has lost significant amounts of its primordial atmosphere to hydrodynamic escape. However, in the case of the \texttt{MESA}-based models, the escape is expected to be more significant compared to the results of the \texttt{PASTA} models, implying that the planet had a more extensive hydrogen envelope at the beginning of its evolution.


\section{Discussion}
\label{sec:discussion}

\subsection{Comparison of different atmospheric evolution models}
In Section \ref{sec_atmosphere}, we have employed two different algorithms to simulate the evolution of the primary atmospheres of the TOI-421 planets. The two models mainly differ in the interior structure model they employ to convert atmospheric mass fractions to planetary radii and vice versa. Additionally, the algorithm using the \texttt{MESA} models does also account for planetary thermal evolution and how this affects mass loss, while \texttt{PASTA} does not. Furthermore, they employ different models to follow the evolution of the stellar rotation rate and XUV luminosity. While both models always use their built-in structure models at every step of the evolution, we also tried to constrain the present-day gas-mass by considering the results of a more sophisticated structure model (see Section \ref{sec_interior}). In these cases, the models fit for the present-day atmospheric mass fraction instead of the observed present-day radius.

In general, we find that the predicted initial atmospheric mass fraction is strongly dependent on the underlying interior structure model used to convert a given planetary radius to an atmospheric mass fraction, and vice versa. When the structure model predicts a higher present-day atmospheric mass fraction, the evolution also results in a higher initial atmospheric mass fraction. This is to be expected as the existence of more gas today presumably implies the existence of more gas at the beginning of the evolution. The relation is however not linear, because a larger amount of gas at any point in time implies a larger planetary radius, and thus a stronger mass loss \citep{kubyshkina2020mesa}. Therefore we observe that the \texttt{MESA}-based models predict consistently larger initial atmospheric mass fractions than the \texttt{PASTA} models, because its corresponding structure model also results in larger atmospheric mass fractions given the same planetary radii.

With respect to using a modelled present-day atmospheric mass fraction instead of the observed planetary radius to constrain the evolution model \citep[e.g.][]{Delrez2021,Cabrera2023}, we find that this approach can lead to significantly over- or underestimated planetary radii in the evolution calculations. Since each model always uses its internal structure model to convert atmospheric mass fractions to planetary radii and vice versa at every step of the evolution, this approach forces the model to assume a radius different from the one actually observed in order to match its prediction of an atmospheric mass fraction with the input value. The calculation of a mass loss rate is however strongly dependent on the planetary radius, implying that this approach does not predict consistent mass loss rates during the evolution. Therefore, to employ this approach it is first necessary to check that the interior structure model used in the evolution model returns a radius consistent with the observed one, when providing the atmospheric mass fraction as input.

When converting the observed planetary radius to an atmospheric mass fraction, the more simplistic interior structure models used within our evolution models are not consistent with the more complex interior structure model presented in Section \ref{sec_interior}. Nevertheless, constraining our evolution models with the results of the more complex interior structure model, also resulted in being inaccurate. Therefore, it appears that none of the model approaches presented in this work are ideal to simulate the atmospheric evolution of this system. Their results may therefore be only considered to be indicative and firm conclusions should be drawn on results that are consistent with all of the approaches. To resolve this issue, new atmospheric evolution models, which use a more complex interior structure model self-consistently, are necessary. Ideally, such a model would also include the possibility to model the presence of volatiles, such as water vapour, in the atmosphere \citep{burn2024_atmospheresWithWater}.

\subsection{Planet formation and evolution scenarios of the TOI-421 planetary system}
\citet{mordasini2020} derived a power-law to compute the envelope mass $M_{\rm e,0}$ accreted by a planet while being embedded in the protoplanetary disk as a function of its core-mass $M_{\rm c}$ and orbital separation $a$

\begin{equation}
\label{eq_mordasini}
    \frac{M_{\rm e,0}}{M_{\oplus}} = 0.024 \left(\frac{M_{\rm c}}{M_{\oplus}}\right)^{2.23} \left( \frac{a}{1 \rm AU} \right)^{0.72}\,.
\end{equation}

This power-law results from fits of corresponding numerical simulations assuming a Sun-like star and a grid of planets with orbital separations ranging from $0.1$ to $1$ AU and core-masses ranging from $1$ to $10\,M_{\oplus}$. Although also being a G-type star, TOI-421 is not exactly Sun-like and with an orbital separation of $\sim0.055$ AU planet b is also not covered by the grid. Furthermore, the amount of accreted gas is dependent on the lifetime of the protoplanetary disk, with longer disk-lifetimes resulting in more accretion. The disk-lifetime was a free parameter in the simulations of \citet{mordasini2020} and the derived power-law was then computed on the whole set of simulation results, which had an average disk-lifetime of $2$ Myrs. However, we do not know the real lifetime of the protoplanetary disk of TOI-421, which could be significantly different from $2$ Myrs. Therefore, Equation (\ref{eq_mordasini}) can just be used to get an order-of-magnitude estimate of the accreted gas-mass. 

Employing Equation (\ref{eq_mordasini}) for both planets using the planetary parameters in Table \ref{tab:fitted_transit_param} and assuming that $M_{\rm c} = M_{\rm p} - M_{\mathrm{gas}}$ with $M_{\mathrm{gas}}$ being the present-day gas-mass derived in Section \ref{sec_interior}, we computed an initial atmospheric mass fraction of $0.030$ and $0.097$ for TOI-421\,b and c, respectively. As previously concluded by both \citet{Carleo2020} and \citet{kubyshkina2022_tois_II}, we find that TOI-421\,b should have lost all of its primary atmosphere to hydrodynamic escape. Even if the planet had accreted a gas-mass an order-of-magnitude larger than estimated by Equation (\ref{eq_mordasini}) and even if the star had been a very slow rotator, the short orbital separation would have led to enough XUV irradiation to completely strip the planet of its atmosphere. Under the assumptions taken by the evolution models considered in this work, the low measured density of planet b remains an unresolved problem. 

In the case of TOI-421\,c, we find that all of our atmospheric evolution models result in the planet having had a larger initial atmospheric mass fraction than that predicted by Equation \ref{eq_mordasini}. This would imply that similarly to planet b, it is unlikely that planet c formed and evolved hosting a hydrogen-dominated atmosphere at its current orbital separation.

As discussed by \citet{kubyshkina2022_tois_II}, the low observed mean density of TOI-421\,b might be explained by migration, which implies that the planet formed at a larger orbital separation, where it could have accreted more volatiles. Therefore, following \citet{kubyshkina_fossati2022}, without migration, a planet born with a primary atmosphere could have the parameters of TOI-421\,b in terms of mass and radius if it formed -- and evolved -- around 0.2--0.3\,AU. In a migration scenario, this distance can be seen as a closest-in orbit at which TOI-421\,b could have been formed, with the real birthplace located likely further away from the star. In particular, if the planet formed beyond the ice line, a large fraction of its atmosphere would be likely formed or replenished by (partial) melting of the ice in its core \citep{Lopez2017}. In this case, atmospheric evolution would likely be rather different from the case of a pure hydrogen-dominated atmosphere atop a rocky core as considered here, because the available large amount of water vapour in the atmosphere would lead to a more compact atmosphere, significantly reducing atmospheric loss. Furthermore, the addition of a significant amount of heavier species in the atmosphere would lead to additional cooling, and thus lower mass loss \citep[e.g.][]{johnstone2020}. Additionally, one would also need to employ a different internal structure model \citep[e.g.][]{Aguichine2021} than the one used in Section \ref{sec_interior}. If TOI-421\,b has been subject to migration, TOI-421\,c most likely has migrated as well. As mentioned before, the majority of atmospheric evolution models would require planet c to have accreted significantly more gas than what is predicted by Equation (\ref{eq_mordasini}). Analogous to planet b, this could also be explained by formation further out compared to the current orbital location and subsequent inward migration of TOI-421\,c.

Alternatively, the low density of TOI-421\,b might also be explained by the presence of high-altitude clouds \citep[e.g.][]{Lammer2016,Cubillos2017,Gao2020}. These clouds would bias the measurement of the planetary radius by artificially inflating it compared to that obtained assuming a clear atmosphere. In the case of the Kepler-bandpass (430--880 nm), which is very similar to the CHEOPS-bandpass, a clear hydrogen-dominated atmosphere is opaque in transmission at $\sim100$ mbar \citep{Lammer2016,Gao2020}, while the presence of aerosols can make the atmosphere opaque already at pressures as low as several nbar. The exact pressure at which the atmosphere becomes opaque is strongly dependent on the atmospheric temperature profile and chemical composition. In order to quantify the possible bias, we use the pressure-temperature profile computed for TOI-421\,b by the hydrodynamic model described in \citet{kubyshkina2018grid} to determine the height of the $1 \mu \rm{bar}$-level, which is a plausible location of high-altitude clouds \citep{Kawashima2018,Kawashima2019a,Kawashima2019b,Lavvas2019,Gao2020}. Assuming that the atmosphere is opaque below this level, this would imply an overestimation of the planetary radius by 34\%. The planetary radius would then be at $\sim1.97\,R_{\oplus}$, which would in turn result in a mean planetary density of $\sim0.88\, \rho_{\oplus}$. Using the interior structure model by \citet{Johnstone2015} built into \texttt{PASTA} this would result in a present-day atmospheric mass fraction of $0.00011$ (i.e. $\sim98.7\%$ less than when using the observed radius). Additionally, assuming identical relative uncertainties on mass and radius as for the actual measurements (see Table \ref{tab:fitted_transit_param}), the planet would also be consistent with no H-He atmosphere at all within $1\sigma$, thus allowing for the planet to have formed at its current location without the need of additional orbital migration. Future transmission spectroscopy observations with the James Webb Space Telescope \citep{Gardner2023} can enable one to explore the atmospheric composition and the existence of high-altitude aerosols, thus constraining the formation and evolution scenarios described above.


\section{Conclusion}
\label{sec_conclusion}

Using three sectors of TESS data, six CHEOPS visits, and a total of 156 archival radial velocity data points obtained by four different instruments, we derived the mean density of TOI-421\,b $0.37 \pm 0.05~\rho_{\oplus}$ and TOI-421\,c $0.107 \pm 0.012~\rho_{\oplus}$, provided updated ephemerides for both planets and a detailed characterisation of the host star. In the process we compared a novel reduction approach for HARPS data, which fits a Skew Normal fit onto the CCFs, with the standard DRS reduced data, finding that results obtained with both approaches are well in agreement with each other. We also attempted to detect TTVs, but concluded that the photometric precision achieved by CHEOPS is not sufficient to statistically significantly detect the expected TTV signal of planet b. In the case of planet c, despite the sufficient precision achieved by CHEOPS, no TTV signal could be identified.

On the basis of the newly retrieved planetary parameters and assuming the presence of a hydrogen-dominated atmosphere, we performed modelling of the interior structure of the planets. We find that under this assumption both planets would have an extensive envelope. Finally, we modelled the evolution of these hydrogen-dominated atmospheres using four different approaches and two different atmospheric evolution models. Independently of the considered model and evolution of the stellar rotation rate, assuming no post-disk migration, we found that TOI-421\,b should have completely lost its primary atmosphere. The observed low mean density can therefore not be explained by a formation and subsequent evolution of planet b at its current orbital position. One possible explanation would be a bias in the measured planetary parameters, particularly the planetary radius, as a result of the presence of high-altitude aerosols. Alternatively, the planet might have also undergone orbital migration, including the possibility of a formation beyond the ice line. In this case the atmosphere would no longer be hydrogen-dominated, but most-likely also include a significant amount of water vapour, which would reduce atmospheric loss. A spectroscopic follow-up observation, for example by JWST, has the potential of resolving this dichotomy.

When comparing the different atmospheric evolution approaches used in this work for TOI-421\,c, we find that the predicted initial atmospheric mass fraction is strongly dependent on the underlying interior structure model. It ranges between $\sim10$ and $\sim35\%$, depending on the considered model and evolution of the stellar rotation rate. While the lower boundary of this range would allow for in situ formation, the majority of solutions would suggest that the planet has also been subject to orbital migration at some point in its evolution. The large range of possible initial atmospheric mass fractions and the fact that no model is clearly more advantageous than the others led us to conclude that results of such models must be reported in ranges rather than in fixed values. It also clearly highlights the necessity of developing new atmospheric evolution models, which use more complex interior structure models self-consistently throughout the modelled evolution, and also account for the presence of volatiles such as water vapour in the atmosphere.

\begin{acknowledgements}
CHEOPS is an ESA mission in partnership with Switzerland with important contributions to the payload and the ground segment from Austria, Belgium, France, Germany, Hungary, Italy, Portugal, Spain, Sweden, and the United Kingdom. The CHEOPS Consortium would like to gratefully acknowledge the support received by all the agencies, offices, universities, and industries involved. Their flexibility and willingness to explore new approaches were essential to the success of this mission.
Based on observations made with the ESO-3.6m telescope at La Silla Observatory under programs ID 1102.C-0923, 0103.C-0874, 0103.C-0759, 0103.C-0442, and 60.A-970.
This work has been carried out within the framework of the NCCR PlanetS supported by the Swiss National Science Foundation under grants 51NF40\_182901 and 51NF40\_205606. 
This project has received funding from the European Research Council (ERC) under the European Union’s Horizon 2020 research and innovation programme (project {\sc Four Aces}. 
grant agreement No 724427). It has also been carried out in the frame of the National Centre for Competence in Research PlanetS supported by the Swiss National Science Foundation (SNSF). DE acknowledges financial support from the Swiss National Science Foundation for project 200021\_200726. 
TWi and ACCa acknowledge support from STFC consolidated grant numbers ST/R000824/1 and ST/V000861/1, and UKSA grant number ST/R003203/1. 
YAl acknowledges support from the Swiss National Science Foundation (SNSF) under grant 200020\_192038. 
The Belgian participation to CHEOPS has been supported by the Belgian Federal Science Policy Office (BELSPO) in the framework of the PRODEX Program, and by the University of Liège through an ARC grant for Concerted Research Actions financed by the Wallonia-Brussels Federation. 
L.D. is an F.R.S.-FNRS Postdoctoral Researcher. 
ML acknowledges support of the Swiss National Science Foundation under grant number PCEFP2\_194576. 
RAl, DBa, EPa, and IRi acknowledge financial support from the Agencia Estatal de Investigación of the Ministerio de Ciencia e Innovación MCIN/AEI/10.13039/501100011033 and the ERDF “A way of making Europe” through projects PID2019-107061GB-C61, PID2019-107061GB-C66, PID2021-125627OB-C31, and PID2021-125627OB-C32, from the Centre of Excellence “Severo Ochoa'' award to the Instituto de Astrofísica de Canarias (CEX2019-000920-S), from the Centre of Excellence “María de Maeztu” award to the Institut de Ciències de l’Espai (CEX2020-001058-M), and from the Generalitat de Catalunya/CERCA programme. 
S.C.C.B. acknowledges support from FCT through FCT contracts nr. IF/01312/2014/CP1215/CT0004. 
XB, SC, DG, MF and JL acknowledge their role as ESA-appointed CHEOPS science team members. 
LBo, VNa, IPa, GPi, RRa, and GSc acknowledge support from CHEOPS ASI-INAF agreement n. 2019-29-HH.0. 
ABr was supported by the SNSA. 
P.E.C. is funded by the Austrian Science Fund (FWF) Erwin Schroedinger Fellowship, program J4595-N. 
This project was supported by the CNES. 
This work was supported by FCT - Fundação para a Ciência e a Tecnologia through national funds and by FEDER through COMPETE2020 - Programa Operacional Competitividade e Internacionalizacão by these grants: UID/FIS/04434/2019, UIDB/04434/2020, UIDP/04434/2020, PTDC/FIS-AST/32113/2017 \& POCI-01-0145-FEDER- 032113, PTDC/FIS-AST/28953/2017 \& POCI-01-0145-FEDER-028953, PTDC/FIS-AST/28987/2017 \& POCI-01-0145-FEDER-028987, O.D.S.D. is supported in the form of work contract (DL 57/2016/CP1364/CT0004) funded by national funds through FCT. V.A. was supported by FCT through national funds by the following grant: 2022.06962.PTDC.
A.C.M.C. acknowledges support from the FCT, Portugal, through the CFisUC projects UIDB/04564/2020 and UIDP/04564/2020, with DOI identifiers 10.54499/UIDB/04564/2020 and 10.54499/UIDP/04564/2020, respectively.
B.-O. D. acknowledges support from the Swiss State Secretariat for Education, Research and Innovation (SERI) under contract number MB22.00046. 
MF and CMP gratefully acknowledge the support of the Swedish National Space Agency (DNR 65/19, 174/18). 
DG gratefully acknowledges financial support from the CRT foundation under Grant No. 2018.2323 'Gaseousor rocky? Unveiling the nature of small worlds'. 
M.G. is an F.R.S.-FNRS Senior Research Associate. 
MNG is the ESA CHEOPS Project Scientist and Mission Representative, and as such also responsible for the Guest Observers (GO) Programme. MNG does not relay proprietary information between the GO and Guaranteed Time Observation (GTO) Programmes, and does not decide on the definition and target selection of the GTO Programme. 
CHe acknowledges support from the European Union H2020-MSCA-ITN-2019 under Grant Agreement no. 860470
(CHAMELEON). 
SH gratefully acknowledges CNES funding through the grant 837319. 
KGI is the ESA CHEOPS Project Scientist and is responsible for the ESA CHEOPS Guest Observers Programme. She does not participate in, or contribute to, the definition of the Guaranteed Time Programme of the CHEOPS mission through which observations described in this paper have been taken, nor to any aspect of target selection for the programme. 
K.W.F.L. was supported by Deutsche Forschungsgemeinschaft grants RA714/14-1 within the DFG Schwerpunkt SPP 1992, Exploring the Diversity of Extrasolar Planets. 
This work was granted access to the HPC resources of MesoPSL financed by the Region Ile de France and the project Equip@Meso (reference ANR-10-EQPX-29-01) of the programme Investissements d'Avenir supervised by the Agence Nationale pour la Recherche. 
PM acknowledges support from STFC research grant number ST/M001040/1. 
This work was also partially supported by a grant from the Simons Foundation (PI Queloz, grant number 327127). 
NCSa acknowledges funding by the European Union (ERC, FIERCE, 101052347). Views and opinions expressed are however those of the author(s) only and do not necessarily reflect those of the European Union or the European Research Council. Neither the European Union nor the granting authority can be held responsible for them. 
S.G.S. acknowledge support from FCT through FCT contract nr. CEECIND/00826/2018 and POPH/FSE (EC). 
GyMSz acknowledges the support of the Hungarian National Research, Development and Innovation Office (NKFIH) grant K-125015, a a PRODEX Experiment Agreement No. 4000137122, the Lend\"ulet LP2018-7/2021 grant of the Hungarian Academy of Science and the support of the city of Szombathely. 
V.V.G. is an F.R.S-FNRS Research Associate. 
NAW acknowledges UKSA grant ST/R004838/1. 
This research has made use of the Spanish Virtual Observatory (https://svo.cab.inta-csic.es) project funded by MCIN/AEI/10.13039/501100011033/ through grant PID2020-112949GB-I00.
JV acknowledges support from the Swiss National Science Foundation (SNSF) under grant PZ00P2\textunderscore208945.
\end{acknowledgements}

\bibliographystyle{aa} 
\bibliography{aa48584-23} 

\begin{thebibliography}{143}
\expandafter\ifx\csname natexlab\endcsname\relax\def\natexlab#1{#1}\fi

\bibitem[{{Adibekyan} {et~al.}(2015){Adibekyan}, {Figueira}, {Santos}, {Sousa}, {Faria}, {Delgado-Mena}, {Oshagh}, {Tsantaki}, {Hakobyan}, {Gonz{\'a}lez Hern{\'a}ndez}, {Su{\'a}rez-Andr{\'e}s}, \& {Israelian}}]{Adibekyan-15}
{Adibekyan}, V., {Figueira}, P., {Santos}, N.~C., {et~al.} 2015, \aap, 583, A94

\bibitem[{{Adibekyan} {et~al.}(2012){Adibekyan}, {Sousa}, {Santos}, {Delgado Mena}, {Gonz{\'a}lez Hern{\'a}ndez}, {Israelian}, {Mayor}, \& {Khachatryan}}]{Adibekyan-12}
{Adibekyan}, V.~Z., {Sousa}, S.~G., {Santos}, N.~C., {et~al.} 2012, \aap, 545, A32

\bibitem[{{Aguichine} {et~al.}(2021){Aguichine}, {Mousis}, {Deleuil}, \& {Marcq}}]{Aguichine2021}
{Aguichine}, A., {Mousis}, O., {Deleuil}, M., \& {Marcq}, E. 2021, \apj, 914, 84

\bibitem[{{Alexander} {et~al.}(2014){Alexander}, {Pascucci}, {Andrews}, {Armitage}, \& {Cieza}}]{Alexander2014}
{Alexander}, R., {Pascucci}, I., {Andrews}, S., {Armitage}, P., \& {Cieza}, L. 2014, in Protostars and Planets VI, ed. H.~{Beuther}, R.~S. {Klessen}, C.~P. {Dullemond}, \& T.~{Henning}, 475--496

\bibitem[{{Alibert} {et~al.}(2013){Alibert}, {Carron}, {Fortier}, {Pfyffer}, {Benz}, {Mordasini}, \& {Swoboda}}]{Alibert2013}
{Alibert}, Y., {Carron}, F., {Fortier}, A., {et~al.} 2013, \aap, 558, A109

\bibitem[{{Barrag{\'a}n} {et~al.}(2019){Barrag{\'a}n}, {Gandolfi}, \& {Antoniciello}}]{barragan2019}
{Barrag{\'a}n}, O., {Gandolfi}, D., \& {Antoniciello}, G. 2019, \mnras, 482, 1017

\bibitem[{{Benz} {et~al.}(2021){Benz}, {Broeg}, {Fortier}, {Rando}, {Beck}, {Beck}, {Queloz}, {Ehrenreich}, {Maxted}, {Isaak}, {Billot}, {Alibert}, {Alonso}, {Ant{\'o}nio}, {Asquier}, {Bandy}, {B{\'a}rczy}, {Barrado}, {Barros}, {Baumjohann}, {Bekkelien}, {Bergomi}, {Biondi}, {Bonfils}, {Borsato}, {Brandeker}, {Busch}, {Cabrera}, {Cessa}, {Charnoz}, {Chazelas}, {Collier Cameron}, {Corral Van Damme}, {Cortes}, {Davies}, {Deleuil}, {Deline}, {Delrez}, {Demangeon}, {Demory}, {Erikson}, {Farinato}, {Fossati}, {Fridlund}, {Futyan}, {Gandolfi}, {Garcia Munoz}, {Gillon}, {Guterman}, {Gutierrez}, {Hasiba}, {Heng}, {Hernandez}, {Hoyer}, {Kiss}, {Kovacs}, {Kuntzer}, {Laskar}, {Lecavelier des Etangs}, {Lendl}, {L{\'o}pez}, {Lora}, {Lovis}, {L{\"u}ftinger}, {Magrin}, {Malvasio}, {Marafatto}, {Michaelis}, {de Miguel}, {Modrego}, {Munari}, {Nascimbeni}, {Olofsson}, {Ottacher}, {Ottensamer}, {Pagano}, {Palacios}, {Pall{\'e}}, {Peter}, {Piazza}, {Piotto}, {Pizarro}, {Pollaco}, {Ragazzoni}, {Ratti}, {Rauer}, {Ribas}, {Rieder},
  {Rohlfs}, {Safa}, {Salatti}, {Santos}, {Scandariato}, {S{\'e}gransan}, {Simon}, {Smith}, {Sordet}, {Sousa}, {Steller}, {Szab{\'o}}, {Szoke}, {Thomas}, {Tschentscher}, {Udry}, {Van Grootel}, {Viotto}, {Walter}, {Walton}, {Wildi}, \& {Wolter}}]{Benz2021}
{Benz}, W., {Broeg}, C., {Fortier}, A., {et~al.} 2021, Experimental Astronomy, 51, 109

\bibitem[{{Berezutsky} {et~al.}(2022){Berezutsky}, {Shaikhislamov}, {Rumenskikh}, {Khodachenko}, {Lammer}, \& {Miroshnichenko}}]{Berezutsky2022}
{Berezutsky}, A.~G., {Shaikhislamov}, I.~F., {Rumenskikh}, M.~S., {et~al.} 2022, \mnras, 515, 706

\bibitem[{{Blackwell} \& {Shallis}(1977)}]{Blackwell1977}
{Blackwell}, D.~E. \& {Shallis}, M.~J. 1977, \mnras, 180, 177

\bibitem[{{Bodenheimer} \& {Pollack}(1986)}]{Bodenheimer1986}
{Bodenheimer}, P. \& {Pollack}, J.~B. 1986, \icarus, 67, 391

\bibitem[{{Bonfanti} {et~al.}(2021{\natexlab{a}}){Bonfanti}, {Delrez}, {Hooton}, {Wilson}, {Fossati}, {Alibert}, {Hoyer}, {Mustill}, {Osborn}, {Adibekyan}, {Gandolfi}, {Salmon}, {Sousa}, {Tuson}, {Van Grootel}, {Cabrera}, {Nascimbeni}, {Maxted}, {Barros}, {Billot}, {Bonfils}, {Borsato}, {Broeg}, {Davies}, {Deleuil}, {Demangeon}, {Fridlund}, {Lacedelli}, {Lendl}, {Persson}, {Santos}, {Scandariato}, {Szab{\'o}}, {Collier Cameron}, {Udry}, {Benz}, {Beck}, {Ehrenreich}, {Fortier}, {Isaak}, {Queloz}, {Alonso}, {Asquier}, {Bandy}, {B{\'a}rczy}, {Barrado}, {Barrag{\'a}n}, {Baumjohann}, {Beck}, {Bekkelien}, {Bergomi}, {Brandeker}, {Busch}, {Cessa}, {Charnoz}, {Chazelas}, {Corral Van Damme}, {Demory}, {Erikson}, {Farinato}, {Futyan}, {Garcia Mu{\~n}oz}, {Gillon}, {Guedel}, {Guterman}, {Hasiba}, {Heng}, {Hernandez}, {Kiss}, {Kuntzer}, {Laskar}, {Lecavelier des Etangs}, {Lovis}, {Magrin}, {Malvasio}, {Marafatto}, {Michaelis}, {Munari}, {Olofsson}, {Ottacher}, {Ottensamer}, {Pagano}, {Pall{\'e}}, {Peter}, {Piazza},
  {Piotto}, {Pollacco}, {Ragazzoni}, {Rando}, {Ratti}, {Rauer}, {Ribas}, {Rieder}, {Rohlfs}, {Safa}, {Salatti}, {S{\'e}gransan}, {Simon}, {Smith}, {Sordet}, {Steller}, {Thomas}, {Tschentscher}, {Van Eylen}, {Viotto}, {Walter}, {Walton}, {Wildi}, \& {Wolter}}]{Bonfanti2021}
{Bonfanti}, A., {Delrez}, L., {Hooton}, M.~J., {et~al.} 2021{\natexlab{a}}, \aap, 646, A157

\bibitem[{{Bonfanti} {et~al.}(2021{\natexlab{b}}){Bonfanti}, {Fossati}, {Kubyshkina}, \& {Cubillos}}]{bonfanti2021_pasta}
{Bonfanti}, A., {Fossati}, L., {Kubyshkina}, D., \& {Cubillos}, P.~E. 2021{\natexlab{b}}, \aap, 656, A157

\bibitem[{{Bonfanti} {et~al.}(2023){Bonfanti}, {Gandolfi}, {Egger}, {Fossati}, {Cabrera}, {Krenn}, {Alibert}, {Benz}, {Billot}, {Flor{\'e}n}, {Lendl}, {Adibekyan}, {Salmon}, {Santos}, {Sousa}, {Wilson}, {Barrag{\'a}n}, {Collier Cameron}, {Delrez}, {Esposito}, {Goffo}, {Osborne}, {Osborn}, {Serrano}, {Van Eylen}, {Alarcon}, {Alonso}, {Anglada}, {B{\'a}rczy}, {Barrado Navascues}, {Barros}, {Baumjohann}, {Beck}, {Beck}, {Bedell}, {Bonfils}, {Borsato}, {Brandeker}, {Broeg}, {Charnoz}, {Corral Van Damme}, {Csizmadia}, {Cubillos}, {Davies}, {Deleuil}, {Demangeon}, {Demory}, {Ehrenreich}, {Erikson}, {Fortier}, {Fridlund}, {Gillon}, {G{\"u}del}, {Hoyer}, {Isaak}, {Kerschbaum}, {Kiss}, {Laskar}, {Lecavelier des Etangs}, {Lorenzo-Oliveira}, {Lovis}, {Magrin}, {Marafatto}, {Maxted}, {Mel{\'e}ndez}, {Mordasini}, {Nascimbeni}, {Olofsson}, {Ottensamer}, {Pagano}, {Pall{\'e}}, {Peter}, {Piazza}, {Piotto}, {Pollacco}, {Queloz}, {Ragazzoni}, {Rando}, {Rauer}, {Ribas}, {Scandariato}, {S{\'e}gransan}, {Simon}, {Smith},
  {Steller}, {Szab{\'o}}, {Thomas}, {Udry}, {Ulmer}, {Van Grootel}, {Venturini}, \& {Walton}}]{Bonfanti2023}
{Bonfanti}, A., {Gandolfi}, D., {Egger}, J.~A., {et~al.} 2023, \aap, 671, L8

\bibitem[{{Bonfanti} \& {Gillon}(2020)}]{Bonfanti2020}
{Bonfanti}, A. \& {Gillon}, M. 2020, \aap, 635, A6

\bibitem[{{Bonfanti} {et~al.}(2016){Bonfanti}, {Ortolani}, \& {Nascimbeni}}]{bonfanti2016}
{Bonfanti}, A., {Ortolani}, S., \& {Nascimbeni}, V. 2016, \aap, 585, A5

\bibitem[{{Bonfanti} {et~al.}(2015){Bonfanti}, {Ortolani}, {Piotto}, \& {Nascimbeni}}]{bonfanti2015}
{Bonfanti}, A., {Ortolani}, S., {Piotto}, G., \& {Nascimbeni}, V. 2015, \aap, 575, A18

\bibitem[{{Brandeker} {et~al.}(2022){Brandeker}, {Heng}, {Lendl}, {Patel}, {Morris}, {Broeg}, {Guterman}, {Beck}, {Maxted}, {Demangeon}, {Delrez}, {Demory}, {Kitzmann}, {Santos}, {Singh}, {Alibert}, {Alonso}, {Anglada}, {B{\'a}rczy}, {Barrado y Navascues}, {Barros}, {Baumjohann}, {Beck}, {Benz}, {Billot}, {Bonfils}, {Bruno}, {Cabrera}, {Charnoz}, {Collier Cameron}, {Corral van Damme}, {Csizmadia}, {Davies}, {Deleuil}, {Deline}, {Ehrenreich}, {Erikson}, {Farinato}, {Fortier}, {Fossati}, {Fridlund}, {Gandolfi}, {Gillon}, {G{\"u}del}, {Hoyer}, {Isaak}, {Kiss}, {Laskar}, {Lecavelier des Etangs}, {Lovis}, {Luntzer}, {Magrin}, {Nascimbeni}, {Olofsson}, {Ottensamer}, {Pagano}, {Pall{\'e}}, {Peter}, {Piotto}, {Pollacco}, {Queloz}, {Ragazzoni}, {Rando}, {Rauer}, {Ribas}, {Scandariato}, {S{\'e}gransan}, {Simon}, {Smith}, {Sousa}, {Steller}, {Szab{\'o}}, {Thomas}, {Udry}, {Van Grootel}, {Walton}, \& {Wolter}}]{Brandeker2022}
{Brandeker}, A., {Heng}, K., {Lendl}, M., {et~al.} 2022, \aap, 659, L4

\bibitem[{{Burn} {et~al.}(2024){Burn}, {Mordasini}, {Mishra}, {Haldemann}, {Venturini}, {Emsenhuber}, \& {Henning}}]{burn2024_atmospheresWithWater}
{Burn}, R., {Mordasini}, C., {Mishra}, L., {et~al.} 2024, Nature Astronomy [\eprint[arXiv]{2401.04380}]

\bibitem[{{Cabrera} {et~al.}(2023){Cabrera}, {Gandolfi}, {Serrano}, {Csizmadia}, {Egger}, {Baumeister}, {Krenn}, {Benz}, {Deline}, {Flor{\'e}n}, {Collier Cameron}, {Adibekyan}, {Alibert}, {Bellomo}, {Delrez}, {Fossati}, {Fortier}, {Grziwa}, {Hoyer}, {Bonfanti}, {Salmon}, {Sousa}, {Wilson}, {Alarc{\'o}n}, {Alonso}, {Anglada Escud{\'e}}, {B{\'a}rczy}, {Barrag{\'a}n}, {Barrado}, {Barros}, {Baumjohann}, {Beck}, {Beck}, {Bernab{\`o}}, {Billot}, {Bonfils}, {Borsato}, {Brandeker}, {Broeg}, {Carri{\'o}n-Gonz{\'a}lez}, {Charnoz}, {Ciardi}, {Cochran}, {Collins}, {Collins}, {Conti}, {Davies}, {Deeg}, {Deleuil}, {Demangeon}, {Demory}, {Ehrenreich}, {Erikson}, {Esposito}, {Fridlund}, {Gillon}, {Goffo}, {G{\"u}del}, {Guenther}, {Harre}, {Heng}, {Hooton}, {Isaak}, {Jenkins}, {Kiss}, {Knudstrup}, {Lam}, {Laskar}, {Lecavelier des Etangs}, {Lendl}, {Lovis}, {Luque}, {Magrin}, {Maxted}, {Muresan}, {Nascimbeni}, {Olofsson}, {Osborn}, {Osborne}, {Ottensamer}, {Pagano}, {Pall{\'e}}, {Persson}, {Peter}, {Piotto}, {Pollacco},
  {Queloz}, {Ragazzoni}, {Rando}, {Rauer}, {Redfield}, {Ribas}, {Ricker}, {Rodler}, {Santos}, {Scandariato}, {Seager}, {S{\'e}gransan}, {Simon}, {Smith}, {Steller}, {Szab{\'o}}, {Thomas}, {Tosi}, {Twicken}, {Udry}, {Van Eylen}, {Van Grootel}, {Walton}, \& {Winn}}]{Cabrera2023}
{Cabrera}, J., {Gandolfi}, D., {Serrano}, L.~M., {et~al.} 2023, \aap, 675, A183

\bibitem[{{Carleo} {et~al.}(2020){Carleo}, {Gandolfi}, {Barrag{\'a}n}, {Livingston}, {Persson}, {Lam}, {Vidotto}, {Lund}, {Villarreal D'Angelo}, {Collins}, {Fossati}, {Howard}, {Kubyshkina}, {Brahm}, {Oklop{\v{c}}i{\'c}}, {Molli{\`e}re}, {Redfield}, {Serrano}, {Dai}, {Fridlund}, {Borsa}, {Korth}, {Esposito}, {D{\'\i}az}, {Dyregaard Nielsen}, {Hellier}, {Mathur}, {Deeg}, {Hatzes}, {Benatti}, {Rodler}, {Alarcon}, {Spina}, {Santos}, {Georgieva}, {Garc{\'\i}a}, {Gonz{\'a}lez-Cuesta}, {Ricker}, {Vanderspek}, {Latham}, {Seager}, {Winn}, {Jenkins}, {Albrecht}, {Batalha}, {Beard}, {Boyd}, {Bouchy}, {Burt}, {Butler}, {Cabrera}, {Chontos}, {Ciardi}, {Cochran}, {Collins}, {Crane}, {Crossfield}, {Csizmadia}, {Dragomir}, {Dressing}, {Eigm{\"u}ller}, {Endl}, {Erikson}, {Espinoza}, {Fausnaugh}, {Feng}, {Flowers}, {Fulton}, {Gonzales}, {Grieves}, {Grziwa}, {Guenther}, {Guerrero}, {Henning}, {Hidalgo}, {Hirano}, {Hjorth}, {Huber}, {Isaacson}, {Jones}, {Jord{\'a}n}, {Kab{\'a}th}, {Kane}, {Knudstrup}, {Lubin}, {Luque},
  {Mireles}, {Narita}, {Nespral}, {Niraula}, {Nowak}, {Palle}, {P{\"a}tzold}, {Petigura}, {Prieto-Arranz}, {Rauer}, {Robertson}, {Rose}, {Roy}, {Sarkis}, {Schlieder}, {S{\'e}gransan}, {Shectman}, {Skarka}, {Smith}, {Smith}, {Stassun}, {Teske}, {Twicken}, {Van Eylen}, {Wang}, {Weiss}, \& {Wyttenbach}}]{Carleo2020}
{Carleo}, I., {Gandolfi}, D., {Barrag{\'a}n}, O., {et~al.} 2020, \aj, 160, 114

\bibitem[{{Carolan} {et~al.}(2021){Carolan}, {Vidotto}, {Hazra}, {Villarreal D'Angelo}, \& {Kubyshkina}}]{Carolan2021}
{Carolan}, S., {Vidotto}, A.~A., {Hazra}, G., {Villarreal D'Angelo}, C., \& {Kubyshkina}, D. 2021, \mnras, 508, 6001

\bibitem[{{Castelli} \& {Kurucz}(2003)}]{Castelli2003}
{Castelli}, F. \& {Kurucz}, R.~L. 2003, in IAU Symposium, Vol. 210, Modelling of Stellar Atmospheres, ed. N.~{Piskunov}, W.~W. {Weiss}, \& D.~F. {Gray}, A20

\bibitem[{{Chachan} {et~al.}(2020){Chachan}, {Jontof-Hutter}, {Knutson}, {Adams}, {Gao}, {Benneke}, {Berta-Thompson}, {Dai}, {Deming}, {Ford}, {Lee}, {Libby-Roberts}, {Madhusudhan}, {Wakeford}, \& {Wong}}]{chachan2020}
{Chachan}, Y., {Jontof-Hutter}, D., {Knutson}, H.~A., {et~al.} 2020, \aj, 160, 201

\bibitem[{{Choi} {et~al.}(2016){Choi}, {Dotter}, {Conroy}, {Cantiello}, {Paxton}, \& {Johnson}}]{Choi2016}
{Choi}, J., {Dotter}, A., {Conroy}, C., {et~al.} 2016, \apj, 823, 102

\bibitem[{{Correia} {et~al.}(2010){Correia}, {Couetdic}, {Laskar}, {Bonfils}, {Mayor}, {Bertaux}, {Bouchy}, {Delfosse}, {Forveille}, {Lovis}, {Pepe}, {Perrier}, {Queloz}, \& {Udry}}]{Correia_etal_2010}
{Correia}, A.~C.~M., {Couetdic}, J., {Laskar}, J., {et~al.} 2010, \aap, 511, A21

\bibitem[{{Correia} {et~al.}(2005){Correia}, {Udry}, {Mayor}, {Laskar}, {Naef}, {Pepe}, {Queloz}, \& {Santos}}]{Correia_etal_2005}
{Correia}, A.~C.~M., {Udry}, S., {Mayor}, M., {et~al.} 2005, \aap, 440, 751

\bibitem[{{Cosentino} {et~al.}(2012){Cosentino}, {Lovis}, {Pepe}, {Collier Cameron}, {Latham}, {Molinari}, {Udry}, {Bezawada}, {Black}, {Born}, {Buchschacher}, {Charbonneau}, {Figueira}, {Fleury}, {Galli}, {Gallie}, {Gao}, {Ghedina}, {Gonzalez}, {Gonzalez}, {Guerra}, {Henry}, {Horne}, {Hughes}, {Kelly}, {Lodi}, {Lunney}, {Maire}, {Mayor}, {Micela}, {Ordway}, {Peacock}, {Phillips}, {Piotto}, {Pollacco}, {Queloz}, {Rice}, {Riverol}, {Riverol}, {San Juan}, {Sasselov}, {Segransan}, {Sozzetti}, {Sosnowska}, {Stobie}, {Szentgyorgyi}, {Vick}, \& {Weber}}]{cosentino2012}
{Cosentino}, R., {Lovis}, C., {Pepe}, F., {et~al.} 2012, in Society of Photo-Optical Instrumentation Engineers (SPIE) Conference Series, Vol. 8446, Ground-based and Airborne Instrumentation for Astronomy IV, ed. I.~S. {McLean}, S.~K. {Ramsay}, \& H.~{Takami}, 84461V

\bibitem[{{Crane} {et~al.}(2006){Crane}, {Shectman}, \& {Butler}}]{Crane2006}
{Crane}, J.~D., {Shectman}, S.~A., \& {Butler}, R.~P. 2006, in Society of Photo-Optical Instrumentation Engineers (SPIE) Conference Series, Vol. 6269, Society of Photo-Optical Instrumentation Engineers (SPIE) Conference Series, ed. I.~S. {McLean} \& M.~{Iye}, 626931

\bibitem[{{Crane} {et~al.}(2010){Crane}, {Shectman}, {Butler}, {Thompson}, {Birk}, {Jones}, \& {Burley}}]{Crane2010}
{Crane}, J.~D., {Shectman}, S.~A., {Butler}, R.~P., {et~al.} 2010, in Society of Photo-Optical Instrumentation Engineers (SPIE) Conference Series, Vol. 7735, Ground-based and Airborne Instrumentation for Astronomy III, ed. I.~S. {McLean}, S.~K. {Ramsay}, \& H.~{Takami}, 773553

\bibitem[{{Crane} {et~al.}(2008){Crane}, {Shectman}, {Butler}, {Thompson}, \& {Burley}}]{Crane2008}
{Crane}, J.~D., {Shectman}, S.~A., {Butler}, R.~P., {Thompson}, I.~B., \& {Burley}, G.~S. 2008, in Society of Photo-Optical Instrumentation Engineers (SPIE) Conference Series, Vol. 7014, Ground-based and Airborne Instrumentation for Astronomy II, ed. I.~S. {McLean} \& M.~M. {Casali}, 701479

\bibitem[{{Cubillos} {et~al.}(2017){Cubillos}, {Erkaev}, {Juvan}, {Fossati}, {Johnstone}, {Lammer}, {Lendl}, {Odert}, \& {Kislyakova}}]{Cubillos2017}
{Cubillos}, P., {Erkaev}, N.~V., {Juvan}, I., {et~al.} 2017, \mnras, 466, 1868

\bibitem[{{Deck} {et~al.}(2014){Deck}, {Agol}, {Holman}, \& {Nesvorn{\'y}}}]{Deck2014}
{Deck}, K.~M., {Agol}, E., {Holman}, M.~J., \& {Nesvorn{\'y}}, D. 2014, \apj, 787, 132

\bibitem[{{Delisle} \& {Laskar}(2014)}]{Delisle_Laskar_2014}
{Delisle}, J.~B. \& {Laskar}, J. 2014, \aap, 570, L7

\bibitem[{{Delrez} {et~al.}(2021){Delrez}, {Ehrenreich}, {Alibert}, {Bonfanti}, {Borsato}, {Fossati}, {Hooton}, {Hoyer}, {Pozuelos}, {Salmon}, {Sulis}, {Wilson}, {Adibekyan}, {Bourrier}, {Brandeker}, {Charnoz}, {Deline}, {Guterman}, {Haldemann}, {Hara}, {Oshagh}, {Sousa}, {Van Grootel}, {Alonso}, {Anglada-Escud{\'e}}, {B{\'a}rczy}, {Barrado}, {Barros}, {Baumjohann}, {Beck}, {Bekkelien}, {Benz}, {Billot}, {Bonfils}, {Broeg}, {Cabrera}, {Collier Cameron}, {Davies}, {Deleuil}, {Delisle}, {Demangeon}, {Demory}, {Erikson}, {Fortier}, {Fridlund}, {Futyan}, {Gandolfi}, {Garcia Mu{\~n}oz}, {Gillon}, {Guedel}, {Heng}, {Kiss}, {Laskar}, {Lecavelier des Etangs}, {Lendl}, {Lovis}, {Maxted}, {Nascimbeni}, {Olofsson}, {Osborn}, {Pagano}, {Pall{\'e}}, {Piotto}, {Pollacco}, {Queloz}, {Rauer}, {Ragazzoni}, {Ribas}, {Santos}, {Scandariato}, {S{\'e}gransan}, {Simon}, {Smith}, {Steller}, {Szab{\'o}}, {Thomas}, {Udry}, \& {Walton}}]{Delrez2021}
{Delrez}, L., {Ehrenreich}, D., {Alibert}, Y., {et~al.} 2021, Nature Astronomy, 5, 775

\bibitem[{{Dorn} {et~al.}(2017){Dorn}, {Venturini}, {Khan}, {Heng}, {Alibert}, {Helled}, {Rivoldini}, \& {Benz}}]{Dorn2017}
{Dorn}, C., {Venturini}, J., {Khan}, A., {et~al.} 2017, \aap, 597, A37

\bibitem[{{Elser} {et~al.}(2013){Elser}, {Grimm}, \& {Stadel}}]{Elser_etal_2013}
{Elser}, S., {Grimm}, S.~L., \& {Stadel}, J.~G. 2013, \mnras, 433, 2194

\bibitem[{{Emsenhuber} {et~al.}(2021){Emsenhuber}, {Mordasini}, {Burn}, {Alibert}, {Benz}, \& {Asphaug}}]{Emenshuber2021}
{Emsenhuber}, A., {Mordasini}, C., {Burn}, R., {et~al.} 2021, \aap, 656, A69

\bibitem[{{Espinoza} \& {Jord{\'a}n}(2015)}]{espinoza2015}
{Espinoza}, N. \& {Jord{\'a}n}, A. 2015, \mnras, 450, 1879

\bibitem[{{Feroz} \& {Hobson}(2008)}]{feroz2008}
{Feroz}, F. \& {Hobson}, M.~P. 2008, \mnras, 384, 449

\bibitem[{{Feroz} {et~al.}(2019){Feroz}, {Hobson}, {Cameron}, \& {Pettitt}}]{feroz2019}
{Feroz}, F., {Hobson}, M.~P., {Cameron}, E., \& {Pettitt}, A.~N. 2019, The Open Journal of Astrophysics, 2, 10

\bibitem[{{Foreman-Mackey} {et~al.}(2017){Foreman-Mackey}, {Agol}, {Ambikasaran}, \& {Angus}}]{Foreman-Mackey2017}
{Foreman-Mackey}, D., {Agol}, E., {Ambikasaran}, S., \& {Angus}, R. 2017, \aj, 154, 220

\bibitem[{{Fossati} {et~al.}(2017){Fossati}, {Marcelja}, {Staab}, {Cubillos}, {France}, {Haswell}, {Ingrassia}, {Jenkins}, {Koskinen}, {Lanza}, {Redfield}, {Youngblood}, \& {Pelzmann}}]{Fossati2017A&A...601A.104F}
{Fossati}, L., {Marcelja}, S.~E., {Staab}, D., {et~al.} 2017, \aap, 601, A104

\bibitem[{{Frandsen} \& {Lindberg}(1999)}]{Fransen1999}
{Frandsen}, S. \& {Lindberg}, B. 1999, in Astrophysics with the NOT, ed. H.~{Karttunen} \& V.~{Piirola}, 71

\bibitem[{{Gaia Collaboration} {et~al.}(2023){Gaia Collaboration}, {Vallenari}, {Brown}, {Prusti}, {de Bruijne}, {Arenou}, {Babusiaux}, {Biermann}, {Creevey}, {Ducourant}, {Evans}, {Eyer}, {Guerra}, {Hutton}, {Jordi}, {Klioner}, {Lammers}, {Lindegren}, {Luri}, {Mignard}, {Panem}, {Pourbaix}, {Randich}, {Sartoretti}, {Soubiran}, {Tanga}, {Walton}, {Bailer-Jones}, {Bastian}, {Drimmel}, {Jansen}, {Katz}, {Lattanzi}, {van Leeuwen}, {Bakker}, {Cacciari}, {Casta{\~n}eda}, {De Angeli}, {Fabricius}, {Fouesneau}, {Fr{\'e}mat}, {Galluccio}, {Guerrier}, {Heiter}, {Masana}, {Messineo}, {Mowlavi}, {Nicolas}, {Nienartowicz}, {Pailler}, {Panuzzo}, {Riclet}, {Roux}, {Seabroke}, {Sordo}, {Th{\'e}venin}, {Gracia-Abril}, {Portell}, {Teyssier}, {Altmann}, {Andrae}, {Audard}, {Bellas-Velidis}, {Benson}, {Berthier}, {Blomme}, {Burgess}, {Busonero}, {Busso}, {C{\'a}novas}, {Carry}, {Cellino}, {Cheek}, {Clementini}, {Damerdji}, {Davidson}, {de Teodoro}, {Nu{\~n}ez Campos}, {Delchambre}, {Dell'Oro}, {Esquej},
  {Fern{\'a}ndez-Hern{\'a}ndez}, {Fraile}, {Garabato}, {Garc{\'\i}a-Lario}, {Gosset}, {Haigron}, {Halbwachs}, {Hambly}, {Harrison}, {Hern{\'a}ndez}, {Hestroffer}, {Hodgkin}, {Holl}, {Jan{\ss}en}, {Jevardat de Fombelle}, {Jordan}, {Krone-Martins}, {Lanzafame}, {L{\"o}ffler}, {Marchal}, {Marrese}, {Moitinho}, {Muinonen}, {Osborne}, {Pancino}, {Pauwels}, {Recio-Blanco}, {Reyl{\'e}}, {Riello}, {Rimoldini}, {Roegiers}, {Rybizki}, {Sarro}, {Siopis}, {Smith}, {Sozzetti}, {Utrilla}, {van Leeuwen}, {Abbas}, {{\'A}brah{\'a}m}, {Abreu Aramburu}, {Aerts}, {Aguado}, {Ajaj}, {Aldea-Montero}, {Altavilla}, {{\'A}lvarez}, {Alves}, {Anders}, {Anderson}, {Anglada Varela}, {Antoja}, {Baines}, {Baker}, {Balaguer-N{\'u}{\~n}ez}, {Balbinot}, {Balog}, {Barache}, {Barbato}, {Barros}, {Barstow}, {Bartolom{\'e}}, {Bassilana}, {Bauchet}, {Becciani}, {Bellazzini}, {Berihuete}, {Bernet}, {Bertone}, {Bianchi}, {Binnenfeld}, {Blanco-Cuaresma}, {Blazere}, {Boch}, {Bombrun}, {Bossini}, {Bouquillon}, {Bragaglia}, {Bramante}, {Breedt},
  {Bressan}, {Brouillet}, {Brugaletta}, {Bucciarelli}, {Burlacu}, {Butkevich}, {Buzzi}, {Caffau}, {Cancelliere}, {Cantat-Gaudin}, {Carballo}, {Carlucci}, {Carnerero}, {Carrasco}, {Casamiquela}, {Castellani}, {Castro-Ginard}, {Chaoul}, {Charlot}, {Chemin}, {Chiaramida}, {Chiavassa}, {Chornay}, {Comoretto}, {Contursi}, {Cooper}, {Cornez}, {Cowell}, {Crifo}, {Cropper}, {Crosta}, {Crowley}, {Dafonte}, {Dapergolas}, {David}, {David}, {de Laverny}, {De Luise}, {De March}, {De Ridder}, {de Souza}, {de Torres}, {del Peloso}, {del Pozo}, {Delbo}, {Delgado}, {Delisle}, {Demouchy}, {Dharmawardena}, {Di Matteo}, {Diakite}, {Diener}, {Distefano}, {Dolding}, {Edvardsson}, {Enke}, {Fabre}, {Fabrizio}, {Faigler}, {Fedorets}, {Fernique}, {Fienga}, {Figueras}, {Fournier}, {Fouron}, {Fragkoudi}, {Gai}, {Garcia-Gutierrez}, {Garcia-Reinaldos}, {Garc{\'\i}a-Torres}, {Garofalo}, {Gavel}, {Gavras}, {Gerlach}, {Geyer}, {Giacobbe}, {Gilmore}, {Girona}, {Giuffrida}, {Gomel}, {Gomez}, {Gonz{\'a}lez-N{\'u}{\~n}ez},
  {Gonz{\'a}lez-Santamar{\'\i}a}, {Gonz{\'a}lez-Vidal}, {Granvik}, {Guillout}, {Guiraud}, {Guti{\'e}rrez-S{\'a}nchez}, {Guy}, {Hatzidimitriou}, {Hauser}, {Haywood}, {Helmer}, {Helmi}, {Sarmiento}, {Hidalgo}, {Hilger}, {H{\l}adczuk}, {Hobbs}, {Holland}, {Huckle}, {Jardine}, {Jasniewicz}, {Jean-Antoine Piccolo}, {Jim{\'e}nez-Arranz}, {Jorissen}, {Juaristi Campillo}, {Julbe}, {Karbevska}, {Kervella}, {Khanna}, {Kontizas}, {Kordopatis}, {Korn}, {K{\'o}sp{\'a}l}, {Kostrzewa-Rutkowska}, {Kruszy{\'n}ska}, {Kun}, {Laizeau}, {Lambert}, {Lanza}, {Lasne}, {Le Campion}, {Lebreton}, {Lebzelter}, {Leccia}, {Leclerc}, {Lecoeur-Taibi}, {Liao}, {Licata}, {Lindstr{\o}m}, {Lister}, {Livanou}, {Lobel}, {Lorca}, {Loup}, {Madrero Pardo}, {Magdaleno Romeo}, {Managau}, {Mann}, {Manteiga}, {Marchant}, {Marconi}, {Marcos}, {Marcos Santos}, {Mar{\'\i}n Pina}, {Marinoni}, {Marocco}, {Marshall}, {Martin Polo}, {Mart{\'\i}n-Fleitas}, {Marton}, {Mary}, {Masip}, {Massari}, {Mastrobuono-Battisti}, {Mazeh}, {McMillan}, {Messina}, {Michalik},
  {Millar}, {Mints}, {Molina}, {Molinaro}, {Moln{\'a}r}, {Monari}, {Mongui{\'o}}, {Montegriffo}, {Montero}, {Mor}, {Mora}, {Morbidelli}, {Morel}, {Morris}, {Muraveva}, {Murphy}, {Musella}, {Nagy}, {Noval}, {Oca{\~n}a}, {Ogden}, {Ordenovic}, {Osinde}, {Pagani}, {Pagano}, {Palaversa}, {Palicio}, {Pallas-Quintela}, {Panahi}, {Payne-Wardenaar}, {Pe{\~n}alosa Esteller}, {Penttil{\"a}}, {Pichon}, {Piersimoni}, {Pineau}, {Plachy}, {Plum}, {Poggio}, {Pr{\v{s}}a}, {Pulone}, {Racero}, {Ragaini}, {Rainer}, {Raiteri}, {Rambaux}, {Ramos}, {Ramos-Lerate}, {Re Fiorentin}, {Regibo}, {Richards}, {Rios Diaz}, {Ripepi}, {Riva}, {Rix}, {Rixon}, {Robichon}, {Robin}, {Robin}, {Roelens}, {Rogues}, {Rohrbasser}, {Romero-G{\'o}mez}, {Rowell}, {Royer}, {Ruz Mieres}, {Rybicki}, {Sadowski}, {S{\'a}ez N{\'u}{\~n}ez}, {Sagrist{\`a} Sell{\'e}s}, {Sahlmann}, {Salguero}, {Samaras}, {Sanchez Gimenez}, {Sanna}, {Santove{\~n}a}, {Sarasso}, {Schultheis}, {Sciacca}, {Segol}, {Segovia}, {S{\'e}gransan}, {Semeux}, {Shahaf}, {Siddiqui}, {Siebert},
  {Siltala}, {Silvelo}, {Slezak}, {Slezak}, {Smart}, {Snaith}, {Solano}, {Solitro}, {Souami}, {Souchay}, {Spagna}, {Spina}, {Spoto}, {Steele}, {Steidelm{\"u}ller}, {Stephenson}, {S{\"u}veges}, {Surdej}, {Szabados}, {Szegedi-Elek}, {Taris}, {Taylor}, {Teixeira}, {Tolomei}, {Tonello}, {Torra}, {Torra}, {Torralba Elipe}, {Trabucchi}, {Tsounis}, {Turon}, {Ulla}, {Unger}, {Vaillant}, {van Dillen}, {van Reeven}, {Vanel}, {Vecchiato}, {Viala}, {Vicente}, {Voutsinas}, {Weiler}, {Wevers}, {Wyrzykowski}, {Yoldas}, {Yvard}, {Zhao}, {Zorec}, {Zucker}, \& {Zwitter}}]{GaiaCollaboration2022}
{Gaia Collaboration}, {Vallenari}, A., {Brown}, A.~G.~A., {et~al.} 2023, \aap, 674, A1

\bibitem[{{Gao} \& {Zhang}(2020)}]{Gao2020}
{Gao}, P. \& {Zhang}, X. 2020, \apj, 890, 93

\bibitem[{{Gardner} {et~al.}(2023){Gardner}, {Mather}, {Abbott}, {Abell}, {Abernathy}, {Abney}, {Abraham}, {Abraham}, {Abul-Huda}, {Acton}, {Adams}, {Adams}, {Adler}, {Adriaensen}, {Aguilar}, {Ahmed}, {Ahmed}, {Ahmed}, {Albat}, {Albert}, {Alberts}, {Aldridge}, {Allen}, {Allen}, {Altenburg}, {Altunc}, {Alvarez}, {{\'A}lvarez-M{\'a}rquez}, {Alves de Oliveira}, {Ambrose}, {Anandakrishnan}, {Andersen}, {Anderson}, {Anderson}, {Anderson}, {Anderson}, {Aprea}, {Archer}, {Arenberg}, {Argyriou}, {Arribas}, {Artigau}, {Arvai}, {Atcheson}, {Atkinson}, {Averbukh}, {Aymergen}, {Bacinski}, {Baggett}, {Bagnasco}, {Baker}, {Balzano}, {Banks}, {Baran}, {Barker}, {Barrett}, {Barringer}, {Barto}, {Bast}, {Baudoz}, {Baum}, {Beatty}, {Beaulieu}, {Bechtold}, {Beck}, {Beddard}, {Beichman}, {Bellagama}, {Bely}, {Berger}, {Bergeron}, {Bernier}, {Bertch}, {Beskow}, {Betz}, {Biagetti}, {Birkmann}, {Bjorklund}, {Blackwood}, {Blazek}, {Blossfeld}, {Bluth}, {Boccaletti}, {Boegner}, {Bohlin}, {Boia}, {B{\"o}ker}, {Bonaventura}, {Bond},
  {Bosley}, {Boucarut}, {Bouchet}, {Bouwman}, {Bower}, {Bowers}, {Bowers}, {Boyce}, {Boyer}, {Boyer}, {Boyer}, {Boyer}, {Bradley}, {Brady}, {Brandl}, {Brannen}, {Breda}, {Bremmer}, {Brennan}, {Bresnahan}, {Bright}, {Broiles}, {Bromenschenkel}, {Brooks}, {Brooks}, {Brown}, {Brown}, {Brown}, {Bruce}, {Bryson}, {Bujanda}, {Bullock}, {Bunker}, {Bureo}, {Burt}, {Bush}, {Bushouse}, {Bussman}, {Cabaud}, {Cale}, {Calhoon}, {Calvani}, {Canipe}, {Caputo}, {Cara}, {Carey}, {Case}, {Cesari}, {Cetorelli}, {Chance}, {Chandler}, {Chaney}, {Chapman}, {Charlot}, {Chayer}, {Cheezum}, {Chen}, {Chen}, {Cherinka}, {Chichester}, {Chilton}, {Chittiraibalan}, {Clampin}, {Clark}, {Clark}, {Clark}, {Claybrooks}, {Cleveland}, {Cohen}, {Cohen}, {Col{\'o}n}, {Coleman}, {Colina}, {Comber}, {Comeau}, {Comer}, {Conde Reis}, {Connolly}, {Conroy}, {Contos}, {Contreras}, {Cook}, {Cooper}, {Cooper}, {Correia}, {Correnti}, {Cossou}, {Costanza}, {Coulais}, {Cox}, {Coyle}, {Cracraft}, {Crew}, {Curtis}, {Cusveller}, {Da Costa Maciel}, {Dailey},
  {Daugeron}, {Davidson}, {Davies}, {Davis}, {Davis}, {Day}, {de Chambure}, {de Jong}, {De Marchi}, {Dean}, {Decker}, {Delisa}, {Dell}, {Dellagatta}, {Dembinska}, {Demosthenes}, {Dencheva}, {Deneu}, {DePriest}, {Deschenes}, {Dethienne}, {Detre}, {Diaz}, {Dicken}, {DiFelice}, {Dillman}, {Disharoon}, {Dixon}, {Doggett}, {Dominguez}, {Donaldson}, {Doria-Warner}, {Santos}, {Doty}, {Douglas}, {Doyon}, {Dressler}, {Driggers}, {Driggers}, {Dunn}, {DuPrie}, {Dupuis}, {Durning}, {Dutta}, {Earl}, {Eccleston}, {Ecobichon}, {Egami}, {Ehrenwinkler}, {Eisenhamer}, {Eisenhower}, {Eisenstein}, {El Hamel}, {Elie}, {Elliott}, {Elliott}, {Engesser}, {Espinoza}, {Etienne}, {Etxaluze}, {Evans}, {Fabreguettes}, {Falcolini}, {Falini}, {Fatig}, {Feeney}, {Feinberg}, {Fels}, {Ferdous}, {Ferguson}, {Ferrarese}, {Ferreira}, {Ferruit}, {Ferry}, {Filippazzo}, {Firre}, {Fix}, {Flagey}, {Flanagan}, {Fleming}, {Florian}, {Flynn}, {Foiadelli}, {Fontaine}, {Fontanella}, {Forshay}, {Fortner}, {Fox}, {Framarini}, {Francisco}, {Franck}, {Franx},
  {Franz}, {Friedman}, {Friend}, {Frost}, {Fu}, {Fullerton}, {Gaillard}, {Galkin}, {Gallagher}, {Galyer}, {Garc{\'\i}a Mar{\'\i}n}, {Gardner}, {Garland}, {Garrett}, {Gasman}, {G{\'a}sp{\'a}r}, {Gastaud}, {Gaudreau}, {Gauthier}, {Geers}, {Geithner}, {Gennaro}, {Gerber}, {Gereau}, {Giampaoli}, {Giardino}, {Gibbons}, {Gilbert}, {Gilman}, {Girard}, {Giuliano}, {Gkountis}, {Glasse}, {Glassmire}, {Glauser}, {Glazer}, {Goldberg}, {Golimowski}, {Gonzaga}, {Gordon}, {Gordon}, {Goudfrooij}, {Gough}, {Graham}, {Grau}, {Green}, {Greene}, {Greene}, {Greenfield}, {Greenhouse}, {Greve}, {Greville}, {Grimaldi}, {Groe}, {Groebner}, {Grumm}, {Grundy}, {G{\"u}del}, {Guillard}, {Guldalian}, {Gunn}, {Gurule}, {Gutman}, {Guy}, {Guyot}, {Hack}, {Haderlein}, {Hagan}, {Hagedorn}, {Hainline}, {Haley}, {Hami}, {Hamilton}, {Hammann}, {Hammel}, {Hanley}, {Hansen}, {Hardy}, {Harnisch}, {Harr}, {Harris}, {Hart}, {Hartig}, {Hasan}, {Hashim}, {Hashimoto}, {Haskins}, {Hawkins}, {Hayden}, {Hayden}, {Healy}, {Hecht}, {Heeg}, {Hejal}, {Helm},
  {Hengemihle}, {Henning}, {Henry}, {Henry}, {Henshaw}, {Hernandez}, {Herrington}, {Heske}, {Hesman}, {Hickey}, {Hilbert}, {Hines}, {Hinz}, {Hirsch}, {Hitcho}, {Hodapp}, {Hodge}, {Hoffman}, {Holfeltz}, {Holler}, {Hoppa}, {Horner}, {Howard}, {Howard}, {Huber}, {Hunkeler}, {Hunter}, {Hunter}, {Hurd}, {Hurst}, {Hutchings}, {Hylan}, {Ignat}, {Illingworth}, {Irish}, {Isaacs}, {Jackson}, {Jaffe}, {Jahic}, {Jahromi}, {Jakobsen}, {James}, {James}, {James}, {Jamieson}, {Jandra}, {Jayawardhana}, {Jedrzejewski}, {Jeffers}, {Jensen}, {Joanne}, {Johns}, {Johnson}, {Johnson}, {Johnson}, {Johnson}, {Johnson}, {Johnson}, {Johnstone}, {Jollet}, {Jones}, {Jones}, {Jones}, {Jones}, {Jones}, {Jordan}, {Jordan}, {Jue}, {Jurkowski}, {Justis}, {Justtanont}, {Kaleida}, {Kalirai}, {Kalmanson}, {Kaltenegger}, {Kammerer}, {Kan}, {Kanarek}, {Kao}, {Karakla}, {Karl}, {Kassin}, {Kauffman}, {Kavanagh}, {Kelley}, {Kelly}, {Kendrew}, {Kennedy}, {Kenny}, {Keski-Kuha}, {Keyes}, {Khan}, {Kidwell}, {Kimble}, {King}, {King}, {Kinzel}, {Kirk},
  {Kirkpatrick}, {Klaassen}, {Klingemann}, {Klintworth}, {Knapp}, {Knight}, {Knollenberg}, {Knutsen}, {Koehler}, {Koekemoer}, {Kofler}, {Kontson}, {Kovacs}, {Kozhurina-Platais}, {Krause}, {Kriss}, {Krist}, {Kristoffersen}, {Krogel}, {Krueger}, {Kulp}, {Kumari}, {Kwan}, {Kyprianou}, {Labador}, {Labiano}, {Lafreni{\`e}re}, {Lagage}, {Laidler}, {Laine}, {Laird}, {Lajoie}, {Lallo}, {Lam}, {LaMassa}, {Lambros}, {Lampenfield}, {Lander}, {Langston}, {Larson}, {Larson}, {LaVerghetta}, {Law}, {Lawrence}, {Lee}, {Lee}, {Lee}, {Leisenring}, {Leveille}, {Levenson}, {Levi}, {Levine}, {Lewis}, {Lewis}, {Lewis}, {Libralato}, {Lidon}, {Liebrecht}, {Lightsey}, {Lilly}, {Lim}, {Lim}, {Ling}, {Link}, {Link}, {Lipinski}, {Liu}, {Lo}, {Lobmeyer}, {Logue}, {Long}, {Long}, {Long}, {Long}, {L{\'o}pez-Caniego}, {Lotz}, {Love-Pruitt}, {Lubskiy}, {Luers}, {Luetgens}, {Luevano}, {Lui}, {Lund}, {Lundquist}, {Lunine}, {L{\"u}tzgendorf}, {Lynch}, {MacDonald}, {MacDonald}, {Macias}, {Macklis}, {Maghami}, {Maharaja}, {Maiolino},
  {Makrygiannis}, {Malla}, {Malumuth}, {Manjavacas}, {Marini}, {Marrione}, {Marston}, {Martel}, {Martin}, {Martin}, {Martinez}, {Maschmann}, {Masci}, {Masetti}, {Maszkiewicz}, {Matthews}, {Matuskey}, {McBrayer}, {McCarthy}, {McCaughrean}, {McClare}, {McClare}, {McCloskey}, {McClurg}, {McCoy}, {McElwain}, {McGregor}, {McGuffey}, {McKay}, {McKenzie}, {McLean}, {McMaster}, {McNeil}, {De Meester}, {Mehalick}, {Meixner}, {Mel{\'e}ndez}, {Menzel}, {Menzel}, {Merz}, {Mesterharm}, {Meyer}, {Meyett}, {Meza}, {Midwinter}, {Milam}, {Miller}, {Miller}, {Miskey}, {Misselt}, {Mitchell}, {Mohan}, {Montoya}, {Moran}, {Morishita}, {Moro-Mart{\'\i}n}, {Morrison}, {Morrison}, {Morse}, {Moschos}, {Moseley}, {Mosier}, {Mosner}, {Mountain}, {Muckenthaler}, {Mueller}, {Mueller}, {Muhiem}, {M{\"u}hlmann}, {Mullally}, {Mullen}, {Munger}, {Murphy}, {Murray}, {Muzerolle}, {Mycroft}, {Myers}, {Myers}, {Myers}, {Myers}, {Myrick}, {Nagle}, {Nayak}, {Naylor}, {Neff}, {Nelan}, {Nella}, {Nguyen}, {Nguyen}, {Nickson}, {Nidhiry}, {Niedner},
  {Nieto-Santisteban}, {Nikolov}, {Nishisaka}, {Noriega-Crespo}, {Nota}, {O'Mara}, {Oboryshko}, {O'Brien}, {Ochs}, {Offenberg}, {Ogle}, {Ohl}, {Olmsted}, {Osborne}, {O'Shaughnessy}, {{\"O}stlin}, {O'Sullivan}, {Otor}, {Ottens}, {Ouellette}, {Outlaw}, {Owens}, {Pacifici}, {Page}, {Paranilam}, {Park}, {Parrish}, {Paschal}, {Patapis}, {Patel}, {Patrick}, {Pattishall}, {Paul}, {Paul}, {Pauly}, {Pavlovsky}, {Pe{\~n}a-Guerrero}, {Pedder}, {Peek}, {Pelham}, {Penanen}, {Perriello}, {Perrin}, {Perrine}, {Perrygo}, {Peslier}, {Petach}, {Peterson}, {Pfarr}, {Pierson}, {Pietraszkiewicz}, {Pilchen}, {Pipher}, {Pirzkal}, {Pitman}, {Player}, {Plesha}, {Plitzke}, {Pohner}, {Poletis}, {Pollizzi}, {Polster}, {Pontius}, {Pontoppidan}, {Porges}, {Potter}, {Prescott}, {Proffitt}, {Pueyo}, {Quispe Neira}, {Radich}, {Rager}, {Rameau}, {Ramey}, {Ramos Alarcon}, {Rampini}, {Rapp}, {Rashford}, {Rauscher}, {Ravindranath}, {Rawle}, {Rawlings}, {Ray}, {Regan}, {Rehm}, {Rehm}, {Reid}, {Reis}, {Renk}, {Reoch}, {Ressler}, {Rest},
  {Reynolds}, {Richon}, {Richon}, {Ridgaway}, {Riedel}, {Rieke}, {Rieke}, {Rifelli}, {Rigby}, {Riggs}, {Ringel}, {Ritchie}, {Rix}, {Robberto}, {Robinson}, {Robinson}, {Robinson}, {Rock}, {Rodriguez}, {Rodr{\'\i}guez del Pino}, {Roellig}, {Rohrbach}, {Roman}, {Romelfanger}, {Romo}, {Rosales}, {Rose}, {Roteliuk}, {Roth}, {Rothwell}, {Rouzaud}, {Rowe}, {Rowlands}, {Roy}, {Royer}, {Rui}, {Rumler}, {Rumpl}, {Russ}, {Ryan}, {Ryan}, {Saad}, {Sabata}, {Sabatino}, {Sabbi}, {Sabelhaus}, {Sabia}, {Sahu}, {Saif}, {Salvignol}, {Samara-Ratna}, {Samuelson}, {Sanders}, {Sappington}, {Sargent}, {Sauer}, {Savadkin}, {Sawicki}, {Schappell}, {Scheffer}, {Scheithauer}, {Scherer}, {Schiff}, {Schlawin}, {Schmeitzky}, {Schmitz}, {Schmude}, {Schneider}, {Schreiber}, {Schroeven-Deceuninck}, {Schultz}, {Schwab}, {Schwartz}, {Scoccimarro}, {Scott}, {Scott}, {Seaton}, {Seely}, {Seery}, {Seidleck}, {Sembach}, {Shanahan}, {Shaughnessy}, {Shaw}, {Shay}, {Sheehan}, {Sheth}, {Shih}, {Shivaei}, {Siegel}, {Sienkiewicz}, {Simmons}, {Simon},
  {Sirianni}, {Sivaramakrishnan}, {Slade}, {Sloan}, {Slocum}, {Slowinski}, {Smith}, {Smith}, {Smith}, {Smith}, {Smith}, {Smith}, {Smolik}, {Soderblom}, {Sohn}, {Sokol}, {Sonneborn}, {Sontag}, {Sooy}, {Soummer}, {Southwood}, {Spain}, {Sparmo}, {Speer}, {Spencer}, {Sprofera}, {Stallcup}, {Stanley}, {Stansberry}, {Stark}, {Starr}, {Stassi}, {Steck}, {Steeley}, {Stephens}, {Stephenson}, {Stewart}, {Stiavelli}, {}, {Strada}, {Straughn}, {Streetman}, {Strickland}, {Strobele}, {Stuhlinger}, {Stys}, {Such}, {Sukhatme}, {Sullivan}, {Sullivan}, {Sumner}, {Sun}, {Sunnquist}, {Swade}, {Swam}, {Swenton}, {Swoish}, {Tam Litten}, {Tamas}, {Tao}, {Taylor}, {Taylor}, {te Plate}, {Van Tea}, {Teague}, {Telfer}, {Temim}, {Texter}, {Thatte}, {Thompson}, {Thompson}, {Thomson}, {Thronson}, {Tierney}, {Tikkanen}, {Tinnin}, {Tippet}, {Todd}, {Tran}, {Trauger}, {Trejo}, {Vinh Truong}, {Tsukamoto}, {Tufail}, {Tumlinson}, {Tustain}, {Tyra}, {Ubeda}, {Underwood}, {Uzzo}, {Vaclavik}, {Valenduc}, {Valenti}, {Van Campen}, {van de Wetering},
  {Van Der Marel}, {van Haarlem}, {Vandenbussche}, {van Dishoeck}, {Vanterpool}, {Vernoy}, {Vila Costas}, {Volk}, {Voorzaat}, {Voyton}, {Vydra}, {Waddy}, {Waelkens}, {Wahlgren}, {Walker}, {Wander}, {Warfield}, {Warner}, {Wasiak}, {Wasiak}, {Wehner}, {Weiler}, {Weilert}, {Weiss}, {Wells}, {Welty}, {Wheate}, {Wheeler}, {White}, {Whitehouse}, {Whiteleather}, {Whitman}, {Williams}, {Willmer}, {Willott}, {Willoughby}, {Wilson}, {Wilson}, {Wilson}, {Windhorst}, {Wislowski}, {Wolfe}, {Wolfe}, {Wolff}, {Wondel}, {Woo}, {Woods}, {Worden}, {Workman}, {Wright}, {Wu}, {Wu}, {Wun}, {Wymer}, {Yadetie}, {Yan}, {Yang}, {Yates}, {Yeager}, {Yerger}, {Young}, {Young}, {Yu}, {Yu}, {Zak}, {Zeidler}, {Zepp}, {Zhou}, {Zincke}, {Zonak}, \& {Zondag}}]{Gardner2023}
{Gardner}, J.~P., {Mather}, J.~C., {Abbott}, R., {et~al.} 2023, \pasp, 135, 068001

\bibitem[{{Gorti} {et~al.}(2016){Gorti}, {Liseau}, {S{\'a}ndor}, \& {Clarke}}]{Gorti2016}
{Gorti}, U., {Liseau}, R., {S{\'a}ndor}, Z., \& {Clarke}, C. 2016, \ssr, 205, 125

\bibitem[{{G{\"u}nther} \& {Daylan}(2021)}]{guenther2021}
{G{\"u}nther}, M.~N. \& {Daylan}, T. 2021, \apjs, 254, 13

\bibitem[{{Hakim} {et~al.}(2018){Hakim}, {Rivoldini}, {Van Hoolst}, {Cottenier}, {Jaeken}, {Chust}, \& {Steinle-Neumann}}]{Hakim2018}
{Hakim}, K., {Rivoldini}, A., {Van Hoolst}, T., {et~al.} 2018, \icarus, 313, 61

\bibitem[{{Haldemann} {et~al.}(2020){Haldemann}, {Alibert}, {Mordasini}, \& {Benz}}]{Haldemann2020}
{Haldemann}, J., {Alibert}, Y., {Mordasini}, C., \& {Benz}, W. 2020, \aap, 643, A105

\bibitem[{{Husser} {et~al.}(2013){Husser}, {Wende-von Berg}, {Dreizler}, {Homeier}, {Reiners}, {Barman}, \& {Hauschildt}}]{husser2013}
{Husser}, T.~O., {Wende-von Berg}, S., {Dreizler}, S., {et~al.} 2013, \aap, 553, A6

\bibitem[{{Izidoro} {et~al.}(2021){Izidoro}, {Bitsch}, \& {Dasgupta}}]{Izidoro2021b}
{Izidoro}, A., {Bitsch}, B., \& {Dasgupta}, R. 2021, \apj, 915, 62

\bibitem[{{Jenkins} {et~al.}(2016){Jenkins}, {Twicken}, {McCauliff}, {Campbell}, {Sanderfer}, {Lung}, {Mansouri-Samani}, {Girouard}, {Tenenbaum}, {Klaus}, {Smith}, {Caldwell}, {Chacon}, {Henze}, {Heiges}, {Latham}, {Morgan}, {Swade}, {Rinehart}, \& {Vanderspek}}]{2016SPIE.9913E..3EJ}
{Jenkins}, J.~M., {Twicken}, J.~D., {McCauliff}, S., {et~al.} 2016, in Society of Photo-Optical Instrumentation Engineers (SPIE) Conference Series, Vol. 9913, Software and Cyberinfrastructure for Astronomy IV, ed. G.~{Chiozzi} \& J.~C. {Guzman}, 99133E

\bibitem[{{Johansen} \& {Lacerda}(2010)}]{Johansen2010}
{Johansen}, A. \& {Lacerda}, P. 2010, \mnras, 404, 475

\bibitem[{{Johansen} {et~al.}(2015){Johansen}, {Mac Low}, {Lacerda}, \& {Bizzarro}}]{Johansen2015}
{Johansen}, A., {Mac Low}, M.-M., {Lacerda}, P., \& {Bizzarro}, M. 2015, Science Advances, 1, 1500109

\bibitem[{{Johnstone}(2020)}]{johnstone2020}
{Johnstone}, C.~P. 2020, \apj, 890, 79

\bibitem[{{Johnstone} {et~al.}(2021){Johnstone}, {Bartel}, \& {G{\"u}del}}]{johnstone2021mors}
{Johnstone}, C.~P., {Bartel}, M., \& {G{\"u}del}, M. 2021, \aap, 649, A96

\bibitem[{{Johnstone} {et~al.}(2015{\natexlab{a}}){Johnstone}, {G{\"u}del}, {Brott}, \& {L{\"u}ftinger}}]{johnstone2015_stII}
{Johnstone}, C.~P., {G{\"u}del}, M., {Brott}, I., \& {L{\"u}ftinger}, T. 2015{\natexlab{a}}, \aap, 577, A28

\bibitem[{{Johnstone} {et~al.}(2015{\natexlab{b}}){Johnstone}, {G{\"u}del}, {St{\"o}kl}, {Lammer}, {Tu}, {Kislyakova}, {L{\"u}ftinger}, {Odert}, {Erkaev}, \& {Dorfi}}]{Johnstone2015}
{Johnstone}, C.~P., {G{\"u}del}, M., {St{\"o}kl}, A., {et~al.} 2015{\natexlab{b}}, \apjl, 815, L12

\bibitem[{{Kawashima} {et~al.}(2019){Kawashima}, {Hu}, \& {Ikoma}}]{Kawashima2019a}
{Kawashima}, Y., {Hu}, R., \& {Ikoma}, M. 2019, \apjl, 876, L5

\bibitem[{{Kawashima} \& {Ikoma}(2018)}]{Kawashima2018}
{Kawashima}, Y. \& {Ikoma}, M. 2018, \apj, 853, 7

\bibitem[{{Kawashima} \& {Ikoma}(2019)}]{Kawashima2019b}
{Kawashima}, Y. \& {Ikoma}, M. 2019, \apj, 877, 109

\bibitem[{{Khodachenko} {et~al.}(2015){Khodachenko}, {Shaikhislamov}, {Lammer}, \& {Prokopov}}]{Khodachenko2015}
{Khodachenko}, M.~L., {Shaikhislamov}, I.~F., {Lammer}, H., \& {Prokopov}, P.~A. 2015, \apj, 813, 50

\bibitem[{{Kimura} {et~al.}(2016){Kimura}, {Kunitomo}, \& {Takahashi}}]{Kimura2016}
{Kimura}, S.~S., {Kunitomo}, M., \& {Takahashi}, S.~Z. 2016, \mnras, 461, 2257

\bibitem[{{Kipping}(2013)}]{Kipping2013}
{Kipping}, D.~M. 2013, \mnras, 435, 2152

\bibitem[{{Kite} {et~al.}(2019){Kite}, {Fegley}, {Schaefer}, \& {Ford}}]{Kite2019}
{Kite}, E.~S., {Fegley}, Bruce, J., {Schaefer}, L., \& {Ford}, E.~B. 2019, \apjl, 887, L33

\bibitem[{{Kubyshkina} {et~al.}(2019{\natexlab{a}}){Kubyshkina}, {Cubillos}, {Fossati}, {Erkaev}, {Johnstone}, {Kislyakova}, {Lammer}, {Lendl}, {Odert}, \& {G{\"u}del}}]{Kubyshkina2019a}
{Kubyshkina}, D., {Cubillos}, P.~E., {Fossati}, L., {et~al.} 2019{\natexlab{a}}, \apj, 879, 26

\bibitem[{{Kubyshkina} \& {Fossati}(2022)}]{kubyshkina_fossati2022}
{Kubyshkina}, D. \& {Fossati}, L. 2022, \aap, 668, A178

\bibitem[{{Kubyshkina} {et~al.}(2018){Kubyshkina}, {Fossati}, {Erkaev}, {Johnstone}, {Cubillos}, {Kislyakova}, {Lammer}, {Lendl}, \& {Odert}}]{kubyshkina2018grid}
{Kubyshkina}, D., {Fossati}, L., {Erkaev}, N.~V., {et~al.} 2018, \aap, 619, A151

\bibitem[{{Kubyshkina} {et~al.}(2019{\natexlab{b}}){Kubyshkina}, {Fossati}, {Mustill}, {Cubillos}, {Davies}, {Erkaev}, {Johnstone}, {Kislyakova}, {Lammer}, {Lendl}, \& {Odert}}]{Kubyshkina2019b}
{Kubyshkina}, D., {Fossati}, L., {Mustill}, A.~J., {et~al.} 2019{\natexlab{b}}, \aap, 632, A65

\bibitem[{{Kubyshkina} \& {Vidotto}(2021)}]{kubyshkina2021mesa}
{Kubyshkina}, D. \& {Vidotto}, A.~A. 2021, \mnras, 504, 2034

\bibitem[{{Kubyshkina} {et~al.}(2020){Kubyshkina}, {Vidotto}, {Fossati}, \& {Farrell}}]{kubyshkina2020mesa}
{Kubyshkina}, D., {Vidotto}, A.~A., {Fossati}, L., \& {Farrell}, E. 2020, \mnras, 499, 77

\bibitem[{{Kubyshkina} {et~al.}(2022{\natexlab{a}}){Kubyshkina}, {Vidotto}, {Villarreal D'Angelo}, {Carolan}, {Hazra}, \& {Carleo}}]{kubyshkina2022_tois_I}
{Kubyshkina}, D., {Vidotto}, A.~A., {Villarreal D'Angelo}, C., {et~al.} 2022{\natexlab{a}}, \mnras, 510, 2111

\bibitem[{{Kubyshkina} {et~al.}(2022{\natexlab{b}}){Kubyshkina}, {Vidotto}, {Villarreal D'Angelo}, {Carolan}, {Hazra}, \& {Carleo}}]{kubyshkina2022_tois_II}
{Kubyshkina}, D., {Vidotto}, A.~A., {Villarreal D'Angelo}, C., {et~al.} 2022{\natexlab{b}}, \mnras, 510, 3039

\bibitem[{{Kubyshkina} \& {Fossati}(2021)}]{kubyshkina2021grid}
{Kubyshkina}, D.~I. \& {Fossati}, L. 2021, Research Notes of the American Astronomical Society, 5, 74

\bibitem[{{Kurucz}(1979)}]{Kurucz1979}
{Kurucz}, R.~L. 1979, \apjs, 40, 1

\bibitem[{{Kurucz}(1993)}]{Kurucz1993}
{Kurucz}, R.~L. 1993, {SYNTHE spectrum synthesis programs and line data} (Astrophysics Source Code Library)

\bibitem[{{Lambrechts} \& {Johansen}(2012)}]{Lambrechts2012}
{Lambrechts}, M. \& {Johansen}, A. 2012, \aap, 544, A32

\bibitem[{{Lammer} {et~al.}(2016){Lammer}, {Erkaev}, {Fossati}, {Juvan}, {Odert}, {Cubillos}, {Guenther}, {Kislyakova}, {Johnstone}, {L{\"u}ftinger}, \& {G{\"u}del}}]{Lammer2016}
{Lammer}, H., {Erkaev}, N.~V., {Fossati}, L., {et~al.} 2016, \mnras, 461, L62

\bibitem[{{Lanza} {et~al.}(2001){Lanza}, {Rodon{\`o}}, {Mazzola}, \& {Messina}}]{Lanza2001}
{Lanza}, A.~F., {Rodon{\`o}}, M., {Mazzola}, L., \& {Messina}, S. 2001, \aap, 376, 1011

\bibitem[{{Laskar}(1990)}]{Laskar_1990}
{Laskar}, J. 1990, \icarus, 88, 266

\bibitem[{{Laskar}(1993)}]{Laskar_1993PD}
{Laskar}, J. 1993, Physica D Nonlinear Phenomena, 67, 257

\bibitem[{{Laskar} \& {Robutel}(2001)}]{Laskar_Robutel_2001}
{Laskar}, J. \& {Robutel}, P. 2001, Celestial Mechanics and Dynamical Astronomy, 80, 39

\bibitem[{{Lavvas} {et~al.}(2019){Lavvas}, {Koskinen}, {Steinrueck}, {Garc{\'\i}a Mu{\~n}oz}, \& {Showman}}]{Lavvas2019}
{Lavvas}, P., {Koskinen}, T., {Steinrueck}, M.~E., {Garc{\'\i}a Mu{\~n}oz}, A., \& {Showman}, A.~P. 2019, \apj, 878, 118

\bibitem[{{Leleu} {et~al.}(2021){Leleu}, {Alibert}, {Hara}, {Hooton}, {Wilson}, {Robutel}, {Delisle}, {Laskar}, {Hoyer}, {Lovis}, {Bryant}, {Ducrot}, {Cabrera}, {Delrez}, {Acton}, {Adibekyan}, {Allart}, {Allende Prieto}, {Alonso}, {Alves}, {Anderson}, {Angerhausen}, {Anglada Escud{\'e}}, {Asquier}, {Barrado}, {Barros}, {Baumjohann}, {Bayliss}, {Beck}, {Beck}, {Bekkelien}, {Benz}, {Billot}, {Bonfanti}, {Bonfils}, {Bouchy}, {Bourrier}, {Bou{\'e}}, {Brandeker}, {Broeg}, {Buder}, {Burdanov}, {Burleigh}, {B{\'a}rczy}, {Cameron}, {Chamberlain}, {Charnoz}, {Cooke}, {Corral Van Damme}, {Correia}, {Cristiani}, {Damasso}, {Davies}, {Deleuil}, {Demangeon}, {Demory}, {Di Marcantonio}, {Di Persio}, {Dumusque}, {Ehrenreich}, {Erikson}, {Figueira}, {Fortier}, {Fossati}, {Fridlund}, {Futyan}, {Gandolfi}, {Garc{\'\i}a Mu{\~n}oz}, {Garcia}, {Gill}, {Gillen}, {Gillon}, {Goad}, {Gonz{\'a}lez Hern{\'a}ndez}, {Guedel}, {G{\"u}nther}, {Haldemann}, {Henderson}, {Heng}, {Hogan}, {Isaak}, {Jehin}, {Jenkins}, {Jord{\'a}n}, {Kiss},
  {Kristiansen}, {Lam}, {Lavie}, {Lecavelier des Etangs}, {Lendl}, {Lillo-Box}, {Lo Curto}, {Magrin}, {Martins}, {Maxted}, {McCormac}, {Mehner}, {Micela}, {Molaro}, {Moyano}, {Murray}, {Nascimbeni}, {Nunes}, {Olofsson}, {Osborn}, {Oshagh}, {Ottensamer}, {Pagano}, {Pall{\'e}}, {Pedersen}, {Pepe}, {Persson}, {Peter}, {Piotto}, {Polenta}, {Pollacco}, {Poretti}, {Pozuelos}, {Queloz}, {Ragazzoni}, {Rando}, {Ratti}, {Rauer}, {Raynard}, {Rebolo}, {Reimers}, {Ribas}, {Santos}, {Scandariato}, {Schneider}, {Sebastian}, {Sestovic}, {Simon}, {Smith}, {Sousa}, {Sozzetti}, {Steller}, {Su{\'a}rez Mascare{\~n}o}, {Szab{\'o}}, {S{\'e}gransan}, {Thomas}, {Thompson}, {Tilbrook}, {Triaud}, {Turner}, {Udry}, {Van Grootel}, {Venus}, {Verrecchia}, {Vines}, {Walton}, {West}, {Wheatley}, {Wolter}, \& {Zapatero Osorio}}]{Leleu2021}
{Leleu}, A., {Alibert}, Y., {Hara}, N.~C., {et~al.} 2021, \aap, 649, A26

\bibitem[{{Lendl} {et~al.}(2020){Lendl}, {Csizmadia}, {Deline}, {Fossati}, {Kitzmann}, {Heng}, {Hoyer}, {Salmon}, {Benz}, {Broeg}, {Ehrenreich}, {Fortier}, {Queloz}, {Bonfanti}, {Brandeker}, {Collier Cameron}, {Delrez}, {Garcia Mu{\~n}oz}, {Hooton}, {Maxted}, {Morris}, {Van Grootel}, {Wilson}, {Alibert}, {Alonso}, {Asquier}, {Bandy}, {B{\'a}rczy}, {Barrado}, {Barros}, {Baumjohann}, {Beck}, {Beck}, {Bekkelien}, {Bergomi}, {Billot}, {Biondi}, {Bonfils}, {Bourrier}, {Busch}, {Cabrera}, {Cessa}, {Charnoz}, {Chazelas}, {Corral Van Damme}, {Davies}, {Deleuil}, {Demangeon}, {Demory}, {Erikson}, {Farinato}, {Fridlund}, {Futyan}, {Gandolfi}, {Gillon}, {Guterman}, {Hasiba}, {Hernandez}, {Isaak}, {Kiss}, {Kuntzer}, {Lecavelier des Etangs}, {L{\"u}ftinger}, {Laskar}, {Lovis}, {Magrin}, {Malvasio}, {Marafatto}, {Michaelis}, {Munari}, {Nascimbeni}, {Olofsson}, {Ottacher}, {Ottensamer}, {Pagano}, {Pall{\'e}}, {Peter}, {Piazza}, {Piotto}, {Pollacco}, {Ratti}, {Rauer}, {Ragazzoni}, {Rando}, {Ribas}, {Rieder}, {Rohlfs},
  {Safa}, {Santos}, {Scandariato}, {S{\'e}gransan}, {Simon}, {Singh}, {Smith}, {Sordet}, {Sousa}, {Steller}, {Szab{\'o}}, {Thomas}, {Tschentscher}, {Udry}, {Viotto}, {Walter}, {Walton}, {Wildi}, \& {Wolter}}]{Lendl2020}
{Lendl}, M., {Csizmadia}, S., {Deline}, A., {et~al.} 2020, \aap, 643, A94

\bibitem[{{Libby-Roberts} {et~al.}(2020){Libby-Roberts}, {Berta-Thompson}, {D{\'e}sert}, {Masuda}, {Morley}, {Lopez}, {Deck}, {Fabrycky}, {Fortney}, {Line}, {Sanchis-Ojeda}, \& {Winn}}]{libby2020}
{Libby-Roberts}, J.~E., {Berta-Thompson}, Z.~K., {D{\'e}sert}, J.-M., {et~al.} 2020, \aj, 159, 57

\bibitem[{{Lindegren} {et~al.}(2021){Lindegren}, {Bastian}, {Biermann}, {Bombrun}, {de Torres}, {Gerlach}, {Geyer}, {Hern{\'a}ndez}, {Hilger}, {Hobbs}, {Klioner}, {Lammers}, {McMillan}, {Ramos-Lerate}, {Steidelm{\"u}ller}, {Stephenson}, \& {van Leeuwen}}]{Lindegren2021}
{Lindegren}, L., {Bastian}, U., {Biermann}, M., {et~al.} 2021, \aap, 649, A4

\bibitem[{{Linsky} {et~al.}(2014){Linsky}, {Fontenla}, \& {France}}]{Linsky2014ApJ...780...61L}
{Linsky}, J.~L., {Fontenla}, J., \& {France}, K. 2014, \apj, 780, 61

\bibitem[{{Linsky} {et~al.}(2013){Linsky}, {France}, \& {Ayres}}]{Linsky2013ApJ...766...69L}
{Linsky}, J.~L., {France}, K., \& {Ayres}, T. 2013, \apj, 766, 69

\bibitem[{{Lissauer} {et~al.}(2011){Lissauer}, {Ragozzine}, {Fabrycky}, {Steffen}, {Ford}, {Jenkins}, {Shporer}, {Holman}, {Rowe}, {Quintana}, {Batalha}, {Borucki}, {Bryson}, {Caldwell}, {Carter}, {Ciardi}, {Dunham}, {Fortney}, {Gautier}, {Howell}, {Koch}, {Latham}, {Marcy}, {Morehead}, \& {Sasselov}}]{Lissauer_etal_2011K}
{Lissauer}, J.~J., {Ragozzine}, D., {Fabrycky}, D.~C., {et~al.} 2011, \apjs, 197, 8

\bibitem[{{Lopez}(2017)}]{Lopez2017}
{Lopez}, E.~D. 2017, \mnras, 472, 245

\bibitem[{{Lopez} \& {Fortney}(2014)}]{LopezFortney2014}
{Lopez}, E.~D. \& {Fortney}, J.~J. 2014, \apj, 792, 1

\bibitem[{{Marboeuf} {et~al.}(2014){Marboeuf}, {Thiabaud}, {Alibert}, {Cabral}, \& {Benz}}]{Marboeuf2014}
{Marboeuf}, U., {Thiabaud}, A., {Alibert}, Y., {Cabral}, N., \& {Benz}, W. 2014, \aap, 570, A36

\bibitem[{{Marigo} {et~al.}(2017){Marigo}, {Girardi}, {Bressan}, {Rosenfield}, {Aringer}, {Chen}, {Dussin}, {Nanni}, {Pastorelli}, {Rodrigues}, {Trabucchi}, {Bladh}, {Dalcanton}, {Groenewegen}, {Montalb{\'a}n}, \& {Wood}}]{marigo2017}
{Marigo}, P., {Girardi}, L., {Bressan}, A., {et~al.} 2017, \apj, 835, 77

\bibitem[{{Maxted} {et~al.}(2021){Maxted}, {Ehrenreich}, {Wilson}, {Alibert}, {Collier Cameron}, {Hoyer}, {Sousa}, {Olofsson}, {Bekkelien}, {Deline}, {Delrez}, {Bonfanti}, {Borsato}, {Alonso}, {Anglada Escud{\'e}}, {Barrado}, {Barros}, {Baumjohann}, {Beck}, {Beck}, {Benz}, {Billot}, {Biondi}, {Bonfils}, {Brandeker}, {Broeg}, {B{\'a}rczy}, {Cabrera}, {Charnoz}, {Corral Van Damme}, {Csizmadia}, {Davies}, {Deleuil}, {Demangeon}, {Demory}, {Erikson}, {Flor{\'e}n}, {Fortier}, {Fossati}, {Fridlund}, {Futyan}, {Gandolfi}, {Gillon}, {Guedel}, {Guterman}, {Heng}, {Isaak}, {Kiss}, {Laskar}, {Lecavelier des Etangs}, {Lendl}, {Lovis}, {Magrin}, {Nascimbeni}, {Ottensamer}, {Pagano}, {Pall{\'e}}, {Peter}, {Piotto}, {Pollacco}, {Pozuelos}, {Queloz}, {Ragazzoni}, {Rando}, {Rauer}, {Reimers}, {Ribas}, {Santos}, {Scandariato}, {Simon}, {Smith}, {Steller}, {Swayne}, {Szab{\'o}}, {S{\'e}gransan}, {Thomas}, {Udry}, {Van Grootel}, \& {Walton}}]{Maxted2021}
{Maxted}, P.~F.~L., {Ehrenreich}, D., {Wilson}, T.~G., {et~al.} 2021, \mnras [\eprint[arXiv]{2111.08828}]

\bibitem[{{Mayor} {et~al.}(2003){Mayor}, {Pepe}, {Queloz}, {Bouchy}, {Rupprecht}, {Lo Curto}, {Avila}, {Benz}, {Bertaux}, {Bonfils}, {Dall}, {Dekker}, {Delabre}, {Eckert}, {Fleury}, {Gilliotte}, {Gojak}, {Guzman}, {Kohler}, {Lizon}, {Longinotti}, {Lovis}, {Megevand}, {Pasquini}, {Reyes}, {Sivan}, {Sosnowska}, {Soto}, {Udry}, {van Kesteren}, {Weber}, \& {Weilenmann}}]{Mayor2003}
{Mayor}, M., {Pepe}, F., {Queloz}, D., {et~al.} 2003, The Messenger, 114, 20

\bibitem[{{Mordasini}(2020)}]{mordasini2020}
{Mordasini}, C. 2020, \aap, 638, A52

\bibitem[{{Morris} {et~al.}(2021){Morris}, {Heng}, {Brandeker}, {Swan}, \& {Lendl}}]{2021A&A...651L..12M}
{Morris}, B.~M., {Heng}, K., {Brandeker}, A., {Swan}, A., \& {Lendl}, M. 2021, \aap, 651, L12

\bibitem[{{Ohno} \& {Tanaka}(2021)}]{ohno2021}
{Ohno}, K. \& {Tanaka}, Y.~A. 2021, \apj, 920, 124

\bibitem[{{Ormel} \& {Klahr}(2010)}]{Ormel2010}
{Ormel}, C.~W. \& {Klahr}, H.~H. 2010, \aap, 520, A43

\bibitem[{{Paxton} {et~al.}(2011){Paxton}, {Bildsten}, {Dotter}, {Herwig}, {Lesaffre}, \& {Timmes}}]{paxton2011}
{Paxton}, B., {Bildsten}, L., {Dotter}, A., {et~al.} 2011, \apjs, 192, 3

\bibitem[{{Paxton} {et~al.}(2013){Paxton}, {Cantiello}, {Arras}, {Bildsten}, {Brown}, {Dotter}, {Mankovich}, {Montgomery}, {Stello}, {Timmes}, \& {Townsend}}]{paxton2013}
{Paxton}, B., {Cantiello}, M., {Arras}, P., {et~al.} 2013, \apjs, 208, 4

\bibitem[{{Paxton} {et~al.}(2018){Paxton}, {Schwab}, {Bauer}, {Bildsten}, {Blinnikov}, {Duffell}, {Farmer}, {Goldberg}, {Marchant}, {Sorokina}, {Thoul}, {Townsend}, \& {Timmes}}]{paxton2018}
{Paxton}, B., {Schwab}, J., {Bauer}, E.~B., {et~al.} 2018, \apjs, 234, 34

\bibitem[{{Paxton} {et~al.}(2019){Paxton}, {Smolec}, {Schwab}, {Gautschy}, {Bildsten}, {Cantiello}, {Dotter}, {Farmer}, {Goldberg}, {Jermyn}, {Kanbur}, {Marchant}, {Thoul}, {Townsend}, {Wolf}, {Zhang}, \& {Timmes}}]{paxton2019}
{Paxton}, B., {Smolec}, R., {Schwab}, J., {et~al.} 2019, \apjs, 243, 10

\bibitem[{{Pepe} {et~al.}(2021){Pepe}, {Cristiani}, {Rebolo}, {Santos}, {Dekker}, {Cabral}, {Di Marcantonio}, {Figueira}, {Lo Curto}, {Lovis}, {Mayor}, {M{\'e}gevand}, {Molaro}, {Riva}, {Zapatero Osorio}, {Amate}, {Manescau}, {Pasquini}, {Zerbi}, {Adibekyan}, {Abreu}, {Affolter}, {Alibert}, {Aliverti}, {Allart}, {Allende Prieto}, {{\'A}lvarez}, {Alves}, {Avila}, {Baldini}, {Bandy}, {Barros}, {Benz}, {Bianco}, {Borsa}, {Bourrier}, {Bouchy}, {Broeg}, {Calderone}, {Cirami}, {Coelho}, {Conconi}, {Coretti}, {Cumani}, {Cupani}, {D'Odorico}, {Damasso}, {Deiries}, {Delabre}, {Demangeon}, {Dumusque}, {Ehrenreich}, {Faria}, {Fragoso}, {Genolet}, {Genoni}, {G{\'e}nova Santos}, {Gonz{\'a}lez Hern{\'a}ndez}, {Hughes}, {Iwert}, {Kerber}, {Knudstrup}, {Landoni}, {Lavie}, {Lillo-Box}, {Lizon}, {Maire}, {Martins}, {Mehner}, {Micela}, {Modigliani}, {Monteiro}, {Monteiro}, {Moschetti}, {Murphy}, {Nunes}, {Oggioni}, {Oliveira}, {Oshagh}, {Pall{\'e}}, {Pariani}, {Poretti}, {Rasilla}, {Rebord{\~a}o}, {Redaelli}, {Santana Tschudi},
  {Santin}, {Santos}, {S{\'e}gransan}, {Schmidt}, {Segovia}, {Sosnowska}, {Sozzetti}, {Sousa}, {Span{\`o}}, {Su{\'a}rez Mascare{\~n}o}, {Tabernero}, {Tenegi}, {Udry}, \& {Zanutta}}]{Pepe2021}
{Pepe}, F., {Cristiani}, S., {Rebolo}, R., {et~al.} 2021, \aap, 645, A96

\bibitem[{{Pollack} {et~al.}(1996){Pollack}, {Hubickyj}, {Bodenheimer}, {Lissauer}, {Podolak}, \& {Greenzweig}}]{Pollack1996}
{Pollack}, J.~B., {Hubickyj}, O., {Bodenheimer}, P., {et~al.} 1996, \icarus, 124, 62

\bibitem[{{Pont} {et~al.}(2011){Pont}, {Aigrain}, \& {Zucker}}]{Pont2011}
{Pont}, F., {Aigrain}, S., \& {Zucker}, S. 2011, \mnras, 411, 1953

\bibitem[{{Quirrenbach} {et~al.}(2014){Quirrenbach}, {Amado}, {Caballero}, {Mundt}, {Reiners}, {Ribas}, {Seifert}, {Abril}, {Aceituno}, {Alonso-Floriano}, {Ammler-von Eiff}, {Antona Jim{\'e}nez}, {Anwand-Heerwart}, {Azzaro}, {Bauer}, {Barrado}, {Becerril}, {B{\'e}jar}, {Ben{\'\i}tez}, {Berdi{\~n}as}, {C{\'a}rdenas}, {Casal}, {Claret}, {Colom{\'e}}, {Cort{\'e}s-Contreras}, {Czesla}, {Doellinger}, {Dreizler}, {Feiz}, {Fern{\'a}ndez}, {Galad{\'\i}}, {G{\'a}lvez-Ortiz}, {Garc{\'\i}a-Piquer}, {Garc{\'\i}a-Vargas}, {Garrido}, {Gesa}, {G{\'o}mez Galera}, {Gonz{\'a}lez {\'A}lvarez}, {Gonz{\'a}lez Hern{\'a}ndez}, {Gr{\"o}zinger}, {Gu{\`a}rdia}, {Guenther}, {de Guindos}, {Guti{\'e}rrez-Soto}, {Hagen}, {Hatzes}, {Hauschildt}, {Helmling}, {Henning}, {Hermann}, {Hern{\'a}ndez Casta{\~n}o}, {Herrero}, {Hidalgo}, {Holgado}, {Huber}, {Huber}, {Jeffers}, {Joergens}, {de Juan}, {Kehr}, {Klein}, {K{\"u}rster}, {Lamert}, {Lalitha}, {Laun}, {Lemke}, {Lenzen}, {L{\'o}pez del Fresno}, {L{\'o}pez Mart{\'\i}}, {L{\'o}pez-Santiago},
  {Mall}, {Mandel}, {Mart{\'\i}n}, {Mart{\'\i}n-Ruiz}, {Mart{\'\i}nez-Rodr{\'\i}guez}, {Marvin}, {Mathar}, {Mirabet}, {Montes}, {Morales Mu{\~n}oz}, {Moya}, {Naranjo}, {Ofir}, {Oreiro}, {Pall{\'e}}, {Panduro}, {Passegger}, {P{\'e}rez-Calpena}, {P{\'e}rez Medialdea}, {Perger}, {Pluto}, {Ram{\'o}n}, {Rebolo}, {Redondo}, {Reffert}, {Reinhardt}, {Rhode}, {Rix}, {Rodler}, {Rodr{\'\i}guez}, {Rodr{\'\i}guez-L{\'o}pez}, {Rodr{\'\i}guez-P{\'e}rez}, {Rohloff}, {Rosich}, {S{\'a}nchez-Blanco}, {S{\'a}nchez Carrasco}, {Sanz-Forcada}, {Sarmiento}, {Sch{\"a}fer}, {Schiller}, {Schmidt}, {Schmitt}, {Solano}, {Stahl}, {Storz}, {St{\"u}rmer}, {Su{\'a}rez}, {Ulbrich}, {Veredas}, {Wagner}, {Winkler}, {Zapatero Osorio}, {Zechmeister}, {Abell{\'a}n de Paco}, {Anglada-Escud{\'e}}, {del Burgo}, {Klutsch}, {Lizon}, {L{\'o}pez-Morales}, {Morales}, {Perryman}, {Tulloch}, \& {Xu}}]{quirrenbach2014}
{Quirrenbach}, A., {Amado}, P.~J., {Caballero}, J.~A., {et~al.} 2014, in Society of Photo-Optical Instrumentation Engineers (SPIE) Conference Series, Vol. 9147, Ground-based and Airborne Instrumentation for Astronomy V, ed. S.~K. {Ramsay}, I.~S. {McLean}, \& H.~{Takami}, 91471F

\bibitem[{{Rajpaul} {et~al.}(2015){Rajpaul}, {Aigrain}, {Osborne}, {Reece}, \& {Roberts}}]{Rajpaul2015}
{Rajpaul}, V., {Aigrain}, S., {Osborne}, M.~A., {Reece}, S., \& {Roberts}, S. 2015, \mnras, 452, 2269

\bibitem[{Rasmussen \& Williams(2005)}]{rasmussen05}
Rasmussen, C.~E. \& Williams, C. K.~I. 2005, Gaussian Processes for Machine Learning (Adaptive Computation and Machine Learning) (The MIT Press)

\bibitem[{{Raymond} {et~al.}(2018){Raymond}, {Boulet}, {Izidoro}, {Esteves}, \& {Bitsch}}]{Raymond2018}
{Raymond}, S.~N., {Boulet}, T., {Izidoro}, A., {Esteves}, L., \& {Bitsch}, B. 2018, \mnras, 479, L81

\bibitem[{{Ricker} {et~al.}(2014){Ricker}, {Winn}, {Vanderspek}, {Latham}, {Bakos}, {Bean}, {Berta-Thompson}, {Brown}, {Buchhave}, {Butler}, {Butler}, {Chaplin}, {Charbonneau}, {Christensen-Dalsgaard}, {Clampin}, {Deming}, {Doty}, {De Lee}, {Dressing}, {Dunham}, {Endl}, {Fressin}, {Ge}, {Henning}, {Holman}, {Howard}, {Ida}, {Jenkins}, {Jernigan}, {Johnson}, {Kaltenegger}, {Kawai}, {Kjeldsen}, {Laughlin}, {Levine}, {Lin}, {Lissauer}, {MacQueen}, {Marcy}, {McCullough}, {Morton}, {Narita}, {Paegert}, {Palle}, {Pepe}, {Pepper}, {Quirrenbach}, {Rinehart}, {Sasselov}, {Sato}, {Seager}, {Sozzetti}, {Stassun}, {Sullivan}, {Szentgyorgyi}, {Torres}, {Udry}, \& {Villasenor}}]{Ricker2014}
{Ricker}, G.~R., {Winn}, J.~N., {Vanderspek}, R., {et~al.} 2014, in Society of Photo-Optical Instrumentation Engineers (SPIE) Conference Series, Vol. 9143, Space Telescopes and Instrumentation 2014: Optical, Infrared, and Millimeter Wave, ed. J.~{Oschmann}, Jacobus~M., M.~{Clampin}, G.~G. {Fazio}, \& H.~A. {MacEwen}, 914320

\bibitem[{{Sakai} {et~al.}(2018){Sakai}, {Seki}, {Terada}, {Shinagawa}, {Tanaka}, \& {Ebihara}}]{Sakai2018}
{Sakai}, S., {Seki}, K., {Terada}, N., {et~al.} 2018, \grl, 45, 9336

\bibitem[{{Salmon} {et~al.}(2021){Salmon}, {Van Grootel}, {Buldgen}, {Dupret}, \& {Eggenberger}}]{salmon2021}
{Salmon}, S.~J.~A.~J., {Van Grootel}, V., {Buldgen}, G., {Dupret}, M.~A., \& {Eggenberger}, P. 2021, \aap, 646, A7

\bibitem[{{Santos} {et~al.}(2013){Santos}, {Sousa}, {Mortier}, {Neves}, {Adibekyan}, {Tsantaki}, {Delgado Mena}, {Bonfils}, {Israelian}, {Mayor}, \& {Udry}}]{Santos-13}
{Santos}, N.~C., {Sousa}, S.~G., {Mortier}, A., {et~al.} 2013, \aap, 556, A150

\bibitem[{{Schanche} {et~al.}(2020){Schanche}, {H{\'e}brard}, {Collier Cameron}, {Dalal}, {Smalley}, {Wilson}, {Boisse}, {Bouchy}, {Brown}, {Demangeon}, {Haswell}, {Hellier}, {Kolb}, {Lopez}, {Maxted}, {Pollacco}, {West}, \& {Wheatley}}]{Schanche2020}
{Schanche}, N., {H{\'e}brard}, G., {Collier Cameron}, A., {et~al.} 2020, \mnras, 499, 428

\bibitem[{{Schwarz}(1978)}]{schwarz1978}
{Schwarz}, G. 1978, Annals of Statistics, 6, 461

\bibitem[{{Scuflaire} {et~al.}(2008){Scuflaire}, {Th{\'e}ado}, {Montalb{\'a}n}, {Miglio}, {Bourge}, {Godart}, {Thoul}, \& {Noels}}]{scuflaire2008}
{Scuflaire}, R., {Th{\'e}ado}, S., {Montalb{\'a}n}, J., {et~al.} 2008, \apss, 316, 83

\bibitem[{{Simola} {et~al.}(2022){Simola}, {Bonfanti}, {Dumusque}, {Cisewski-Kehe}, {Kaski}, \& {Corander}}]{simola2022}
{Simola}, U., {Bonfanti}, A., {Dumusque}, X., {et~al.} 2022, \aap, 664, A127

\bibitem[{{Simola} {et~al.}(2019){Simola}, {Dumusque}, \& {Cisewski-Kehe}}]{simola2019}
{Simola}, U., {Dumusque}, X., \& {Cisewski-Kehe}, J. 2019, \aap, 622, A131

\bibitem[{{Skrutskie} {et~al.}(2006){Skrutskie}, {Cutri}, {Stiening}, {Weinberg}, {Schneider}, {Carpenter}, {Beichman}, {Capps}, {Chester}, {Elias}, {Huchra}, {Liebert}, {Lonsdale}, {Monet}, {Price}, {Seitzer}, {Jarrett}, {Kirkpatrick}, {Gizis}, {Howard}, {Evans}, {Fowler}, {Fullmer}, {Hurt}, {Light}, {Kopan}, {Marsh}, {McCallon}, {Tam}, {Van Dyk}, \& {Wheelock}}]{Skrutskie2006}
{Skrutskie}, M.~F., {Cutri}, R.~M., {Stiening}, R., {et~al.} 2006, \aj, 131, 1163

\bibitem[{{Smith} {et~al.}(2012){Smith}, {Stumpe}, {Van Cleve}, {Jenkins}, {Barclay}, {Fanelli}, {Girouard}, {Kolodziejczak}, {McCauliff}, {Morris}, \& {Twicken}}]{Smith2012}
{Smith}, J.~C., {Stumpe}, M.~C., {Van Cleve}, J.~E., {et~al.} 2012, \pasp, 124, 1000

\bibitem[{{Sneden}(1973)}]{Sneden-73}
{Sneden}, C.~A. 1973, PhD thesis, THE UNIVERSITY OF TEXAS AT AUSTIN.

\bibitem[{{Sotin} {et~al.}(2007){Sotin}, {Grasset}, \& {Mocquet}}]{Sotin2007}
{Sotin}, C., {Grasset}, O., \& {Mocquet}, A. 2007, \icarus, 191, 337

\bibitem[{{Sousa}(2014)}]{Sousa-14}
{Sousa}, S.~G. 2014, [arXiv:1407.5817] [\eprint[arXiv]{1407.5817}]

\bibitem[{{Sousa} {et~al.}(2021){Sousa}, {Adibekyan}, {Delgado-Mena}, {Santos}, {Rojas-Ayala}, {Soares}, {Legoinha}, {Ulmer-Moll}, {Camacho}, {Barros}, {Demangeon}, {Hoyer}, {Israelian}, {Mortier}, {Tsantaki}, \& {Monteiro}}]{Sousa-21}
{Sousa}, S.~G., {Adibekyan}, V., {Delgado-Mena}, E., {et~al.} 2021, \aap, 656, A53

\bibitem[{{Sousa} {et~al.}(2015){Sousa}, {Santos}, {Adibekyan}, {Delgado-Mena}, \& {Israelian}}]{Sousa-15}
{Sousa}, S.~G., {Santos}, N.~C., {Adibekyan}, V., {Delgado-Mena}, E., \& {Israelian}, G. 2015, \aap, 577, A67

\bibitem[{{Sousa} {et~al.}(2007){Sousa}, {Santos}, {Israelian}, {Mayor}, \& {Monteiro}}]{Sousa-07}
{Sousa}, S.~G., {Santos}, N.~C., {Israelian}, G., {Mayor}, M., \& {Monteiro}, M.~J.~P.~F.~G. 2007, A\&A, 469, 783

\bibitem[{{Sousa} {et~al.}(2008){Sousa}, {Santos}, {Mayor}, {Udry}, {Casagrande}, {Israelian}, {Pepe}, {Queloz}, \& {Monteiro}}]{Sousa-08}
{Sousa}, S.~G., {Santos}, N.~C., {Mayor}, M., {et~al.} 2008, \aap, 487, 373

\bibitem[{{Spada} {et~al.}(2013){Spada}, {Demarque}, {Kim}, \& {Sills}}]{spada2013}
{Spada}, F., {Demarque}, P., {Kim}, Y.~C., \& {Sills}, A. 2013, \apj, 776, 87

\bibitem[{Speagle(2020)}]{speagle}
Speagle, J.~S. 2020, Monthly Notices of the Royal Astronomical Society, 493, 3132

\bibitem[{{Stumpe} {et~al.}(2014){Stumpe}, {Smith}, {Catanzarite}, {Van Cleve}, {Jenkins}, {Twicken}, \& {Girouard}}]{Stumpe2014}
{Stumpe}, M.~C., {Smith}, J.~C., {Catanzarite}, J.~H., {et~al.} 2014, \pasp, 126, 100

\bibitem[{{Szab{\'o}} {et~al.}(2021){Szab{\'o}}, {Gandolfi}, {Brandeker}, {Csizmadia}, {Garai}, {Billot}, {Broeg}, {Ehrenreich}, {Fortier}, {Fossati}, {Hoyer}, {Kiss}, {Lecavelier des Etangs}, {Maxted}, {Ribas}, {Alibert}, {Alonso}, {Anglada Escud{\'e}}, {B{\'a}rczy}, {Barros}, {Barrado}, {Baumjohann}, {Beck}, {Beck}, {Bekkelien}, {Bonfils}, {Benz}, {Borsato}, {Busch}, {Cabrera}, {Charnoz}, {Collier Cameron}, {Van Damme}, {Davies}, {Delrez}, {Deleuil}, {Demangeon}, {Demory}, {Erikson}, {Fridlund}, {Futyan}, {Garc{\'\i}a Mu{\~n}oz}, {Gillon}, {Guedel}, {Guterman}, {Heng}, {Isaak}, {Lacedelli}, {Laskar}, {Lendl}, {Lovis}, {Luntzer}, {Magrin}, {Nascimbeni}, {Olofsson}, {Osborn}, {Ottensamer}, {Pagano}, {Pall{\'e}}, {Peter}, {Piazza}, {Piotto}, {Pollacco}, {Queloz}, {Ragazzoni}, {Rando}, {Rauer}, {Santos}, {Scandariato}, {S{\'e}gransan}, {Serrano}, {Sicilia}, {Simon}, {Smith}, {Sousa}, {Steller}, {Thomas}, {Udry}, {Van Grootel}, {Walton}, \& {Wilson}}]{2021A&A...654A.159S}
{Szab{\'o}}, G.~M., {Gandolfi}, D., {Brandeker}, A., {et~al.} 2021, \aap, 654, A159

\bibitem[{{Telting} {et~al.}(2014){Telting}, {Avila}, {Buchhave}, {Frandsen}, {Gandolfi}, {Lindberg}, {Stempels}, {Prins}, \& {NOT staff}}]{Telting2014}
{Telting}, J.~H., {Avila}, G., {Buchhave}, L., {et~al.} 2014, Astronomische Nachrichten, 335, 41

\bibitem[{{Thiabaud} {et~al.}(2014){Thiabaud}, {Marboeuf}, {Alibert}, {Cabral}, {Leya}, \& {Mezger}}]{Thiabaud2014}
{Thiabaud}, A., {Marboeuf}, U., {Alibert}, Y., {et~al.} 2014, \aap, 562, A27

\bibitem[{{Thiabaud} {et~al.}(2015){Thiabaud}, {Marboeuf}, {Alibert}, {Leya}, \& {Mezger}}]{Thiabaud2015}
{Thiabaud}, A., {Marboeuf}, U., {Alibert}, Y., {Leya}, I., \& {Mezger}, K. 2015, \aap, 574, A138

\bibitem[{{Tu} {et~al.}(2015){Tu}, {Johnstone}, {G{\"u}del}, \& {Lammer}}]{tu2015}
{Tu}, L., {Johnstone}, C.~P., {G{\"u}del}, M., \& {Lammer}, H. 2015, \aap, 577, L3

\bibitem[{{Tuomi} {et~al.}(2014){Tuomi}, {Anglada-Escude}, {Jenkins}, \& {Jones}}]{Tuomi2014}
{Tuomi}, M., {Anglada-Escude}, G., {Jenkins}, J.~S., \& {Jones}, H. R.~A. 2014, arXiv e-prints, arXiv:1405.2016

\bibitem[{{Venturini} {et~al.}(2020){Venturini}, {Guilera}, {Haldemann}, {Ronco}, \& {Mordasini}}]{Venturini2020}
{Venturini}, J., {Guilera}, O.~M., {Haldemann}, J., {Ronco}, M.~P., \& {Mordasini}, C. 2020, \aap, 643, L1

\bibitem[{{Vogt} {et~al.}(1994){Vogt}, {Allen}, {Bigelow}, {Bresee}, {Brown}, {Cantrall}, {Conrad}, {Couture}, {Delaney}, {Epps}, {Hilyard}, {Hilyard}, {Horn}, {Jern}, {Kanto}, {Keane}, {Kibrick}, {Lewis}, {Osborne}, {Pardeilhan}, {Pfister}, {Ricketts}, {Robinson}, {Stover}, {Tucker}, {Ward}, \& {Wei}}]{Vogt1994}
{Vogt}, S.~S., {Allen}, S.~L., {Bigelow}, B.~C., {et~al.} 1994, in Society of Photo-Optical Instrumentation Engineers (SPIE) Conference Series, Vol. 2198, Instrumentation in Astronomy VIII, ed. D.~L. {Crawford} \& E.~R. {Craine}, 362

\bibitem[{{Wang} \& {Dai}(2019)}]{wang2019}
{Wang}, L. \& {Dai}, F. 2019, \apjl, 873, L1

\bibitem[{{Wright} {et~al.}(2010){Wright}, {Eisenhardt}, {Mainzer}, {Ressler}, {Cutri}, {Jarrett}, {Kirkpatrick}, {Padgett}, {McMillan}, {Skrutskie}, {Stanford}, {Cohen}, {Walker}, {Mather}, {Leisawitz}, {Gautier}, {McLean}, {Benford}, {Lonsdale}, {Blain}, {Mendez}, {Irace}, {Duval}, {Liu}, {Royer}, {Heinrichsen}, {Howard}, {Shannon}, {Kendall}, {Walsh}, {Larsen}, {Cardon}, {Schick}, {Schwalm}, {Abid}, {Fabinsky}, {Naes}, \& {Tsai}}]{Wright2010}
{Wright}, E.~L., {Eisenhardt}, P. R.~M., {Mainzer}, A.~K., {et~al.} 2010, \aj, 140, 1868

\end{thebibliography}

\begin{appendix}
\onecolumn
\section{Results of internal structure modelling}

\begin{figure}[h!]
	\centering
	\includegraphics[width=0.95\textwidth]{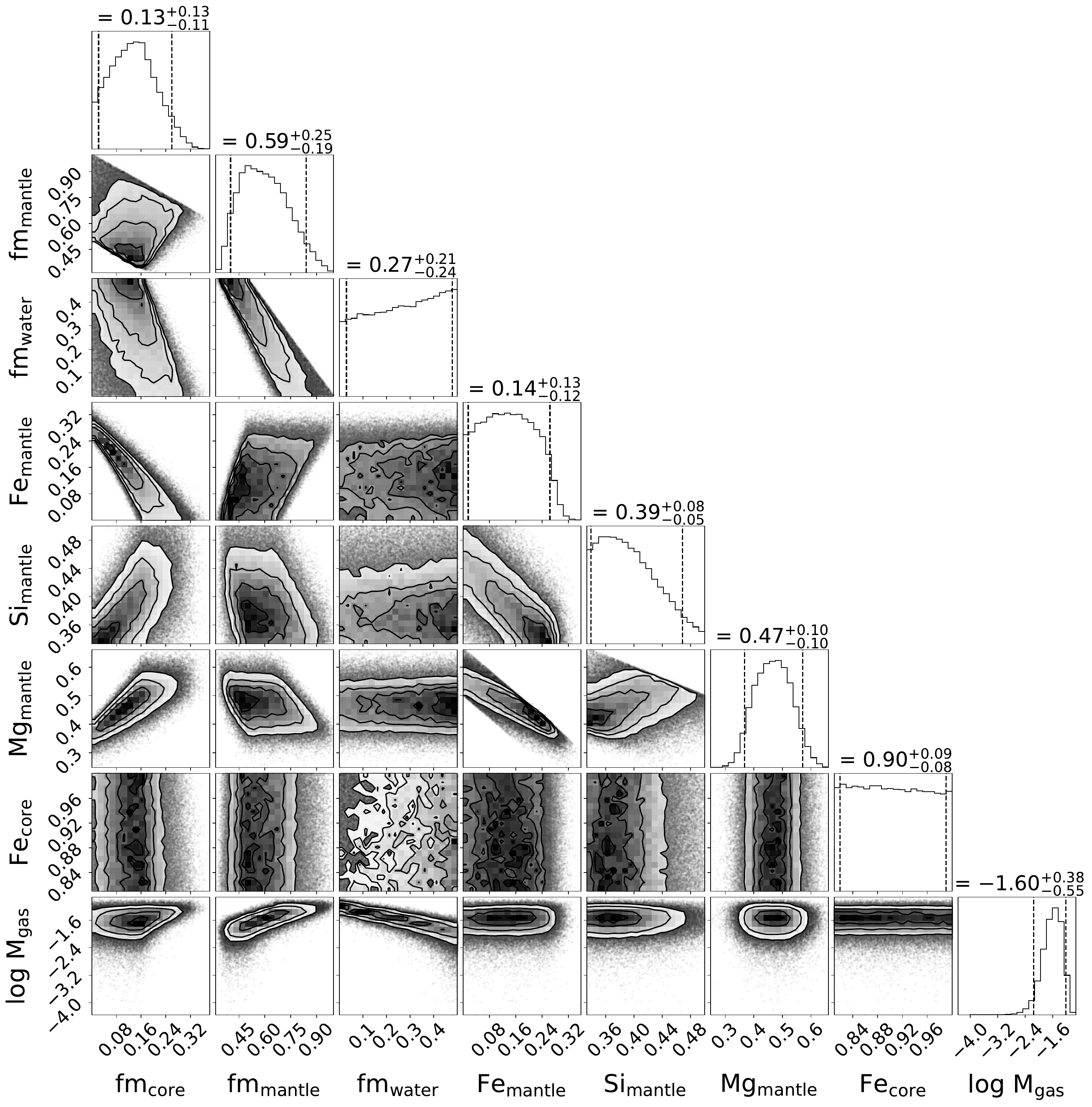}
	\caption{Posteriors of the internal structure parameters obtained for TOI-421\,b, namely the inner iron core, silicate mantle, and water layer mass fractions with respect to the condensed part of the planet without the H-He envelope, the molar fractions of Fe, Si, and Mg in the mantle and Fe in the inner core, and the logarithm of the mass of the H-He layer in Earth masses. Each column is labelled with the median of the corresponding posterior, with the 5$^\textrm{th}$ and 95$^\textrm{th}$ percentiles as uncertainties.}
	\label{internalstructure_b}
\end{figure}

\begin{figure}[h!]
	\centering
	\includegraphics[width=0.95\textwidth]{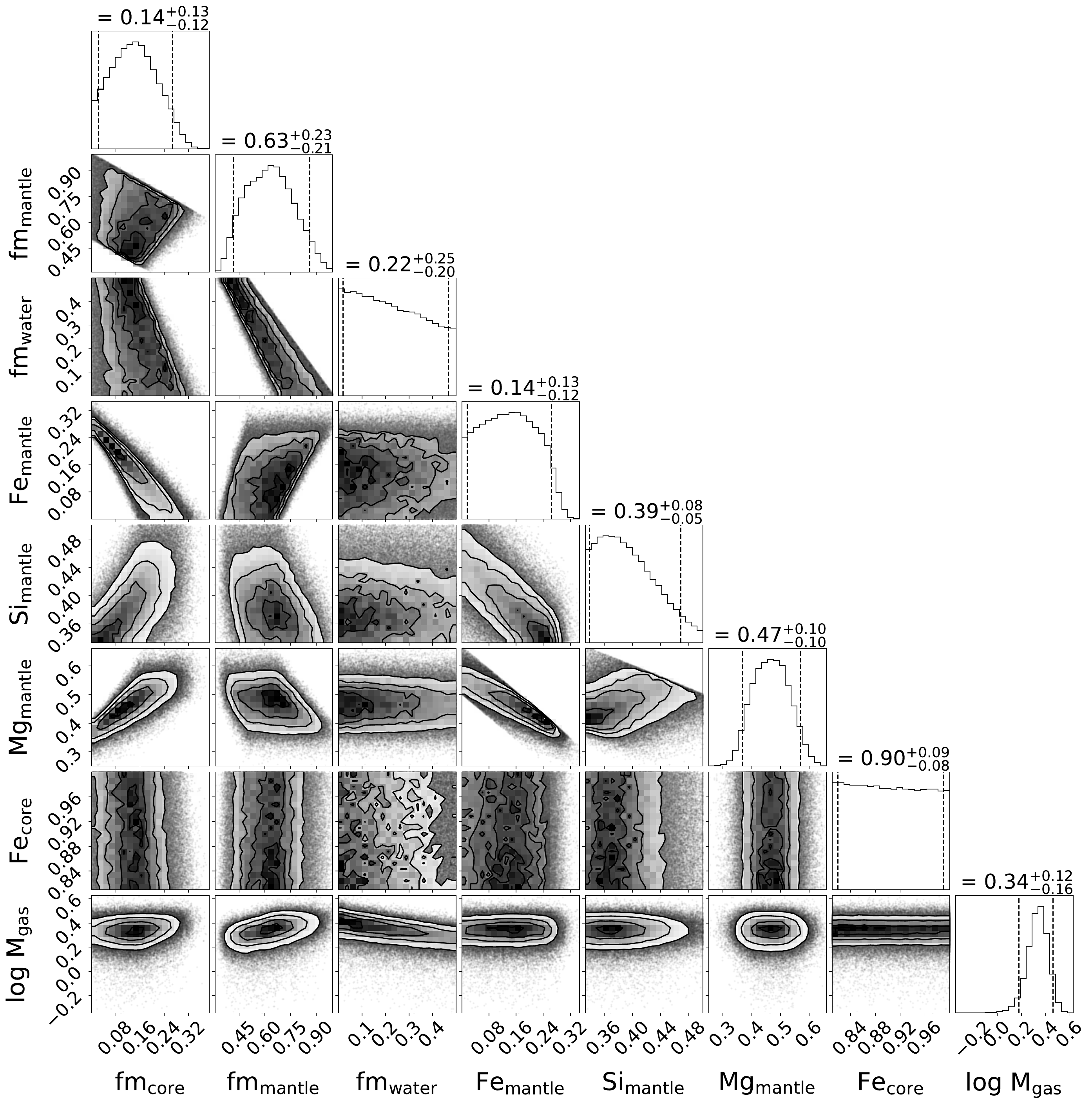}
	\caption{Same as Figure \ref{internalstructure_b}, but for TOI-421\,c.}
	\label{internalstructure_c}
\end{figure}


    
\end{appendix}

\end{document}